\documentclass[%draft,
reprint,superscriptaddress,amsmath,amssymb,aps,prb,floatfix]{revtex4-1}
\usepackage[margin=1.5cm]{geometry}
\usepackage{xcolor} %for colored text in comments
\usepackage{graphicx}% Include figure files
\usepackage{dcolumn}% Align table columns on decimal point
\usepackage{upgreek}% Allows text micron
\usepackage{bm}% bold math
\usepackage{hyperref}% add hypertext capabilities
\usepackage{floatrow}
\usepackage[mathlines]{lineno}% Enable numbering of text and display math
\newcommand*{\citen}[1]{%
	\begingroup
	\romannumeral-`\x % remove space at the beginning of \setcitestyle
	\setcitestyle{numbers}%
	\cite{#1}%
	\endgroup   
}

\begin{document}

\title{A Schr\"odinger Cat Living in Two Boxes}

\author{Chen Wang}
\email{chen.wang@yale.edu}
\author{Yvonne Y.~Gao}
\author{Philip Reinhold}
\author{R.~W.~Heeres}
\author{Nissim Ofek}
\author{Kevin Chou}
\author{Christopher Axline}
\author{Matthew Reagor}
\author{Jacob Blumoff}
\author{K.~M.~Sliwa}
\author{L.~Frunzio}
\author{S.~M.~Girvin}
\author{Liang Jiang}
\affiliation{Department of Applied Physics and Physics, Yale University, New Haven, Connecticut 06511, USA}
\author{M.~Mirrahimi}
\affiliation{Department of Applied Physics and Physics, Yale University, New Haven, Connecticut 06511, USA}
\affiliation{INRIA Paris-Rocquencourt, Domaine de Voluceau, B.~P.~105, 78153 Le Chesnay cedex, France}
\author{M.~H.~Devoret}
\author{R.~J.~Schoelkopf}
\email{robert.schoelkopf@yale.edu}
\affiliation{Department of Applied Physics and Physics, Yale University, New Haven, Connecticut 06511, USA}

\date{\today}

\begin{abstract}
%Quantum superposition of distinct coherent states, known as a ``cat state", has been an elegant demonstration of Schrodinger's famous cat paradox, abstracted in the world of a single-mode harmonic oscillator.  Here, we realize a two-mode cat state of microwave fields in two cavities bridged by a superconducting artificial atom.  This complex cat state extends in a two-cavity Hilbert space of more than 100 effective dimensions, and can also be interpreted as an entangled pair of single-mode cat states.  We present full quantum state tomography of this state via quantum non-demolition measurements of the joint photon number parity using dispersive interaction with three atomic levels.  The ability to manipulate multi-cavity quantum states paves the way for logical operations between redundantly encoded qubits for fault-tolerant quantum computation and communication.
Quantum superpositions of distinct coherent states in a single-mode harmonic oscillator, known as ``cat states", have been an elegant demonstration of Schr\"odinger's famous cat paradox.  Here, we realize a two-mode cat state of electromagnetic fields in two microwave cavities bridged by a superconducting artificial atom, which can also be viewed as an entangled pair of single-cavity cat states.  We present full quantum state tomography of this complex cat state over a Hilbert space exceeding 100 dimensions via quantum non-demolition measurements of the joint photon number parity. %using dispersive interaction with three atomic levels.  This complex cat state exhibits strong Wigner negativity and non-classical correlations between two distinct pairs of coherent states.  
The ability to manipulate such multi-cavity quantum states paves the way for logical operations between redundantly encoded qubits for fault-tolerant quantum computation and communication.

%Alternative abstract? 
%Quantum superpositions of coherent states, known as 'cat states', have been shown to be a valuable resource in quantum information. In particular, the creation and manipulation of such states in single superconducting resonator has shown huge potential... Here, we present a two-mode cat state housed in two spatially separated cavities bridged by a transmon qubit, exponentially extending the accessible/available Hilbert space up to 256. Furthermore, such a state exhibits strong non-classical correlations, which can be viewed as (is a signature of) entanglement between two distinct photonic cat states (two modes containing superpositions of  distinct coherent fields/states). Utilizing both the g-e and e-f level transitions of the transmon qubit, we can fully characterize both the complex cat state as well as the individual modes via high fidelity measurements of the joint photon number parity. The ability to manipulate multi-cavity quantum states paves the way for logical operations between redundantly encoded qubits for fault-tolerant quantum computation and communication.
\end{abstract}

\maketitle

%\section{Introduction}

Rapid progress in controlling individual quantum systems over the past twenty years~\cite{haroche_nobel_2013,wineland_nobel_2013} has opened a wide range of possibilities of quantum information processing.  Potential applications ranging from universal quantum computation to long-distance quantum communication share the central theme of exploiting quantum superpositions within a large Hilbert space.  Further stimulated by curiosity about the quantum-classical boundary, there has been growing interest in generating superpositions of ``macroscopically-distinguishable" states that are far apart in phase space.  The canonical example is superpositions of coherent states of a harmonic oscillator, \textit{i.e.}~$\mathcal{N}(|\alpha\rangle+|-\alpha\rangle)$ with $\mathcal{N}\approx 1/\sqrt2$ at large $|\alpha|$, known as ``cat states".  The two components correspond to distinct quasi-classical wave-packets% separated by a distance of $4|\alpha|^2$
, in analogy to Schr\"odinger's \textit{gedankenexperiment} of an unfortunate cat inside a closed box being simultaneously dead and alive.  Cat states have so far been realized with single-mode optical~\cite{ourjoumtsev_generation_2007} or microwave fields~\cite{haroche_nobel_2013, brune_observing_1996} with up to about 100 photons~\cite{vlastakis_deterministically_2013}, but are increasingly susceptible to decoherence at large size.

Manipulating a large number of excitations in such harmonic oscillator states is one of two possible approaches to expand the information capacity of fully-controlled quantum systems.  Cat states, which span a Hilbert space whose dimension grows linearly with the number of photons, are an attractive approach for redundantly encoding quantum information for error correction~\cite{chuang_bosonic_1997, mirrahimi_dynamically_2014,ofek_demonstrating_2016}.  The other more traditional way to scale up a quantum system is to build many modes of excitations, each operated as a two-level qubit, so that the Hilbert space dimension increases exponentially with the number of modes~\cite{monz_14-qubit_2011,kelly_state_2015}.  Is it possible to combine the benefits of both approaches by creating a cat state that lives in more than a single mode or box?  The idea of non-local or multi-mode cat dates back to the early days of cavity quantum electrodynamics (QED)%, where a Rydberg atom was envisioned to sequentially traverse two cavities, creating a non-local cat composes of coherent states of both cavities
~\cite{davidovich_quantum_1993}, but experimental demonstration has remained a formidable challenge.

Here, we deterministically create a two-mode cat state of microwave fields in two superconducting cavities, using the strong dispersive interaction with a Josephson-junction-based artificial atom.  This state can be expressed as:
\begin{equation}
|\psi_{\pm}\rangle = \mathcal{N}\big(|\alpha\rangle_A|\alpha\rangle_B\pm|-\alpha\rangle_A|-\alpha\rangle_B\big)
\end{equation}
where $|\pm\alpha\rangle_A$ and $|\pm\alpha\rangle_B$ are coherent states of two  microwave eigenmodes (Alice and Bob) at different frequencies, whose amplitudes are prepared equal for simplicity.  Each of the two modes are predominantly localized in one of the two cavities that are weakly connected.  Despite a nonzero (but small) spatial overlap of the two modes, we will refer for
convenience to the state of each mode as the state of each cavity. 
Quantum superpositions of the form $|\psi_{\pm}\rangle$ have been previously realized in the optical domain~\cite{ourjoumtsev_preparation_2009} but were limited to small and non-orthogonal coherent states ($|\alpha|^2=0.65$).  For larger $|\alpha|$ (\textit{i.e.}~$|\alpha|^2\gtrsim2$), $|\psi_{\pm}\rangle$ can be considered a single cat state living in two boxes whose superposed components are coherent states in a hybridized mode involving both Alice and Bob.
%wavefunction overlap between the two superposed components vanishes exponentially, making $|\psi_{\pm}\rangle$ a cat state whose life or death is reflected simultaneously in two distinct symptoms.  Such interpretation of a single cat living in two boxes stems from the fact that $|\alpha\rangle_A|\alpha\rangle_B$ can be synthesized into a complex coherent state via a change of mode basis.  
Alternatively, in the more natural eigenmode basis, $|\psi_{\pm}\rangle$ has been known as the entangled coherent states in theoretical studies~\cite{sanders_review_2012}, and may also be understood as two single-cavity cat states that are entangled with each other.

The two-mode cat state is an eigenstate of the joint photon number parity operator $P_J$:
\begin{equation}
P_J=P_A P_B=e^{i\pi a^{\dagger}a}e^{i\pi b^{\dagger}b}
\end{equation}
where $a(a^{\dagger})$ and $b(b^{\dagger})$ are the annihilation (creation) operators of photons in Alice and Bob, and $P_A$ and $P_B$ are the photon number parity operators in individual cavities.  Remarkably, $|\psi_{+}\rangle$ (or $|\psi_{-}\rangle$) has definitively an even (or odd) number of photons in the two cavities combined, while the photon number parity in each cavity separately is maximally uncertain.  Quantum non-demolition measurements of such parity operators not only illustrate the highly non-classical properties of the state, but also are instrumental for quantum error correction in general.

\begin{figure*}
\floatbox[{\capbeside\thisfloatsetup{capbesideposition={right,center},capbesidewidth=6.9cm}}]{figure}[\FBwidth]
{\caption{\textbf{Sketch (not to scale) of device architecture and experimental protocol.} \textbf{(A)} A three dimensional schematic of the device consisting of two coaxial cavities (Alice and Bob), a Y-shaped transmon with a single Josephson junction (marked by ``$\times$"), and a stripline readout resonator.  All components are housed inside a single piece of bulk high-purity aluminum, with artificial windows drawn for illustration purposes. \textbf{(B)} A top view of the same device, showing the relative position of the sapphire chip, center posts of the coaxial cavities, transmon antenna, and the readout resonator.  \textbf{(C)} The microwave control sequences for generating the two-mode cat state and performing Wigner tomography.  $D_{\beta}$ represents cavity displacement by $\beta$, and a superscript $g$ is added if the displacement is conditional on the ancilla being in $|g\rangle$.  $R^{ge}_\theta$ or $R^{ef}_\theta$ represents ancilla rotation by angle $\theta$ (around an axis in the X-Y plane) in the $|g\rangle$-$|e\rangle$ Bloch sphere or $|e\rangle$-$|f\rangle$ Bloch sphere.  $R^{00}_\pi$ is an ancilla $|g\rangle$-$|e\rangle$ rotation conditional on the cavities being in $|0\rangle_A|0\rangle_B$.  $C_{\phi}$ represents cavity phase shift of $\phi$ conditional on the ancilla being in an excited state.  By choosing $\phi_i+\phi'_i=\pi$ or $2\pi$, we can measure photon number parity of Alice ($P_A$), Bob ($P_B$), or the two combined ($P_J$), to perform Wigner tomography of individual cavities or the joint Wigner tomography.}} 
{\includegraphics[width=11cm]{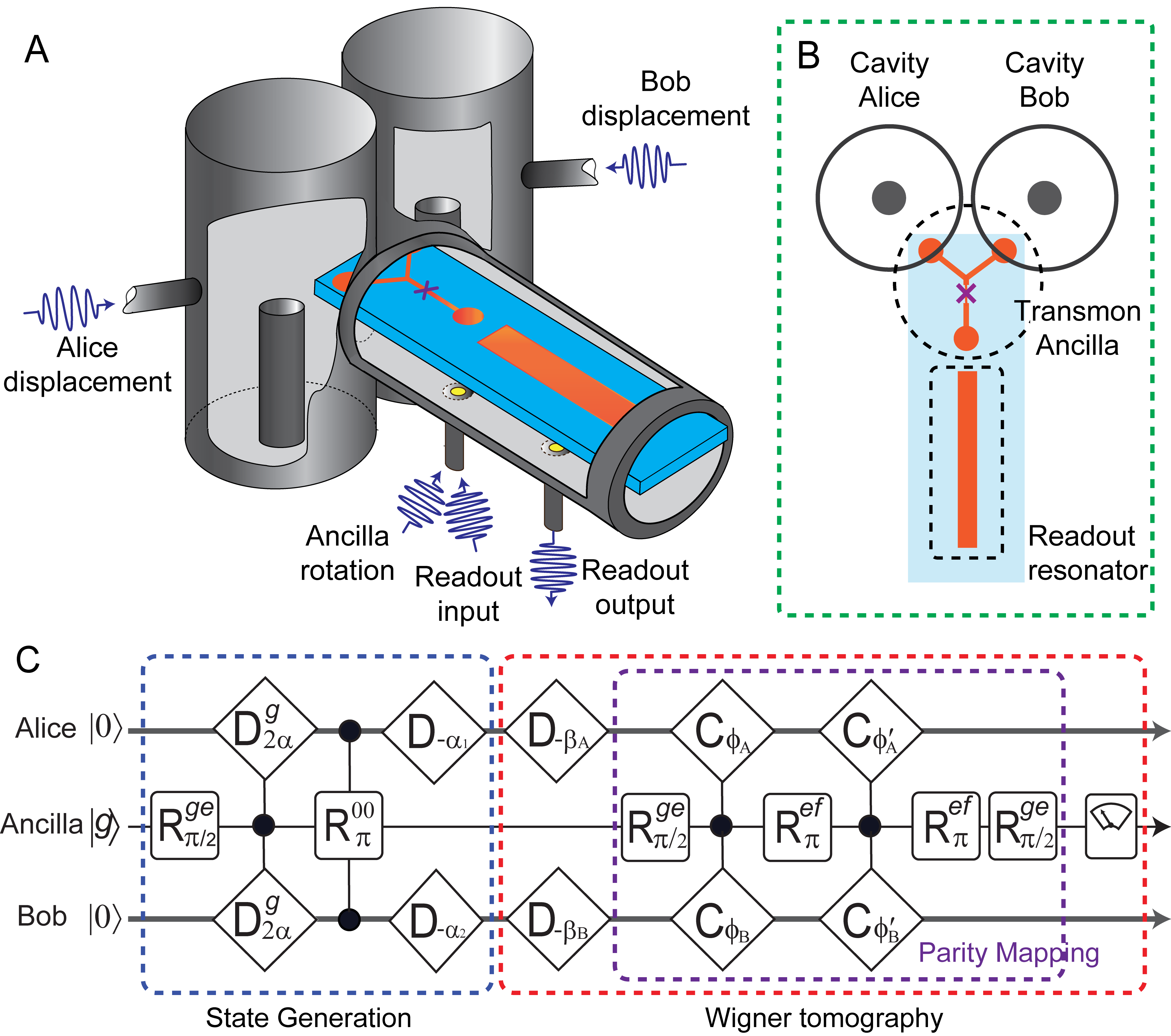}}
\end{figure*}

We realize measurements of the joint photon number parity and single-mode parities using the dispersive interaction with three energy levels of an artificial atom.  Based on joint parity measurements, we further demonstrate full quantum state tomography of the two-cavity system~\cite{milman_proposal_2005}.  %spanning a Hilbert space of more than 100 dimensions.  
This is obtained in the form of the joint Wigner function $W_J(\beta_A,\beta_B)$, which is a continuous-variable representation of the quantum state with $\beta_A$ and $\beta_B$ being complex variables in Alice and Bob respectively.  %The joint parity measurement provides direct single-shot sampling of $W_J$ in the phase space.  
%With long coherence times and near quantum-limited amplification, 
Without correcting for the infidelity of the joint parity measurement operator, we observe quantum state fidelity %(without adjusting for measurement visibility) 
of 81\% for a two-mode cat state with $\alpha=1.92$.  The high-quality and high-dimensional quantum control is further manifested by the presence of entanglement exceeding classical bounds in a CHSH-style inequality for two continuous-variable systems~\cite{milman_proposal_2005}.  Finally, our two-cavity space effectively encodes two coupled logical qubits in the coherent state basis, and we present efficient two-qubit tomography in this encoded space. %, an important step towards continuous-variable quantum computation.

%\section{Experimental Architecture}

Our experimental setup uses a three-dimensional (3D) circuit QED architecture~\cite{paik_observation_2011}, where two high-Q 3D cavities and a quasi-planar readout resonator simultaneously couple to a fixed-frequency transmon-type superconducting qubit (Fig.~1A,B)~\cite{see_supplementary_????}. The two cavities that host the cat state of microwave photons are implementations of the longest-lived quantum memory in circuit QED to date~\cite{reagor_quantum_2015}.  The transmon, while usually considered a qubit, behaves as an artificial atom with multiple energy levels.  We use the transmon as an ancilla to manipulate the multi-photon states in the two cavities, and its lowest three levels, $|g\rangle$, $|e\rangle$ and $|f\rangle$, are accessed in this experiment.  The device is cooled down to 20 mK in a dilution refrigerator, and microwave transmission through the readout resonator is used to projectively measure the ancilla state with a heterodyne detection at room temperature after multiple stages of amplification.

\begin{table}[b]
\caption{Hamiltonian parameters and coherence times of the two storage cavities and the transmon ancilla, including transition frequencies ($\omega/2\pi$), dispersive shifts between each cavity and each transmon transition ($\chi$), energy relaxation time ($T_1$), and Ramsey decoherence time ($T^{*}_2$). The cavity frequencies are given with a precision of $\pm100$ Hz and are stable over the course of several months.}
\centering  
\begin{tabular}{c c c c c} % centered columns (4 columns)
\hline\hline\\[-2ex]
	&		& $\omega/2\pi$	& $T_1$		& $T_2^*$\\
\hline\\[-2ex]
Cavities:	& Alice	& \,4.2196612 GHz\,	& \,2.2-3.3 ms\,	& \,0.8-1.1 ms\,\\
			&	Bob		& 5.4467679 GHz	& 1.2-1.7 ms	& 0.6-0.8 ms\\
Transmon:	& $|e\rangle\rightarrow|g\rangle$	& 4.87805 GHz	& 65-75 $\mu$s	& 30-45 $\mu$s\\
(Ancilla)	& $|f\rangle\rightarrow|e\rangle$	& 4.76288 GHz	& 28-32 $\mu$s	& 12-24 $\mu$s\\[0.5ex]
% [1ex] adds vertical space
\hline
\\
\end{tabular}
\begin{tabular}{c c c} % centered columns (3 columns)
\hline\hline\\[-2ex]
$\chi/2\pi$		& Alice	& Bob\\
\hline\\[-2ex]
\,$\chi^{ge}$\,	&\,0.71 MHz\,	&\,1.41 MHz\,\\
$\chi^{ef}$		& 1.54 MHz		& 0.93 MHz\\
%$|f\rangle$	vs.~$|g\rangle$		& 2.25 MHz		& 2.34 MHz\\[0.5ex]
\hline 
\end{tabular}
\end{table}

We consider the Hamiltonian of the system including two harmonic cavity modes, a three-level atom, and their dispersive interaction (with parameters listed in Table I):
\begin{align}
H/\hbar = & \omega_{A}a^{\dagger}a+\omega_{B}b^{\dagger}b +\omega_{ge}|e\rangle\langle e|+(\omega_{ge}+\omega_{ef})|f\rangle\langle f|\nonumber\\
		& -\chi_A^{ge} a^{\dagger}a|e\rangle\langle e|-(\chi_A^{ge}+\chi_A^{ef}) a^{\dagger}a|f\rangle\langle f|\nonumber\\
        & -\chi_B^{ge} b^{\dagger}b|e\rangle\langle e|-(\chi_B^{ge}+\chi_B^{ef}) b^{\dagger}b|f\rangle\langle f|
\end{align}
where $\omega_A$ and $\omega_B$ are the angular frequencies of the two cavities (Alice and Bob), $\omega_{ge}$ and $\omega_{ef}$ are the $|e\rangle\rightarrow|g\rangle$ and $|f\rangle\rightarrow|e\rangle$ transition frequencies of the ancilla, $\chi_i^{ge}$ and $\chi_i^{ef}$ ($i$ = $A$ or $B$) represent the dispersive frequency shifts of cavity $i$ associated with the two ancilla transitions.  The readout resonator and small high-order nonlinearities are neglected for simplicity~\cite{}.  Using time-dependent external classical drives in the form of microwave pulses% (not included in Eq.~(3))
, we can perform arbitrary ancilla rotations in both $|g\rangle$-$|e\rangle$ and $|e\rangle$-$|f\rangle$ manifolds, and arbitrary cavity state displacements in Alice ($D_{\beta_A} = e^{\beta_A a^\dagger-\beta_A^{*}a}$) and Bob ($D_{\beta_B} = e^{\beta_B b^\dagger-\beta_B^{*}b}$) independently.  More importantly, the state-dependent frequency shifts ($\chi$'s) allow cavity state manipulations conditioned on the ancilla level or vice versa using spectrally-selective control pulses, thus realizing atom-photon quantum logic gates~\cite{vlastakis_deterministically_2013}.  It can be further shown that with separate drives on the two cavities and a drive on the ancilla, this Hamiltonian permits universal quantum control of the entire system~\cite{michael_new_2016}.

We generate the two-mode cat state $|\psi_{\pm}\rangle$ deterministically using a series of logic gates as shown in Fig.~1C~\cite{leghtas_deterministic_2013}. %are the key tools for deterministic generation of the two-mode cat state $|\psi_{\pm}\rangle$.  This is realized by transferring the quantum superposition in a prepared ancilla state into cavity coherent-state basis
 %The mutually-dependent frequencies allow cavity state manipulations conditioned on ancilla level or vice versa using spectrally-selective microwave pulses, thus realizing atom-photon quantum logic gates~\cite{vlastakis_deterministically_2013}. 
In particular, we implement effective displacements ($D^g_{2\alpha}$) of both Alice and Bob conditional on ancilla being in $|g\rangle$~\cite{see_supplementary_????}, which realizes a three-way entangling gate, $\frac{1}{\sqrt2}\big(|g\rangle+|e\rangle\big)|0\rangle_A|0\rangle_B\rightarrow \mathcal{N}\big(|g\rangle|0\rangle_A|0\rangle_B+|e\rangle|2\alpha\rangle_A|2\alpha\rangle_B\big)$.  Then an ancilla rotation ($R^{00}_{\pi}$) conditional on the cavity state $|0\rangle_A|0\rangle_B$ disentangles the ancilla, and subsequent cavity displacements leave the cavities in a two-mode cat state.  The rotation axis controls the sign (or more generally, phase angle) of the cat state superposition.

\begin{figure}[tbp]
    \centering
    \includegraphics[width=5.5cm]{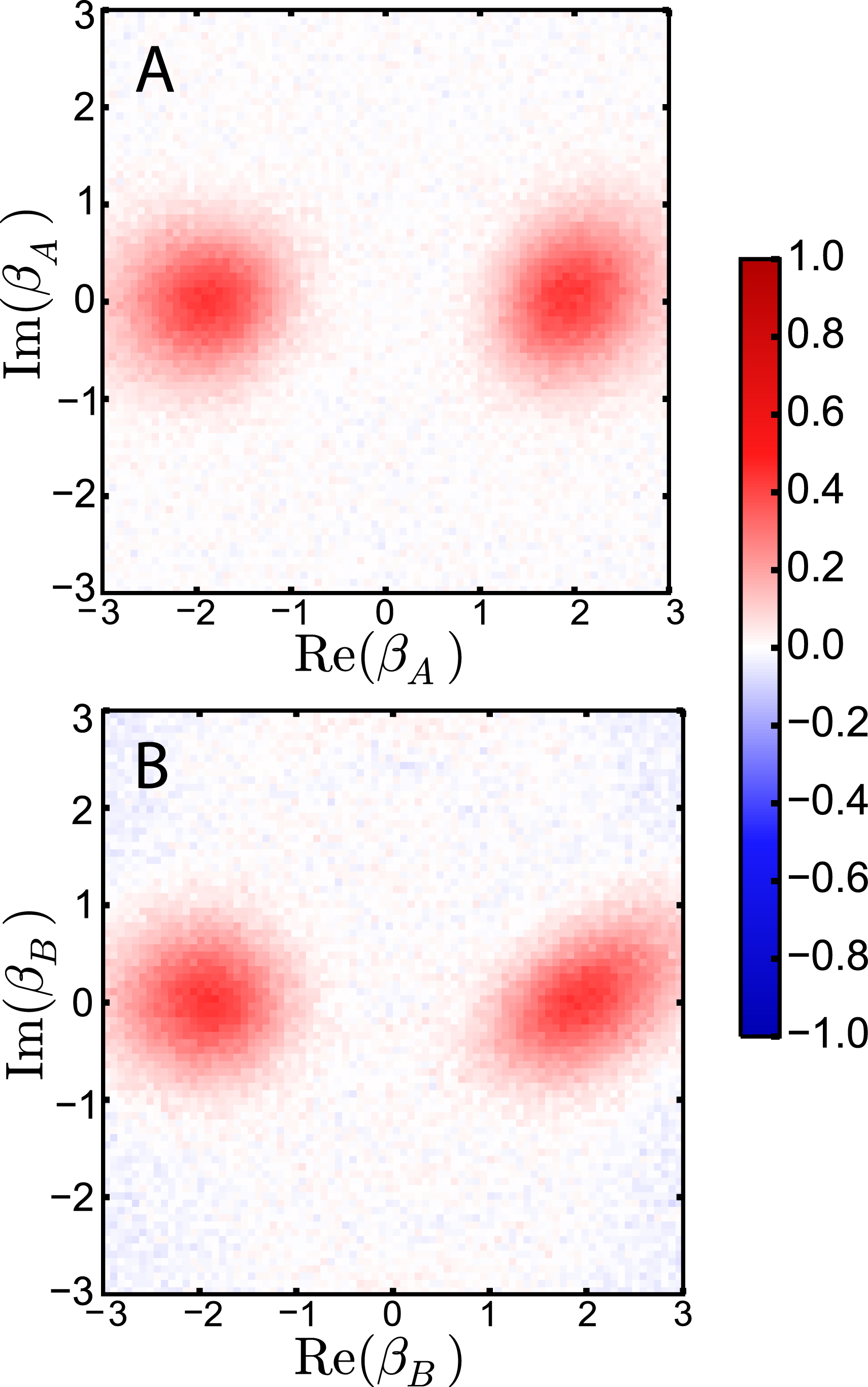}
    \caption{\textbf{Wigner tomography of individual cavities.}  Measured single-cavity scaled Wigner function of \textbf{(A)} Alice ($\frac{\pi}{2}W_A(\beta_A)$) and \textbf{(B)} Bob ($\frac{\pi}{2}W_B(\beta_B)$) respectively for the two-mode cats state $|\psi_{-}\rangle$, each plotted in the complex plane of Re$(\beta_i)$ and Im$(\beta_i)$ ($i$=A or B).  For either cavity, no interference fringes are observed in its Wigner function, indicating a statistical mixture of two coherent states, as opposed to a single-cavity cat state, after tracing out the quantum state of the other cavity.  The distortion of the coherent states is due to higher order Hamiltonian terms (see supplementary).  The photon number parity within each cavity is close to 0, reflected by the value of respective Wigner functions near the origin.}
\end{figure}

We probe the cat state by quantum non-demolition (QND) measurements of the photon number parity.  Parity measurement of a single cavity using a dispersively coupled ancilla qubit has been previously demonstrated \cite{bertet_direct_2002, sun_tracking_2014}, where a conditional cavity phase shift~\cite{brune_observing_1996}, $C_{\phi}=\mathbb{I}\otimes|g\rangle\langle g| + e^{i\phi a^{\dagger}a}\otimes|e\rangle\langle e|$, of $\phi=\pi$ allows the cavity states with even or odd photon numbers to be mapped to $|g\rangle$ or $|e\rangle$ of the qubit for subsequent readout.  In our multi-cavity architecture, measuring the joint photon number parity requires $C_{\pi}$ in both Alice and Bob, which is difficult to achieve simultaneously with existing techniques~\cite{sun_tracking_2014} unless $\chi^{ge}_A=\chi^{ge}_B$.
%This method can in principle be implemented here to measure the joint photon number parity if $\chi_A^{ge}=\chi_B^{ge}$ so that photons in the two cavities are indistinguishable from the ancilla's perspective.
We overcome this challenge by exploiting the $|f\rangle$-level of the transmon. By designing the frequency of the ancilla to be between those of the two cavities, the $|e\rangle\rightarrow|g\rangle$ transition shows stronger interaction with Bob ($\chi_B^{ge}>\chi_A^{ge}$), while the $|f\rangle\rightarrow|e\rangle$ transition shows stronger interaction with Alice ($\chi_A^{ef}>\chi_B^{ef}$).   Manipulating the ancilla in different superposition states among the three levels allows us to concatenate conditional phase gates associated with $\chi_i^{ge}$ and $\chi_i^{ef}$ with arbitrary weights~\cite{see_supplementary_????}.  This additional degree of freedom not only allows for joint parity measurement $P_J$ (applying $C_\pi$ to both cavities), but also enables parity measurement of each cavity $P_A$ or $P_B$ individually without affecting the other (applying $C_\pi$ and $C_{2\pi}$ to the two cavities respectively).  %All three parity measurements are close to quantum non-demolition (QND), leaving the cavity state in the respective parity eigenstate. 

%\section{Wigner Tomography}

Based on single-cavity parity measurements, we can measure the Wigner function of individual cavities, $W_i(\beta_i)=\frac{2}{\pi}Tr\big[\rho D_{\beta_i} P_i D^{\dagger}_{\beta_i}\big]$ ($i$=A or B)~\cite{bertet_direct_2002,deleglise_reconstruction_2008}.  The Wigner function is a standard method to fully determine the quantum state of a single-continuous-variable system, which represents the quasi-probability distribution of photons in the quadrature space (Re$(\beta)$-Im$(\beta)$). Our measured $W_A$ and $W_B$ for a two-mode cat state $|\psi_{-}\rangle$ with $\alpha=1.92$ (Fig.~2) illustrates that the quantum state of either Alice or Bob on its own is a statistical mixture of two clearly-separated coherent states with no coherence between them.  In other words, each cavity does not contain a regular (single-mode) cat state% of the form $\frac{1}{\sqrt2}\big(|\beta\rangle\pm|-\beta\rangle\big)$
, which would contain characteristic interference fringes in the Wigner function~\cite{deleglise_reconstruction_2008, vlastakis_deterministically_2013} (and can also be straightforwardly generated in our experiment~\cite{see_supplementary_????}).  However, for a state involving inter-cavity entanglement like $|\psi_{-}\rangle$, single-cavity Wigner functions are insufficient for characterizing the global quantum state.  Such entanglement can be inferred from measurement of the joint photon number parity, $\langle P_J\rangle=-0.81\pm0.01$, even though each cavity alone shows mean photon number parity of $\langle P_A\rangle\approx\langle P_B\rangle\approx0$.  %This confirms there are almost always odd number of photons in two cavities combined despite uncertain parity in each cavity. 
Additional evidence of the joint parity can be seen in a spectroscopy measurement~\cite{see_supplementary_????}.

%the appearance of a mixed state in each cavity can be merely a result of tracing out the quantum state of the other cavity. Indeed we expect this is the case for an entangled state like $|\psi_{-}\rangle$.  A strong signature of the entanglement can be inferred from measuring parity correlations.  Although both cavities show mean photon number parity of $\langle P_a\rangle\approx\langle P_b\rangle\approx0$ (as shown at the origin in Fig.~2A or Fig.~2B), by sequentially measuring $P_a$ and $P_b$ for the prepared state we find $\langle P_a P_b\rangle=-0.??\pm 0.01$ \textit{(to be updated)}.  A more powerful capability in our experiment is single-shot measurement of the joint parity, which avoids the decoherence and the Kerr effect (due to higher-order terms in Hamiltonian) incurred during the time span of sequential measurements, showing $\langle P_j\rangle=-0.81\pm0.01$.  Both methods confirm there are almost always odd number of photons in two cavities combined despite uncertain parity in each cavity.  An additional evidence of the joint photon number parity can be seen in a spectroscopy measurement shown in the Supplementary Material. 

\begin{figure*}[tbp]
\floatbox[{\capbeside\thisfloatsetup{capbesideposition={right,center},capbesidewidth=6cm}}]{figure}[\FBwidth]
{\caption{\textbf{Joint Wigner tomography.}  \textbf{(A, B)} Two-dimensional plane-cut along (A) axes Re$(\beta_A)$-Re$(\beta_B)$ and (B) axes Im$(\beta_A)$-Im$(\beta_B)$ of the calculated 4D scaled joint Wigner function $\langle P_J(\beta_A,\beta_B)\rangle$ of the ideal odd-parity two-mode cat state $|\psi_{-}\rangle$ with $\alpha=1.92$.  The red features in (A) represent the probability distribution of the two coherent states components.  The central blue feature in (A) and fringes in (B) demonstrate quantum interference between the two components. \textbf{(C, D)} The corresponding Re$(\beta_A)$-Re$(\beta_B)$ and Im$(\beta_A)$-Im$(\beta_B)$ plane-cuts of the measured joint Wigner function of $|\psi_{-}\rangle$, to be compared with the ideal results in (A) and (B) respectively.  Data are taken in a 81$\times$81 grid, where every point represents an average of about 2000 binary joint parity measurements. \textbf{(E)} Diagonal line-cuts of the data shown in (A) and (C), corresponding to 1D plots of the calculated (black) and measured (red) scaled joint Wigner function along Re$(\beta_A)$ = Re$(\beta_B)$ with Im$(\beta_A)=$ Im$(\beta_B)=0$.  \textbf{(F)} Diagonal line-cuts of the data shown in (B) and (D), corresponding to 1D plots of the calculated (black) and measured (red) scaled joint Wigner function along Im$(\beta_A)=$ Im$(\beta_B)$ with Re$(\beta_A)=$ Re$(\beta_B)=0$.}}
{\includegraphics[width=10.8cm]{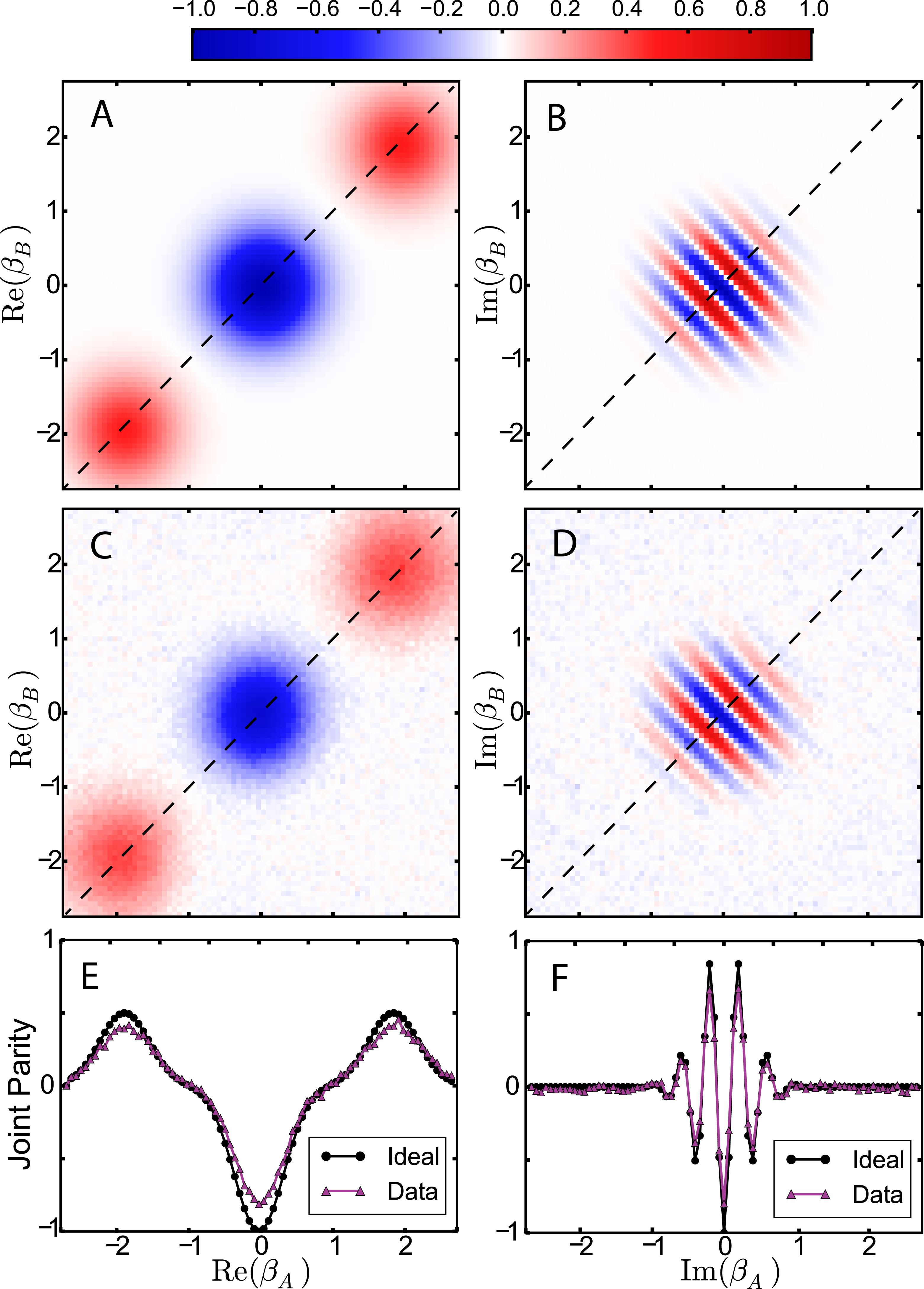}}
\end{figure*}

A full quantum state tomography of the two-cavity system can be realized by measuring the joint Wigner function~\cite{cahill_density_1969}:
\begin{align}
W_J(\beta_A,\beta_B)&=\frac{4}{\pi^2}\langle P_J(\beta_A,\beta_B)\rangle\nonumber\\
&=\frac{4}{\pi^2}Tr\big[\rho D_{\beta_A}D_{\beta_B}P_J D^{\dagger}_{\beta_B}D^{\dagger}_{\beta_A}\big]
\end{align}
$W_J$ is a function in the four-dimensional (4D) phase space, whose value at each point (Re$(\beta_A)$, Im$(\beta_A)$, Re$(\beta_B)$, Im$(\beta_B)$), after rescaling by $\pi^2/4$, is equal to the expectation value of the joint parity after independent displacements in Alice and Bob~\cite{milman_proposal_2005}.  For simplicity, we will therefore use the scaled joint Wigner function, or ``displaced joint parity" $\langle P_J(\beta_A,\beta_B)\rangle$ to represent the cavity state. $\langle P_J\rangle$ at any given point $(\beta_A,\beta_B)$ is directly measured by averaging single-shot readout outcomes and takes values between -1 and +1.  To illustrate the core features in this 4D Wigner function of the state $|\psi_{-}\rangle$, we show its two-dimensional (2D) cuts along the Re$(\beta_A)$-Re$(\beta_B)$ plane and Im$(\beta_A)$-Im$(\beta_B)$ plane for both the calculated ideal state (Fig.~3A, B, also see Ref.~\citen{milman_proposal_2005}) and the measured data (Fig.~3C, D).  The Wigner function contains two positively-valued Gaussian hyperspheres representing the probability distribution of the two coherent-state components, and an interference structure around the origin with strong negativity. 
%The Wigner function of the ideal state $|\psi_-\rangle$ contains three non-zero regions.  Two hyper-spheres, representing the phase-space locations of the two coherent-state components ($|\beta_a\rangle|\beta_b\rangle$ and $|-\beta_a\rangle|-\beta_b\rangle$, the red circles in Fig.~3A), %are centered at $(\alpha_a=\beta_a, \alpha_b=\beta_b)$ and $(\alpha_a=-\beta_a, \alpha_b=-\beta_b)$ respectively, have maxima of $\langle P_j\rangle=0.5$ at their respective centers and decay in Gaussian functions in all four dimensions.  An interference structure emerges at the origin with minima of $\langle P_j\rangle=-1$, which features an oscillation along the diagonal axis $Im(\alpha_a)=Im(\alpha_b)$, and has Gaussian envelopes in all dimensions (the blue circle in Fig.~3A and fringes in Fig.~3B).  
%Our measured raw data of the Wigner function 2D-cuts (Fig.~3C, D, with $\beta_a=\beta_b=1.92$) is in 
Excellent agreement is achieved between measurement and theory, with the raw data showing an overall 81\% contrast of the ideal Wigner function. %without normalizing against measurement infidelity. 
Comprehensive measurements of $\langle P_J\rangle$ in the entire 4D parameter space further allow us to reconstruct the density matrix of the quantum state, which shows a total fidelity of also about 81\% against the ideal $|\psi_-\rangle$ state.  The actual state fidelity may be significantly higher if various errors associated with tomography are removed~\cite{see_supplementary_????}. Additional visualization of the Wigner function data is presented in a Supplementary movie~\cite{see_supplementary_????}. 

\begin{figure*}
\floatbox[{\capbeside\thisfloatsetup{capbesideposition={right,center},capbesidewidth=6cm}}]{figure}[\FBwidth]
{\caption{\textbf{Encoded two-qubit tomography.} \textbf{(A)} Red bars show tomography of two logical qubits encoded in the coherent state basis of two cavities, with the prepared state being an even-parity two-mode cat state, $|\psi_{+}\rangle$ with $\alpha=1.92$.  Gray bars represent the ideal state.  Insets show the Re$(\beta_A)$-Re$(\beta_B)$ and Im$(\beta_A)$-Im$(\beta_B)$ plane-cuts of the measured scaled joint Wigner function of the same state.  The measured identity operator differs from 1 as a result of the parity measurement infidelity and leakage out of the code space.  \textbf{(B)} Encoded two-qubit tomography of an approximate product state of single-cavity cat states, $\mathcal{N'}\big(|\alpha\rangle-|-\alpha\rangle\big)_A\otimes\big(|\alpha\rangle-|-\alpha\rangle\big)_B$ also with $\alpha=1.92$.  Insets show plane-cuts of the measured scaled joint Wigner function of the same state.  Given familiarity with single-cavity cat states, these Wigner function patterns can be understood by considering $W_J(\beta_A,\beta_B)=W_A(\beta_A)W_B(\beta_B)$ for separable states.  For example, the ``checkerboard" patterns in the Im$(\beta_A)$-Im$(\beta_B)$ plane-cuts can be understood by multiplying orthogonal fringes from the two independent cat states.
}}
{\includegraphics[width=10cm]{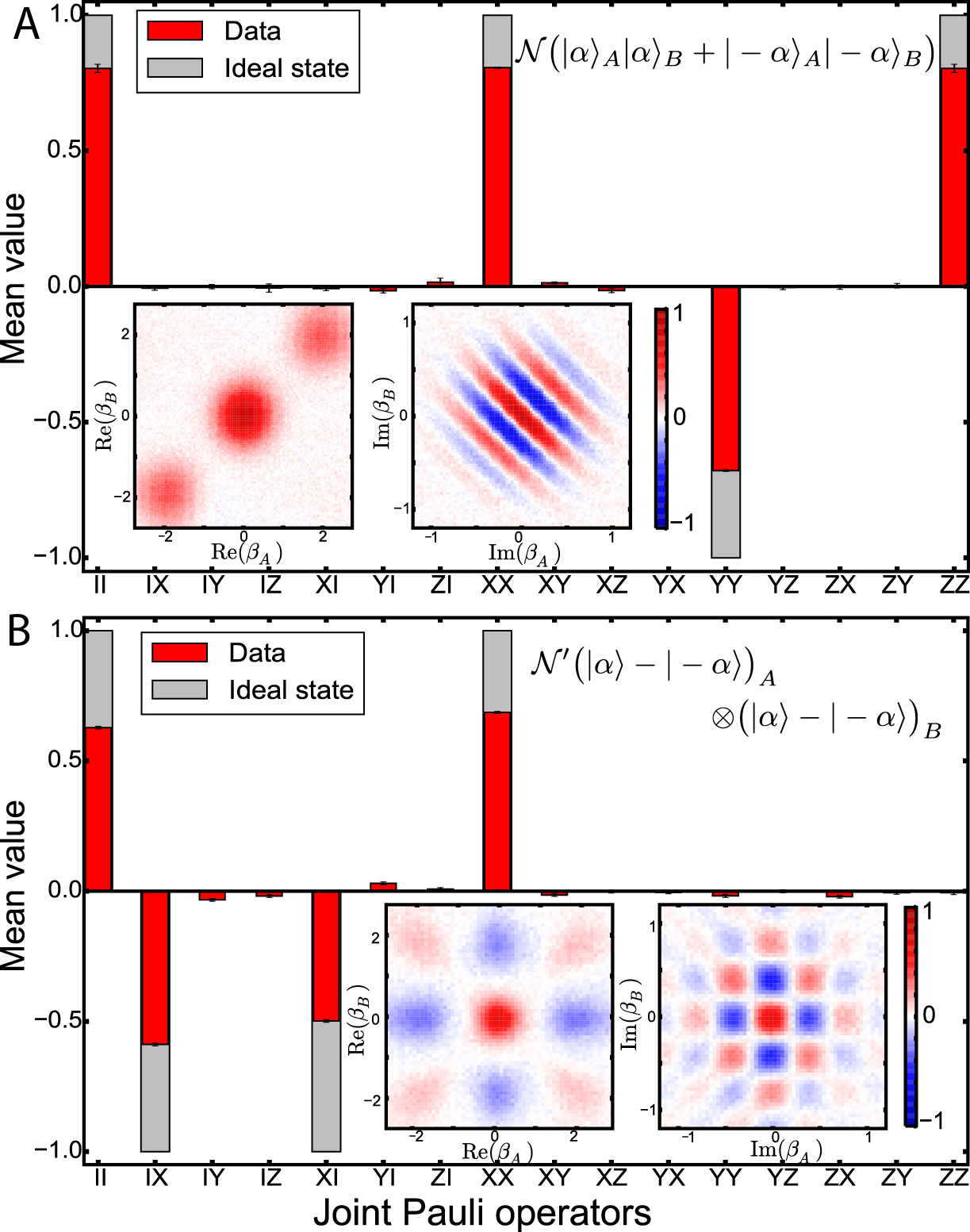}}
\end{figure*}

%We have measured a characteristic decay time of about 155 $\mu s$ for the joint parity of the two-mode cat state, manifesting long-lived entanglement between the two mesoscopic systems.  This coherence time is in fact longer than any superconducting qubits reported so far, and consistent with the combined photon loss rate of the two high-Q superconducting cavities (see Supplementary for details). 
% Precision measurements of the joint parity decay at higher photon numbers may allow investigation of possible mesoscopic decoherence in addition to single-photon loss, although testing the existing gravitational decoherence theories~\cite{pikovski_universal_2015} appears difficult due to the small photon mass.

Analyzed within the energy eigen-mode basis (Alice and Bob), the two-mode cat state is a manifestation of quantum entanglement between two quasi-classical systems.  The entanglement can be tested against a CHSH-style Bell's inequality constructed from displaced joint parity at 4 points in the phase space% [$(\beta_{A}, \beta_{B})$, $(\beta_{A}, \beta'_{B})$, $(\beta'_{A}, \beta_{B})$, $(\beta'_{A}, \beta'_{B})$]
~\cite{milman_proposal_2005}.  %With a close-to-optimal choice of $\alpha_{a}=\alpha_{b}=-0.102i$, $\alpha'_{a}=\alpha'_{b}=0.307i$, 
We measure a Bell signal~\cite{see_supplementary_????} of $2.17\pm 0.01$ for the state in Fig.~3, exceeding the classical bound of 2.  %We note that the two modes in our experiment are not completely separated spatially since their interactions with the ancilla persist throughout the measurement process.  Therefore, 
Without complete spatial separation and fully independent readout of the two modes, the violation should be considered a demonstration of the fidelity of the entanglement and the measurement rather than a true test of non-locality.  Nevertheless, various schemes exist to further separate the two modes such as converting the cavity fields into itinerant microwave signals and/or optical photons~\cite{andrews_quantum-enabled_2015}.

Compared with other reported quantum states of two harmonic oscillators, a striking property of the two-mode cat state is that its underlying compositions are highly-distinguishable.  Two-mode squeezed states~\cite{ou_realization_1992,julsgaard_experimental_2001,eichler_observation_2011,flurin_generating_2012} have shown strong entanglement, but are Gaussian states without the Wigner negativity and the phase space separation as in a cat state.  The ``$N00N$" state, an entangled state in the discrete Fock state basis, typically requires quantum operations of $N$ photons one by one and so far has been realized with up to 5 photons~\cite{afek_high-noon_2010,wang_deterministic_2011}.  The two components of the cat state in Fig.~3 have a phase space separation of $|\alpha-(-\alpha)|=\sqrt{15}$ in each cavity, giving an action distance of $\sqrt{30}$ in the 4D phase space, or a cat size~\cite{deleglise_reconstruction_2008} of 30 photons.  Our technique in principle allows generation of two-mode cat states with arbitrary size using the same operation.  So far we have measured cat sizes of up to 80 photons~\cite{see_supplementary_????}, and more macroscopic states can be achieved by implementing numerically optimized control pulses~\cite{khaneja_optimal_2005} and engineering more favorable Hamiltonian parameters.

%Since information capacity increases linearly with the photon number in an oscillator but exponentially with the number of modes, the introduction of the second mode to cat states is a necessary and important step towards quantum information processing using harmonic systems.  

Compared with single-cavity quantum states, the addition of the second cavity mode increases the quantum information capacity significantly.  Despite the modest mean photon numbers, a full tomography of the two-mode cat state (partly shown in Fig.~3) requires a Hilbert space of at least 100 dimensions to be described (capturing 99\% of the population), comparable to a 6 or 7 qubit GHZ state.
%10 Fock state basis in each cavity to capture 99\% of probability distribution, or 14 Fock state basis in each cavity to include all wavefunction components with amplitude $>0.01$.
Our conservatively estimated quantum state fidelity is comparable to that reported for an 8-qubit GHZ state in trapped ions~\cite{monz_14-qubit_2011} and the largest GHZ state in superconducting circuits~\cite{kelly_state_2015} (5 qubits).  In addition, a great advantage of continuous-variable quantum control is illustrated by our hardware-efficient quantum state tomography protocol that covers an enormous Hilbert space by simply varying two complex variables of cavity displacements. %, realizing an equivalent 7 or 8 qubit quantum state tomography which otherwise is a challenging task in any quantum information experiments.

%\section{Logical encoding}

An important motivation for creating multi-cavity cat states is to implement a promising paradigm towards fault-tolerant quantum computation~\cite{cochrane_macroscopically_1999,mirrahimi_dynamically_2014}, where information is redundantly encoded in the coherent state basis~\cite{vlastakis_deterministically_2013}.  This approach has recently led to the first realization of quantum error correction of a logical qubit achieving the break-even point~\cite{ofek_demonstrating_2016}.  In this context, our experiment realizes an architecture of two coupled logical qubits.  
The two-mode cat state can be considered a two-qubit Bell state $\frac{1}{\sqrt2}\big(|0\rangle|0\rangle\pm|1\rangle|1\rangle\big)$, where the quasi-orthogonal coherent states $|\pm\alpha\rangle$ in each of the two cavities represent $|0\rangle$ and $|1\rangle$ of a logical qubit. %, following a promising paradigm of quantum computation~\cite{mirrahimi_dynamically_2014,petrenko_demonstrating_2015}. The large cavity Hilbert space allows hardware-efficient redundant encoding of logical qubits, and QND measurements of photon number parity allows active quantum error-correction against the dominant decoherence channel, single photon loss~\cite{sun_tracking_2014}.  
%Despite being a multi-photon state, $|\psi_{\pm}\rangle$ carries only one unit (``e-bit") of entanglement, but in a redundant and robust way that allows quantum error correction, a crucial advantage of coherent-state based quantum information processing.  Photon loss in either cavity converts $|\psi_{\pm}\rangle$ into $|\psi_{\mp}\rangle$, which is tractable by repetitive QND parity measurements~\cite{sun_tracking_2014}. 

For any two-qubit logical state encoded in this subspace, we can perform efficient tomography without extensive measurement of the joint Wigner function. %, which is an important ingredient of implementing coherent-state based computation schemes.  
This is carried out by measuring $\langle P_J\rangle$ at 16 selected points of the phase space $(\beta_A, \beta_B)$~\cite{see_supplementary_????}.  The encoded two-qubit tomography of a state $|\psi_{+}\rangle$ with $\alpha=1.92$ is shown in Fig.~4A, providing a direct fidelity estimation~\cite{flammia_direct_2011} of $\frac{1}{4}\big(\langle II\rangle+\langle XX\rangle-\langle YY\rangle+\langle ZZ\rangle\big)$ = 78\% against the ideal Bell state, surpassing the 50\% bound for classical correlation.  As a comparison, Fig.~4B illustrates a product state of single-mode cat states in Alice and Bob, which is identified as $|-X\rangle_A|-X\rangle_B$ in the logical space.  For both states illustrated here, the two-qubit tomography suggests that errors within the encoded space are quite small.  The reduced contrast compared to the ideal state is mostly due to infidelity of the joint parity measurement and leakage out of the code space (due to higher-order Hamiltonian terms).

%Our joint parity measurement is quantum non-demolition (QND) by design, making it well-positioned for tracking error syndromes~\cite{sun_tracking_2014} and facilitating concurrent remote entanglement~\cite{roy_remote_2015}.

%\section{Discussion}

In this Report, we have demonstrated a Schr\"odinger's cat that lives in two cavities.  This two-mode cat state is not only a beautiful manifestation of mesoscopic superposition and entanglement constructed from quasi-classical states~\cite{sanders_review_2012}, but also a highly-desirable resource for quantum metrology~\cite{joo_quantum_2011}, quantum networks and teleportation~\cite{van_enk_entangled_2001}.  Moreover, the demonstration of high-fidelity quantum control over the large two-cavity Hilbert space has important implications for continuous-variable-based quantum computation.  The measurement of the joint photon number parity realized here is QND by design, and will play a central role in quantum error correction~\cite{ofek_demonstrating_2016,chuang_bosonic_1997,michael_new_2016} and facilitating concurrent remote entanglement~\cite{roy_remote_2015} in a modular architecture of quantum computation. 

We thank A.~Petrenko, Z.~Leghtas, B.~Vlastakis, W.~Pfaff, M.~Silveri and R.~T.~Brierley for helpful discussions, and M.~J.~Hatridge and A.~Narla for assistance on the Josephson parametric amplifier.  This research was supported by the U.S.~Army Research Office (W911NF-14-1-0011), NSF DMR-1301798 and the Multidisciplinary University Research Initiatives program (MURI) through the Air Force Office of Scientific Research (FA9550-14-1-0052). Facilities use was supported by the Yale Institute for Nanoscience and Quantum Engineering (YINQE), the Yale SEAS cleanroom, and the NSF (MRSECDMR 1119826).  Y.~Y.~Gao acknowledges support from an A*STAR NSS Fellowship.

\bibliography{Zotero}

%merlin.mbs apsrev4-1.bst 2010-07-25 4.21a (PWD, AO, DPC) hacked
%Control: key (0)
%Control: author (8) initials jnrlst
%Control: editor formatted (1) identically to author
%Control: production of article title (-1) disabled
%Control: page (0) single
%Control: year (1) truncated
%Control: production of eprint (0) enabled
\begin{thebibliography}{36}%
\makeatletter
\providecommand \@ifxundefined [1]{%
 \@ifx{#1\undefined}
}%
\providecommand \@ifnum [1]{%
 \ifnum #1\expandafter \@firstoftwo
 \else \expandafter \@secondoftwo
 \fi
}%
\providecommand \@ifx [1]{%
 \ifx #1\expandafter \@firstoftwo
 \else \expandafter \@secondoftwo
 \fi
}%
\providecommand \natexlab [1]{#1}%
\providecommand \enquote  [1]{``#1''}%
\providecommand \bibnamefont  [1]{#1}%
\providecommand \bibfnamefont [1]{#1}%
\providecommand \citenamefont [1]{#1}%
\providecommand \href@noop [0]{\@secondoftwo}%
\providecommand \href [0]{\begingroup \@sanitize@url \@href}%
\providecommand \@href[1]{\@@startlink{#1}\@@href}%
\providecommand \@@href[1]{\endgroup#1\@@endlink}%
\providecommand \@sanitize@url [0]{\catcode `\\12\catcode `\$12\catcode
  `\&12\catcode `\#12\catcode `\^12\catcode `\_12\catcode `\%12\relax}%
\providecommand \@@startlink[1]{}%
\providecommand \@@endlink[0]{}%
\providecommand \url  [0]{\begingroup\@sanitize@url \@url }%
\providecommand \@url [1]{\endgroup\@href {#1}{\urlprefix }}%
\providecommand \urlprefix  [0]{URL }%
\providecommand \Eprint [0]{\href }%
\providecommand \doibase [0]{http://dx.doi.org/}%
\providecommand \selectlanguage [0]{\@gobble}%
\providecommand \bibinfo  [0]{\@secondoftwo}%
\providecommand \bibfield  [0]{\@secondoftwo}%
\providecommand \translation [1]{[#1]}%
\providecommand \BibitemOpen [0]{}%
\providecommand \bibitemStop [0]{}%
\providecommand \bibitemNoStop [0]{.\EOS\space}%
\providecommand \EOS [0]{\spacefactor3000\relax}%
\providecommand \BibitemShut  [1]{\csname bibitem#1\endcsname}%
\let\auto@bib@innerbib\@empty
%</preamble>
\bibitem [{\citenamefont {Haroche}(2013)}]{haroche_nobel_2013}%
  \BibitemOpen
  \bibfield  {author} {\bibinfo {author} {\bibfnamefont {S.}~\bibnamefont
  {Haroche}},\ }\href {\doibase 10.1103/RevModPhys.85.1083} {\bibfield
  {journal} {\bibinfo  {journal} {Reviews of Modern Physics}\ }\textbf
  {\bibinfo {volume} {85}},\ \bibinfo {pages} {1083} (\bibinfo {year}
  {2013})}\BibitemShut {NoStop}%
\bibitem [{\citenamefont {Wineland}(2013)}]{wineland_nobel_2013}%
  \BibitemOpen
  \bibfield  {author} {\bibinfo {author} {\bibfnamefont {D.~J.}\ \bibnamefont
  {Wineland}},\ }\href {\doibase 10.1103/RevModPhys.85.1103} {\bibfield
  {journal} {\bibinfo  {journal} {Reviews of Modern Physics}\ }\textbf
  {\bibinfo {volume} {85}},\ \bibinfo {pages} {1103} (\bibinfo {year}
  {2013})}\BibitemShut {NoStop}%
\bibitem [{\citenamefont {Ourjoumtsev}\ \emph {et~al.}(2007)\citenamefont
  {Ourjoumtsev}, \citenamefont {Jeong}, \citenamefont {Tualle-Brouri},\ and\
  \citenamefont {Grangier}}]{ourjoumtsev_generation_2007}%
  \BibitemOpen
  \bibfield  {author} {\bibinfo {author} {\bibfnamefont {A.}~\bibnamefont
  {Ourjoumtsev}}, \bibinfo {author} {\bibfnamefont {H.}~\bibnamefont {Jeong}},
  \bibinfo {author} {\bibfnamefont {R.}~\bibnamefont {Tualle-Brouri}}, \ and\
  \bibinfo {author} {\bibfnamefont {P.}~\bibnamefont {Grangier}},\ }\href
  {\doibase 10.1038/nature06054} {\bibfield  {journal} {\bibinfo  {journal}
  {Nature}\ }\textbf {\bibinfo {volume} {448}},\ \bibinfo {pages} {784}
  (\bibinfo {year} {2007})}\BibitemShut {NoStop}%
\bibitem [{\citenamefont {Brune}\ \emph {et~al.}(1996)\citenamefont {Brune},
  \citenamefont {Hagley}, \citenamefont {Dreyer}, \citenamefont {Maître},
  \citenamefont {Maali}, \citenamefont {Wunderlich}, \citenamefont {Raimond},\
  and\ \citenamefont {Haroche}}]{brune_observing_1996}%
  \BibitemOpen
  \bibfield  {author} {\bibinfo {author} {\bibfnamefont {M.}~\bibnamefont
  {Brune}}, \bibinfo {author} {\bibfnamefont {E.}~\bibnamefont {Hagley}},
  \bibinfo {author} {\bibfnamefont {J.}~\bibnamefont {Dreyer}}, \bibinfo
  {author} {\bibfnamefont {X.}~\bibnamefont {Maître}}, \bibinfo {author}
  {\bibfnamefont {A.}~\bibnamefont {Maali}}, \bibinfo {author} {\bibfnamefont
  {C.}~\bibnamefont {Wunderlich}}, \bibinfo {author} {\bibfnamefont {J.~M.}\
  \bibnamefont {Raimond}}, \ and\ \bibinfo {author} {\bibfnamefont
  {S.}~\bibnamefont {Haroche}},\ }\href {\doibase 10.1103/PhysRevLett.77.4887}
  {\bibfield  {journal} {\bibinfo  {journal} {Physical Review Letters}\
  }\textbf {\bibinfo {volume} {77}},\ \bibinfo {pages} {4887} (\bibinfo {year}
  {1996})}\BibitemShut {NoStop}%
\bibitem [{\citenamefont {Vlastakis}\ \emph {et~al.}(2013)\citenamefont
  {Vlastakis}, \citenamefont {Kirchmair}, \citenamefont {Leghtas},
  \citenamefont {Nigg}, \citenamefont {Frunzio}, \citenamefont {Girvin},
  \citenamefont {Mirrahimi}, \citenamefont {Devoret},\ and\ \citenamefont
  {Schoelkopf}}]{vlastakis_deterministically_2013}%
  \BibitemOpen
  \bibfield  {author} {\bibinfo {author} {\bibfnamefont {B.}~\bibnamefont
  {Vlastakis}}, \bibinfo {author} {\bibfnamefont {G.}~\bibnamefont
  {Kirchmair}}, \bibinfo {author} {\bibfnamefont {Z.}~\bibnamefont {Leghtas}},
  \bibinfo {author} {\bibfnamefont {S.~E.}\ \bibnamefont {Nigg}}, \bibinfo
  {author} {\bibfnamefont {L.}~\bibnamefont {Frunzio}}, \bibinfo {author}
  {\bibfnamefont {S.~M.}\ \bibnamefont {Girvin}}, \bibinfo {author}
  {\bibfnamefont {M.}~\bibnamefont {Mirrahimi}}, \bibinfo {author}
  {\bibfnamefont {M.~H.}\ \bibnamefont {Devoret}}, \ and\ \bibinfo {author}
  {\bibfnamefont {R.~J.}\ \bibnamefont {Schoelkopf}},\ }\href {\doibase
  10.1126/science.1243289} {\bibfield  {journal} {\bibinfo  {journal}
  {Science}\ }\textbf {\bibinfo {volume} {342}},\ \bibinfo {pages} {607}
  (\bibinfo {year} {2013})}\BibitemShut {NoStop}%
\bibitem [{\citenamefont {Chuang}\ \emph {et~al.}(1997)\citenamefont {Chuang},
  \citenamefont {Leung},\ and\ \citenamefont {Yamamoto}}]{chuang_bosonic_1997}%
  \BibitemOpen
  \bibfield  {author} {\bibinfo {author} {\bibfnamefont {I.~L.}\ \bibnamefont
  {Chuang}}, \bibinfo {author} {\bibfnamefont {D.~W.}\ \bibnamefont {Leung}}, \
  and\ \bibinfo {author} {\bibfnamefont {Y.}~\bibnamefont {Yamamoto}},\ }\href
  {\doibase 10.1103/PhysRevA.56.1114} {\bibfield  {journal} {\bibinfo
  {journal} {Physical Review A}\ }\textbf {\bibinfo {volume} {56}},\ \bibinfo
  {pages} {1114} (\bibinfo {year} {1997})}\BibitemShut {NoStop}%
\bibitem [{\citenamefont {Mirrahimi}\ \emph {et~al.}(2014)\citenamefont
  {Mirrahimi}, \citenamefont {Leghtas}, \citenamefont {Albert}, \citenamefont
  {Touzard}, \citenamefont {Schoelkopf}, \citenamefont {Jiang},\ and\
  \citenamefont {Devoret}}]{mirrahimi_dynamically_2014}%
  \BibitemOpen
  \bibfield  {author} {\bibinfo {author} {\bibfnamefont {M.}~\bibnamefont
  {Mirrahimi}}, \bibinfo {author} {\bibfnamefont {Z.}~\bibnamefont {Leghtas}},
  \bibinfo {author} {\bibfnamefont {V.~V.}\ \bibnamefont {Albert}}, \bibinfo
  {author} {\bibfnamefont {S.}~\bibnamefont {Touzard}}, \bibinfo {author}
  {\bibfnamefont {R.~J.}\ \bibnamefont {Schoelkopf}}, \bibinfo {author}
  {\bibfnamefont {L.}~\bibnamefont {Jiang}}, \ and\ \bibinfo {author}
  {\bibfnamefont {M.~H.}\ \bibnamefont {Devoret}},\ }\href {\doibase
  10.1088/1367-2630/16/4/045014} {\bibfield  {journal} {\bibinfo  {journal}
  {New Journal of Physics}\ }\textbf {\bibinfo {volume} {16}},\ \bibinfo
  {pages} {045014} (\bibinfo {year} {2014})}\BibitemShut {NoStop}%
\bibitem [{\citenamefont {Ofek}\ \emph {et~al.}(2016)\citenamefont {Ofek},
  \citenamefont {Petrenko}, \citenamefont {Heeres}, \citenamefont {Reinhold},
  \citenamefont {Leghtas}, \citenamefont {Vlastakis}, \citenamefont {Liu},
  \citenamefont {Frunzio}, \citenamefont {Girvin}, \citenamefont {Jiang},
  \citenamefont {Mirrahimi}, \citenamefont {Devoret},\ and\ \citenamefont
  {Schoelkopf}}]{ofek_demonstrating_2016}%
  \BibitemOpen
  \bibfield  {author} {\bibinfo {author} {\bibfnamefont {N.}~\bibnamefont
  {Ofek}}, \bibinfo {author} {\bibfnamefont {A.}~\bibnamefont {Petrenko}},
  \bibinfo {author} {\bibfnamefont {R.~W.}\ \bibnamefont {Heeres}}, \bibinfo
  {author} {\bibfnamefont {P.}~\bibnamefont {Reinhold}}, \bibinfo {author}
  {\bibfnamefont {Z.}~\bibnamefont {Leghtas}}, \bibinfo {author} {\bibfnamefont
  {B.}~\bibnamefont {Vlastakis}}, \bibinfo {author} {\bibfnamefont
  {Y.}~\bibnamefont {Liu}}, \bibinfo {author} {\bibfnamefont {L.}~\bibnamefont
  {Frunzio}}, \bibinfo {author} {\bibfnamefont {S.~M.}\ \bibnamefont {Girvin}},
  \bibinfo {author} {\bibfnamefont {L.}~\bibnamefont {Jiang}}, \bibinfo
  {author} {\bibfnamefont {M.}~\bibnamefont {Mirrahimi}}, \bibinfo {author}
  {\bibfnamefont {M.~H.}\ \bibnamefont {Devoret}}, \ and\ \bibinfo {author}
  {\bibfnamefont {R.~J.}\ \bibnamefont {Schoelkopf}},\ }\href@noop {}
  {\bibfield  {journal} {\bibinfo  {journal} {in preparation}\ } (\bibinfo
  {year} {2016})}\BibitemShut {NoStop}%
\bibitem [{\citenamefont {Monz}\ \emph {et~al.}(2011)\citenamefont {Monz},
  \citenamefont {Schindler}, \citenamefont {Barreiro}, \citenamefont {Chwalla},
  \citenamefont {Nigg}, \citenamefont {Coish}, \citenamefont {Harlander},
  \citenamefont {Hänsel}, \citenamefont {Hennrich},\ and\ \citenamefont
  {Blatt}}]{monz_14-qubit_2011}%
  \BibitemOpen
  \bibfield  {author} {\bibinfo {author} {\bibfnamefont {T.}~\bibnamefont
  {Monz}}, \bibinfo {author} {\bibfnamefont {P.}~\bibnamefont {Schindler}},
  \bibinfo {author} {\bibfnamefont {J.~T.}\ \bibnamefont {Barreiro}}, \bibinfo
  {author} {\bibfnamefont {M.}~\bibnamefont {Chwalla}}, \bibinfo {author}
  {\bibfnamefont {D.}~\bibnamefont {Nigg}}, \bibinfo {author} {\bibfnamefont
  {W.~A.}\ \bibnamefont {Coish}}, \bibinfo {author} {\bibfnamefont
  {M.}~\bibnamefont {Harlander}}, \bibinfo {author} {\bibfnamefont
  {W.}~\bibnamefont {Hänsel}}, \bibinfo {author} {\bibfnamefont
  {M.}~\bibnamefont {Hennrich}}, \ and\ \bibinfo {author} {\bibfnamefont
  {R.}~\bibnamefont {Blatt}},\ }\href {\doibase 10.1103/PhysRevLett.106.130506}
  {\bibfield  {journal} {\bibinfo  {journal} {Physical Review Letters}\
  }\textbf {\bibinfo {volume} {106}},\ \bibinfo {pages} {130506} (\bibinfo
  {year} {2011})}\BibitemShut {NoStop}%
\bibitem [{\citenamefont {Kelly}\ \emph {et~al.}(2015)\citenamefont {Kelly},
  \citenamefont {Barends}, \citenamefont {Fowler}, \citenamefont {Megrant},
  \citenamefont {Jeffrey}, \citenamefont {White}, \citenamefont {Sank},
  \citenamefont {Mutus}, \citenamefont {Campbell}, \citenamefont {Chen},
  \citenamefont {Chen}, \citenamefont {Chiaro}, \citenamefont {Dunsworth},
  \citenamefont {Hoi}, \citenamefont {Neill}, \citenamefont {O’Malley},
  \citenamefont {Quintana}, \citenamefont {Roushan}, \citenamefont
  {Vainsencher}, \citenamefont {Wenner}, \citenamefont {Cleland},\ and\
  \citenamefont {Martinis}}]{kelly_state_2015}%
  \BibitemOpen
  \bibfield  {author} {\bibinfo {author} {\bibfnamefont {J.}~\bibnamefont
  {Kelly}}, \bibinfo {author} {\bibfnamefont {R.}~\bibnamefont {Barends}},
  \bibinfo {author} {\bibfnamefont {A.~G.}\ \bibnamefont {Fowler}}, \bibinfo
  {author} {\bibfnamefont {A.}~\bibnamefont {Megrant}}, \bibinfo {author}
  {\bibfnamefont {E.}~\bibnamefont {Jeffrey}}, \bibinfo {author} {\bibfnamefont
  {T.~C.}\ \bibnamefont {White}}, \bibinfo {author} {\bibfnamefont
  {D.}~\bibnamefont {Sank}}, \bibinfo {author} {\bibfnamefont {J.~Y.}\
  \bibnamefont {Mutus}}, \bibinfo {author} {\bibfnamefont {B.}~\bibnamefont
  {Campbell}}, \bibinfo {author} {\bibfnamefont {Y.}~\bibnamefont {Chen}},
  \bibinfo {author} {\bibfnamefont {Z.}~\bibnamefont {Chen}}, \bibinfo {author}
  {\bibfnamefont {B.}~\bibnamefont {Chiaro}}, \bibinfo {author} {\bibfnamefont
  {A.}~\bibnamefont {Dunsworth}}, \bibinfo {author} {\bibfnamefont {I.-C.}\
  \bibnamefont {Hoi}}, \bibinfo {author} {\bibfnamefont {C.}~\bibnamefont
  {Neill}}, \bibinfo {author} {\bibfnamefont {P.~J.~J.}\ \bibnamefont
  {O’Malley}}, \bibinfo {author} {\bibfnamefont {C.}~\bibnamefont
  {Quintana}}, \bibinfo {author} {\bibfnamefont {P.}~\bibnamefont {Roushan}},
  \bibinfo {author} {\bibfnamefont {A.}~\bibnamefont {Vainsencher}}, \bibinfo
  {author} {\bibfnamefont {J.}~\bibnamefont {Wenner}}, \bibinfo {author}
  {\bibfnamefont {A.~N.}\ \bibnamefont {Cleland}}, \ and\ \bibinfo {author}
  {\bibfnamefont {J.~M.}\ \bibnamefont {Martinis}},\ }\href {\doibase
  10.1038/nature14270} {\bibfield  {journal} {\bibinfo  {journal} {Nature}\
  }\textbf {\bibinfo {volume} {519}},\ \bibinfo {pages} {66} (\bibinfo {year}
  {2015})}\BibitemShut {NoStop}%
\bibitem [{\citenamefont {Davidovich}\ \emph {et~al.}(1993)\citenamefont
  {Davidovich}, \citenamefont {Maali}, \citenamefont {Brune}, \citenamefont
  {Raimond},\ and\ \citenamefont {Haroche}}]{davidovich_quantum_1993}%
  \BibitemOpen
  \bibfield  {author} {\bibinfo {author} {\bibfnamefont {L.}~\bibnamefont
  {Davidovich}}, \bibinfo {author} {\bibfnamefont {A.}~\bibnamefont {Maali}},
  \bibinfo {author} {\bibfnamefont {M.}~\bibnamefont {Brune}}, \bibinfo
  {author} {\bibfnamefont {J.~M.}\ \bibnamefont {Raimond}}, \ and\ \bibinfo
  {author} {\bibfnamefont {S.}~\bibnamefont {Haroche}},\ }\href {\doibase
  10.1103/PhysRevLett.71.2360} {\bibfield  {journal} {\bibinfo  {journal}
  {Physical Review Letters}\ }\textbf {\bibinfo {volume} {71}},\ \bibinfo
  {pages} {2360} (\bibinfo {year} {1993})}\BibitemShut {NoStop}%
\bibitem [{\citenamefont {Ourjoumtsev}\ \emph {et~al.}(2009)\citenamefont
  {Ourjoumtsev}, \citenamefont {Ferreyrol}, \citenamefont {Tualle-Brouri},\
  and\ \citenamefont {Grangier}}]{ourjoumtsev_preparation_2009}%
  \BibitemOpen
  \bibfield  {author} {\bibinfo {author} {\bibfnamefont {A.}~\bibnamefont
  {Ourjoumtsev}}, \bibinfo {author} {\bibfnamefont {F.}~\bibnamefont
  {Ferreyrol}}, \bibinfo {author} {\bibfnamefont {R.}~\bibnamefont
  {Tualle-Brouri}}, \ and\ \bibinfo {author} {\bibfnamefont {P.}~\bibnamefont
  {Grangier}},\ }\href {\doibase 10.1038/nphys1199} {\bibfield  {journal}
  {\bibinfo  {journal} {Nature Physics}\ }\textbf {\bibinfo {volume} {5}},\
  \bibinfo {pages} {189} (\bibinfo {year} {2009})}\BibitemShut {NoStop}%
\bibitem [{\citenamefont {Sanders}(2012)}]{sanders_review_2012}%
  \BibitemOpen
  \bibfield  {author} {\bibinfo {author} {\bibfnamefont {B.~C.}\ \bibnamefont
  {Sanders}},\ }\href {\doibase 10.1088/1751-8113/45/24/244002} {\bibfield
  {journal} {\bibinfo  {journal} {Journal of Physics A: Mathematical and
  Theoretical}\ }\textbf {\bibinfo {volume} {45}},\ \bibinfo {pages} {244002}
  (\bibinfo {year} {2012})}\BibitemShut {NoStop}%
\bibitem [{\citenamefont {Milman}\ \emph {et~al.}(2005)\citenamefont {Milman},
  \citenamefont {Auffeves}, \citenamefont {Yamaguchi}, \citenamefont {Brune},
  \citenamefont {Raimond},\ and\ \citenamefont
  {Haroche}}]{milman_proposal_2005}%
  \BibitemOpen
  \bibfield  {author} {\bibinfo {author} {\bibfnamefont {P.}~\bibnamefont
  {Milman}}, \bibinfo {author} {\bibfnamefont {A.}~\bibnamefont {Auffeves}},
  \bibinfo {author} {\bibfnamefont {F.}~\bibnamefont {Yamaguchi}}, \bibinfo
  {author} {\bibfnamefont {M.}~\bibnamefont {Brune}}, \bibinfo {author}
  {\bibfnamefont {J.~M.}\ \bibnamefont {Raimond}}, \ and\ \bibinfo {author}
  {\bibfnamefont {S.}~\bibnamefont {Haroche}},\ }\href {\doibase
  10.1140/epjd/e2004-00171-6} {\bibfield  {journal} {\bibinfo  {journal} {The
  European Physical Journal D - Atomic, Molecular, Optical and Plasma Physics}\
  }\textbf {\bibinfo {volume} {32}},\ \bibinfo {pages} {233} (\bibinfo {year}
  {2005})}\BibitemShut {NoStop}%
\bibitem [{\citenamefont {Paik}\ \emph {et~al.}(2011)\citenamefont {Paik},
  \citenamefont {Schuster}, \citenamefont {Bishop}, \citenamefont {Kirchmair},
  \citenamefont {Catelani}, \citenamefont {Sears}, \citenamefont {Johnson},
  \citenamefont {Reagor}, \citenamefont {Frunzio}, \citenamefont {Glazman},
  \citenamefont {Girvin}, \citenamefont {Devoret},\ and\ \citenamefont
  {Schoelkopf}}]{paik_observation_2011}%
  \BibitemOpen
  \bibfield  {author} {\bibinfo {author} {\bibfnamefont {H.}~\bibnamefont
  {Paik}}, \bibinfo {author} {\bibfnamefont {D.~I.}\ \bibnamefont {Schuster}},
  \bibinfo {author} {\bibfnamefont {L.~S.}\ \bibnamefont {Bishop}}, \bibinfo
  {author} {\bibfnamefont {G.}~\bibnamefont {Kirchmair}}, \bibinfo {author}
  {\bibfnamefont {G.}~\bibnamefont {Catelani}}, \bibinfo {author}
  {\bibfnamefont {A.~P.}\ \bibnamefont {Sears}}, \bibinfo {author}
  {\bibfnamefont {B.~R.}\ \bibnamefont {Johnson}}, \bibinfo {author}
  {\bibfnamefont {M.~J.}\ \bibnamefont {Reagor}}, \bibinfo {author}
  {\bibfnamefont {L.}~\bibnamefont {Frunzio}}, \bibinfo {author} {\bibfnamefont
  {L.~I.}\ \bibnamefont {Glazman}}, \bibinfo {author} {\bibfnamefont {S.~M.}\
  \bibnamefont {Girvin}}, \bibinfo {author} {\bibfnamefont {M.~H.}\
  \bibnamefont {Devoret}}, \ and\ \bibinfo {author} {\bibfnamefont {R.~J.}\
  \bibnamefont {Schoelkopf}},\ }\href {\doibase 10.1103/PhysRevLett.107.240501}
  {\bibfield  {journal} {\bibinfo  {journal} {Physical Review Letters}\
  }\textbf {\bibinfo {volume} {107}},\ \bibinfo {pages} {240501} (\bibinfo
  {year} {2011})}\BibitemShut {NoStop}%
\bibitem [{\citenamefont {{See}}()}]{see_supplementary_????}%
  \BibitemOpen
  \bibfield  {author} {\bibinfo {author} {\bibnamefont {{See}}},\ }\href@noop
  {} {\ }\bibinfo {note} {Supplementary materials for additional
  information}\BibitemShut {NoStop}%
\bibitem [{\citenamefont {Reagor}\ \emph {et~al.}(2015)\citenamefont {Reagor},
  \citenamefont {Pfaff}, \citenamefont {Axline}, \citenamefont {Heeres},
  \citenamefont {Ofek}, \citenamefont {Sliwa}, \citenamefont {Holland},
  \citenamefont {Wang}, \citenamefont {Blumoff}, \citenamefont {Chou},
  \citenamefont {Hatridge}, \citenamefont {Frunzio}, \citenamefont {Devoret},
  \citenamefont {Jiang},\ and\ \citenamefont
  {Schoelkopf}}]{reagor_quantum_2015}%
  \BibitemOpen
  \bibfield  {author} {\bibinfo {author} {\bibfnamefont {M.}~\bibnamefont
  {Reagor}}, \bibinfo {author} {\bibfnamefont {W.}~\bibnamefont {Pfaff}},
  \bibinfo {author} {\bibfnamefont {C.}~\bibnamefont {Axline}}, \bibinfo
  {author} {\bibfnamefont {R.~W.}\ \bibnamefont {Heeres}}, \bibinfo {author}
  {\bibfnamefont {N.}~\bibnamefont {Ofek}}, \bibinfo {author} {\bibfnamefont
  {K.}~\bibnamefont {Sliwa}}, \bibinfo {author} {\bibfnamefont
  {E.}~\bibnamefont {Holland}}, \bibinfo {author} {\bibfnamefont
  {C.}~\bibnamefont {Wang}}, \bibinfo {author} {\bibfnamefont {J.}~\bibnamefont
  {Blumoff}}, \bibinfo {author} {\bibfnamefont {K.}~\bibnamefont {Chou}},
  \bibinfo {author} {\bibfnamefont {M.~J.}\ \bibnamefont {Hatridge}}, \bibinfo
  {author} {\bibfnamefont {L.}~\bibnamefont {Frunzio}}, \bibinfo {author}
  {\bibfnamefont {M.~H.}\ \bibnamefont {Devoret}}, \bibinfo {author}
  {\bibfnamefont {L.}~\bibnamefont {Jiang}}, \ and\ \bibinfo {author}
  {\bibfnamefont {R.~J.}\ \bibnamefont {Schoelkopf}},\ }\href
  {http://arxiv.org/abs/1508.05882} {\bibfield  {journal} {\bibinfo  {journal}
  {arXiv:1508.05882 [cond-mat, physics:quant-ph]}\ } (\bibinfo {year}
  {2015})}\BibitemShut {NoStop}%
\bibitem [{\citenamefont {Michael}\ \emph {et~al.}(2016)\citenamefont
  {Michael}, \citenamefont {Silveri}, \citenamefont {Brierley}, \citenamefont
  {Albert}, \citenamefont {Salmilehto}, \citenamefont {Jiang},\ and\
  \citenamefont {Girvin}}]{michael_new_2016}%
  \BibitemOpen
  \bibfield  {author} {\bibinfo {author} {\bibfnamefont {M.}~\bibnamefont
  {Michael}}, \bibinfo {author} {\bibfnamefont {M.}~\bibnamefont {Silveri}},
  \bibinfo {author} {\bibfnamefont {R.~T.}\ \bibnamefont {Brierley}}, \bibinfo
  {author} {\bibfnamefont {V.~V.}\ \bibnamefont {Albert}}, \bibinfo {author}
  {\bibfnamefont {J.}~\bibnamefont {Salmilehto}}, \bibinfo {author}
  {\bibfnamefont {L.}~\bibnamefont {Jiang}}, \ and\ \bibinfo {author}
  {\bibfnamefont {S.~M.}\ \bibnamefont {Girvin}},\ }\href@noop {} {\bibfield
  {journal} {\bibinfo  {journal} {in preparation}\ } (\bibinfo {year}
  {2016})}\BibitemShut {NoStop}%
\bibitem [{\citenamefont {Leghtas}\ \emph {et~al.}(2013)\citenamefont
  {Leghtas}, \citenamefont {Kirchmair}, \citenamefont {Vlastakis},
  \citenamefont {Devoret}, \citenamefont {Schoelkopf},\ and\ \citenamefont
  {Mirrahimi}}]{leghtas_deterministic_2013}%
  \BibitemOpen
  \bibfield  {author} {\bibinfo {author} {\bibfnamefont {Z.}~\bibnamefont
  {Leghtas}}, \bibinfo {author} {\bibfnamefont {G.}~\bibnamefont {Kirchmair}},
  \bibinfo {author} {\bibfnamefont {B.}~\bibnamefont {Vlastakis}}, \bibinfo
  {author} {\bibfnamefont {M.~H.}\ \bibnamefont {Devoret}}, \bibinfo {author}
  {\bibfnamefont {R.~J.}\ \bibnamefont {Schoelkopf}}, \ and\ \bibinfo {author}
  {\bibfnamefont {M.}~\bibnamefont {Mirrahimi}},\ }\href {\doibase
  10.1103/PhysRevA.87.042315} {\bibfield  {journal} {\bibinfo  {journal}
  {Physical Review A}\ }\textbf {\bibinfo {volume} {87}},\ \bibinfo {pages}
  {042315} (\bibinfo {year} {2013})}\BibitemShut {NoStop}%
\bibitem [{\citenamefont {Bertet}\ \emph {et~al.}(2002)\citenamefont {Bertet},
  \citenamefont {Auffeves}, \citenamefont {Maioli}, \citenamefont {Osnaghi},
  \citenamefont {Meunier}, \citenamefont {Brune}, \citenamefont {Raimond},\
  and\ \citenamefont {Haroche}}]{bertet_direct_2002}%
  \BibitemOpen
  \bibfield  {author} {\bibinfo {author} {\bibfnamefont {P.}~\bibnamefont
  {Bertet}}, \bibinfo {author} {\bibfnamefont {A.}~\bibnamefont {Auffeves}},
  \bibinfo {author} {\bibfnamefont {P.}~\bibnamefont {Maioli}}, \bibinfo
  {author} {\bibfnamefont {S.}~\bibnamefont {Osnaghi}}, \bibinfo {author}
  {\bibfnamefont {T.}~\bibnamefont {Meunier}}, \bibinfo {author} {\bibfnamefont
  {M.}~\bibnamefont {Brune}}, \bibinfo {author} {\bibfnamefont {J.~M.}\
  \bibnamefont {Raimond}}, \ and\ \bibinfo {author} {\bibfnamefont
  {S.}~\bibnamefont {Haroche}},\ }\href {\doibase
  10.1103/PhysRevLett.89.200402} {\bibfield  {journal} {\bibinfo  {journal}
  {Physical Review Letters}\ }\textbf {\bibinfo {volume} {89}},\ \bibinfo
  {pages} {200402} (\bibinfo {year} {2002})}\BibitemShut {NoStop}%
\bibitem [{\citenamefont {Sun}\ \emph {et~al.}(2014)\citenamefont {Sun},
  \citenamefont {Petrenko}, \citenamefont {Leghtas}, \citenamefont {Vlastakis},
  \citenamefont {Kirchmair}, \citenamefont {Sliwa}, \citenamefont {Narla},
  \citenamefont {Hatridge}, \citenamefont {Shankar}, \citenamefont {Blumoff},
  \citenamefont {Frunzio}, \citenamefont {Mirrahimi}, \citenamefont {Devoret},\
  and\ \citenamefont {Schoelkopf}}]{sun_tracking_2014}%
  \BibitemOpen
  \bibfield  {author} {\bibinfo {author} {\bibfnamefont {L.}~\bibnamefont
  {Sun}}, \bibinfo {author} {\bibfnamefont {A.}~\bibnamefont {Petrenko}},
  \bibinfo {author} {\bibfnamefont {Z.}~\bibnamefont {Leghtas}}, \bibinfo
  {author} {\bibfnamefont {B.}~\bibnamefont {Vlastakis}}, \bibinfo {author}
  {\bibfnamefont {G.}~\bibnamefont {Kirchmair}}, \bibinfo {author}
  {\bibfnamefont {K.~M.}\ \bibnamefont {Sliwa}}, \bibinfo {author}
  {\bibfnamefont {A.}~\bibnamefont {Narla}}, \bibinfo {author} {\bibfnamefont
  {M.}~\bibnamefont {Hatridge}}, \bibinfo {author} {\bibfnamefont
  {S.}~\bibnamefont {Shankar}}, \bibinfo {author} {\bibfnamefont
  {J.}~\bibnamefont {Blumoff}}, \bibinfo {author} {\bibfnamefont
  {L.}~\bibnamefont {Frunzio}}, \bibinfo {author} {\bibfnamefont
  {M.}~\bibnamefont {Mirrahimi}}, \bibinfo {author} {\bibfnamefont {M.~H.}\
  \bibnamefont {Devoret}}, \ and\ \bibinfo {author} {\bibfnamefont {R.~J.}\
  \bibnamefont {Schoelkopf}},\ }\href {\doibase 10.1038/nature13436} {\bibfield
   {journal} {\bibinfo  {journal} {Nature}\ }\textbf {\bibinfo {volume}
  {511}},\ \bibinfo {pages} {444} (\bibinfo {year} {2014})}\BibitemShut
  {NoStop}%
\bibitem [{\citenamefont {Deléglise}\ \emph {et~al.}(2008)\citenamefont
  {Deléglise}, \citenamefont {Dotsenko}, \citenamefont {Sayrin}, \citenamefont
  {Bernu}, \citenamefont {Brune}, \citenamefont {Raimond},\ and\ \citenamefont
  {Haroche}}]{deleglise_reconstruction_2008}%
  \BibitemOpen
  \bibfield  {author} {\bibinfo {author} {\bibfnamefont {S.}~\bibnamefont
  {Deléglise}}, \bibinfo {author} {\bibfnamefont {I.}~\bibnamefont
  {Dotsenko}}, \bibinfo {author} {\bibfnamefont {C.}~\bibnamefont {Sayrin}},
  \bibinfo {author} {\bibfnamefont {J.}~\bibnamefont {Bernu}}, \bibinfo
  {author} {\bibfnamefont {M.}~\bibnamefont {Brune}}, \bibinfo {author}
  {\bibfnamefont {J.-M.}\ \bibnamefont {Raimond}}, \ and\ \bibinfo {author}
  {\bibfnamefont {S.}~\bibnamefont {Haroche}},\ }\href {\doibase
  10.1038/nature07288} {\bibfield  {journal} {\bibinfo  {journal} {Nature}\
  }\textbf {\bibinfo {volume} {455}},\ \bibinfo {pages} {510} (\bibinfo {year}
  {2008})}\BibitemShut {NoStop}%
\bibitem [{\citenamefont {Cahill}\ and\ \citenamefont
  {Glauber}(1969)}]{cahill_density_1969}%
  \BibitemOpen
  \bibfield  {author} {\bibinfo {author} {\bibfnamefont {K.~E.}\ \bibnamefont
  {Cahill}}\ and\ \bibinfo {author} {\bibfnamefont {R.~J.}\ \bibnamefont
  {Glauber}},\ }\href {\doibase 10.1103/PhysRev.177.1882} {\bibfield  {journal}
  {\bibinfo  {journal} {Physical Review}\ }\textbf {\bibinfo {volume} {177}},\
  \bibinfo {pages} {1882} (\bibinfo {year} {1969})}\BibitemShut {NoStop}%
\bibitem [{\citenamefont {Andrews}\ \emph {et~al.}(2015)\citenamefont
  {Andrews}, \citenamefont {Reed}, \citenamefont {Cicak}, \citenamefont
  {Teufel},\ and\ \citenamefont {Lehnert}}]{andrews_quantum-enabled_2015}%
  \BibitemOpen
  \bibfield  {author} {\bibinfo {author} {\bibfnamefont {R.~W.}\ \bibnamefont
  {Andrews}}, \bibinfo {author} {\bibfnamefont {A.~P.}\ \bibnamefont {Reed}},
  \bibinfo {author} {\bibfnamefont {K.}~\bibnamefont {Cicak}}, \bibinfo
  {author} {\bibfnamefont {J.~D.}\ \bibnamefont {Teufel}}, \ and\ \bibinfo
  {author} {\bibfnamefont {K.~W.}\ \bibnamefont {Lehnert}},\ }\href {\doibase
  10.1038/ncomms10021} {\bibfield  {journal} {\bibinfo  {journal} {Nature
  Communications}\ }\textbf {\bibinfo {volume} {6}},\ \bibinfo {pages} {10021}
  (\bibinfo {year} {2015})}\BibitemShut {NoStop}%
\bibitem [{\citenamefont {Ou}\ \emph {et~al.}(1992)\citenamefont {Ou},
  \citenamefont {Pereira}, \citenamefont {Kimble},\ and\ \citenamefont
  {Peng}}]{ou_realization_1992}%
  \BibitemOpen
  \bibfield  {author} {\bibinfo {author} {\bibfnamefont {Z.~Y.}\ \bibnamefont
  {Ou}}, \bibinfo {author} {\bibfnamefont {S.~F.}\ \bibnamefont {Pereira}},
  \bibinfo {author} {\bibfnamefont {H.~J.}\ \bibnamefont {Kimble}}, \ and\
  \bibinfo {author} {\bibfnamefont {K.~C.}\ \bibnamefont {Peng}},\ }\href
  {\doibase 10.1103/PhysRevLett.68.3663} {\bibfield  {journal} {\bibinfo
  {journal} {Physical Review Letters}\ }\textbf {\bibinfo {volume} {68}},\
  \bibinfo {pages} {3663} (\bibinfo {year} {1992})}\BibitemShut {NoStop}%
\bibitem [{\citenamefont {Julsgaard}\ \emph {et~al.}(2001)\citenamefont
  {Julsgaard}, \citenamefont {Kozhekin},\ and\ \citenamefont
  {Polzik}}]{julsgaard_experimental_2001}%
  \BibitemOpen
  \bibfield  {author} {\bibinfo {author} {\bibfnamefont {B.}~\bibnamefont
  {Julsgaard}}, \bibinfo {author} {\bibfnamefont {A.}~\bibnamefont {Kozhekin}},
  \ and\ \bibinfo {author} {\bibfnamefont {E.~S.}\ \bibnamefont {Polzik}},\
  }\href {\doibase 10.1038/35096524} {\bibfield  {journal} {\bibinfo  {journal}
  {Nature}\ }\textbf {\bibinfo {volume} {413}},\ \bibinfo {pages} {400}
  (\bibinfo {year} {2001})}\BibitemShut {NoStop}%
\bibitem [{\citenamefont {Eichler}\ \emph {et~al.}(2011)\citenamefont
  {Eichler}, \citenamefont {Bozyigit}, \citenamefont {Lang}, \citenamefont
  {Baur}, \citenamefont {Steffen}, \citenamefont {Fink}, \citenamefont
  {Filipp},\ and\ \citenamefont {Wallraff}}]{eichler_observation_2011}%
  \BibitemOpen
  \bibfield  {author} {\bibinfo {author} {\bibfnamefont {C.}~\bibnamefont
  {Eichler}}, \bibinfo {author} {\bibfnamefont {D.}~\bibnamefont {Bozyigit}},
  \bibinfo {author} {\bibfnamefont {C.}~\bibnamefont {Lang}}, \bibinfo {author}
  {\bibfnamefont {M.}~\bibnamefont {Baur}}, \bibinfo {author} {\bibfnamefont
  {L.}~\bibnamefont {Steffen}}, \bibinfo {author} {\bibfnamefont {J.~M.}\
  \bibnamefont {Fink}}, \bibinfo {author} {\bibfnamefont {S.}~\bibnamefont
  {Filipp}}, \ and\ \bibinfo {author} {\bibfnamefont {A.}~\bibnamefont
  {Wallraff}},\ }\href {\doibase 10.1103/PhysRevLett.107.113601} {\bibfield
  {journal} {\bibinfo  {journal} {Physical Review Letters}\ }\textbf {\bibinfo
  {volume} {107}},\ \bibinfo {pages} {113601} (\bibinfo {year}
  {2011})}\BibitemShut {NoStop}%
\bibitem [{\citenamefont {Flurin}\ \emph {et~al.}(2012)\citenamefont {Flurin},
  \citenamefont {Roch}, \citenamefont {Mallet}, \citenamefont {Devoret},\ and\
  \citenamefont {Huard}}]{flurin_generating_2012}%
  \BibitemOpen
  \bibfield  {author} {\bibinfo {author} {\bibfnamefont {E.}~\bibnamefont
  {Flurin}}, \bibinfo {author} {\bibfnamefont {N.}~\bibnamefont {Roch}},
  \bibinfo {author} {\bibfnamefont {F.}~\bibnamefont {Mallet}}, \bibinfo
  {author} {\bibfnamefont {M.~H.}\ \bibnamefont {Devoret}}, \ and\ \bibinfo
  {author} {\bibfnamefont {B.}~\bibnamefont {Huard}},\ }\href {\doibase
  10.1103/PhysRevLett.109.183901} {\bibfield  {journal} {\bibinfo  {journal}
  {Physical Review Letters}\ }\textbf {\bibinfo {volume} {109}},\ \bibinfo
  {pages} {183901} (\bibinfo {year} {2012})}\BibitemShut {NoStop}%
\bibitem [{\citenamefont {Afek}\ \emph {et~al.}(2010)\citenamefont {Afek},
  \citenamefont {Ambar},\ and\ \citenamefont
  {Silberberg}}]{afek_high-noon_2010}%
  \BibitemOpen
  \bibfield  {author} {\bibinfo {author} {\bibfnamefont {I.}~\bibnamefont
  {Afek}}, \bibinfo {author} {\bibfnamefont {O.}~\bibnamefont {Ambar}}, \ and\
  \bibinfo {author} {\bibfnamefont {Y.}~\bibnamefont {Silberberg}},\ }\href
  {\doibase 10.1126/science.1188172} {\bibfield  {journal} {\bibinfo  {journal}
  {Science}\ }\textbf {\bibinfo {volume} {328}},\ \bibinfo {pages} {879}
  (\bibinfo {year} {2010})}\BibitemShut {NoStop}%
\bibitem [{\citenamefont {Wang}\ \emph {et~al.}(2011)\citenamefont {Wang},
  \citenamefont {Mariantoni}, \citenamefont {Bialczak}, \citenamefont
  {Lenander}, \citenamefont {Lucero}, \citenamefont {Neeley}, \citenamefont
  {O’Connell}, \citenamefont {Sank}, \citenamefont {Weides}, \citenamefont
  {Wenner}, \citenamefont {Yamamoto}, \citenamefont {Yin}, \citenamefont
  {Zhao}, \citenamefont {Martinis},\ and\ \citenamefont
  {Cleland}}]{wang_deterministic_2011}%
  \BibitemOpen
  \bibfield  {author} {\bibinfo {author} {\bibfnamefont {H.}~\bibnamefont
  {Wang}}, \bibinfo {author} {\bibfnamefont {M.}~\bibnamefont {Mariantoni}},
  \bibinfo {author} {\bibfnamefont {R.~C.}\ \bibnamefont {Bialczak}}, \bibinfo
  {author} {\bibfnamefont {M.}~\bibnamefont {Lenander}}, \bibinfo {author}
  {\bibfnamefont {E.}~\bibnamefont {Lucero}}, \bibinfo {author} {\bibfnamefont
  {M.}~\bibnamefont {Neeley}}, \bibinfo {author} {\bibfnamefont {A.~D.}\
  \bibnamefont {O’Connell}}, \bibinfo {author} {\bibfnamefont
  {D.}~\bibnamefont {Sank}}, \bibinfo {author} {\bibfnamefont {M.}~\bibnamefont
  {Weides}}, \bibinfo {author} {\bibfnamefont {J.}~\bibnamefont {Wenner}},
  \bibinfo {author} {\bibfnamefont {T.}~\bibnamefont {Yamamoto}}, \bibinfo
  {author} {\bibfnamefont {Y.}~\bibnamefont {Yin}}, \bibinfo {author}
  {\bibfnamefont {J.}~\bibnamefont {Zhao}}, \bibinfo {author} {\bibfnamefont
  {J.~M.}\ \bibnamefont {Martinis}}, \ and\ \bibinfo {author} {\bibfnamefont
  {A.~N.}\ \bibnamefont {Cleland}},\ }\href {\doibase
  10.1103/PhysRevLett.106.060401} {\bibfield  {journal} {\bibinfo  {journal}
  {Physical Review Letters}\ }\textbf {\bibinfo {volume} {106}},\ \bibinfo
  {pages} {060401} (\bibinfo {year} {2011})}\BibitemShut {NoStop}%
\bibitem [{\citenamefont {Khaneja}\ \emph {et~al.}(2005)\citenamefont
  {Khaneja}, \citenamefont {Reiss}, \citenamefont {Kehlet}, \citenamefont
  {Schulte-Herbrüggen},\ and\ \citenamefont {Glaser}}]{khaneja_optimal_2005}%
  \BibitemOpen
  \bibfield  {author} {\bibinfo {author} {\bibfnamefont {N.}~\bibnamefont
  {Khaneja}}, \bibinfo {author} {\bibfnamefont {T.}~\bibnamefont {Reiss}},
  \bibinfo {author} {\bibfnamefont {C.}~\bibnamefont {Kehlet}}, \bibinfo
  {author} {\bibfnamefont {T.}~\bibnamefont {Schulte-Herbrüggen}}, \ and\
  \bibinfo {author} {\bibfnamefont {S.~J.}\ \bibnamefont {Glaser}},\ }\href
  {\doibase 10.1016/j.jmr.2004.11.004} {\bibfield  {journal} {\bibinfo
  {journal} {Journal of Magnetic Resonance}\ }\textbf {\bibinfo {volume}
  {172}},\ \bibinfo {pages} {296} (\bibinfo {year} {2005})}\BibitemShut
  {NoStop}%
\bibitem [{\citenamefont {Cochrane}\ \emph {et~al.}(1999)\citenamefont
  {Cochrane}, \citenamefont {Milburn},\ and\ \citenamefont
  {Munro}}]{cochrane_macroscopically_1999}%
  \BibitemOpen
  \bibfield  {author} {\bibinfo {author} {\bibfnamefont {P.~T.}\ \bibnamefont
  {Cochrane}}, \bibinfo {author} {\bibfnamefont {G.~J.}\ \bibnamefont
  {Milburn}}, \ and\ \bibinfo {author} {\bibfnamefont {W.~J.}\ \bibnamefont
  {Munro}},\ }\href {\doibase 10.1103/PhysRevA.59.2631} {\bibfield  {journal}
  {\bibinfo  {journal} {Physical Review A}\ }\textbf {\bibinfo {volume} {59}},\
  \bibinfo {pages} {2631} (\bibinfo {year} {1999})}\BibitemShut {NoStop}%
\bibitem [{\citenamefont {Flammia}\ and\ \citenamefont
  {Liu}(2011)}]{flammia_direct_2011}%
  \BibitemOpen
  \bibfield  {author} {\bibinfo {author} {\bibfnamefont {S.~T.}\ \bibnamefont
  {Flammia}}\ and\ \bibinfo {author} {\bibfnamefont {Y.-K.}\ \bibnamefont
  {Liu}},\ }\href {\doibase 10.1103/PhysRevLett.106.230501} {\bibfield
  {journal} {\bibinfo  {journal} {Physical Review Letters}\ }\textbf {\bibinfo
  {volume} {106}},\ \bibinfo {pages} {230501} (\bibinfo {year}
  {2011})}\BibitemShut {NoStop}%
\bibitem [{\citenamefont {Joo}\ \emph {et~al.}(2011)\citenamefont {Joo},
  \citenamefont {Munro},\ and\ \citenamefont {Spiller}}]{joo_quantum_2011}%
  \BibitemOpen
  \bibfield  {author} {\bibinfo {author} {\bibfnamefont {J.}~\bibnamefont
  {Joo}}, \bibinfo {author} {\bibfnamefont {W.~J.}\ \bibnamefont {Munro}}, \
  and\ \bibinfo {author} {\bibfnamefont {T.~P.}\ \bibnamefont {Spiller}},\
  }\href {\doibase 10.1103/PhysRevLett.107.083601} {\bibfield  {journal}
  {\bibinfo  {journal} {Physical Review Letters}\ }\textbf {\bibinfo {volume}
  {107}},\ \bibinfo {pages} {083601} (\bibinfo {year} {2011})}\BibitemShut
  {NoStop}%
\bibitem [{\citenamefont {van Enk}\ and\ \citenamefont
  {Hirota}(2001)}]{van_enk_entangled_2001}%
  \BibitemOpen
  \bibfield  {author} {\bibinfo {author} {\bibfnamefont {S.~J.}\ \bibnamefont
  {van Enk}}\ and\ \bibinfo {author} {\bibfnamefont {O.}~\bibnamefont
  {Hirota}},\ }\href {\doibase 10.1103/PhysRevA.64.022313} {\bibfield
  {journal} {\bibinfo  {journal} {Physical Review A}\ }\textbf {\bibinfo
  {volume} {64}},\ \bibinfo {pages} {022313} (\bibinfo {year}
  {2001})}\BibitemShut {NoStop}%
\bibitem [{\citenamefont {Roy}\ \emph {et~al.}(2015)\citenamefont {Roy},
  \citenamefont {Jiang}, \citenamefont {Stone},\ and\ \citenamefont
  {Devoret}}]{roy_remote_2015}%
  \BibitemOpen
  \bibfield  {author} {\bibinfo {author} {\bibfnamefont {A.}~\bibnamefont
  {Roy}}, \bibinfo {author} {\bibfnamefont {L.}~\bibnamefont {Jiang}}, \bibinfo
  {author} {\bibfnamefont {A.~D.}\ \bibnamefont {Stone}}, \ and\ \bibinfo
  {author} {\bibfnamefont {M.}~\bibnamefont {Devoret}},\ }\href {\doibase
  10.1103/PhysRevLett.115.150503} {\bibfield  {journal} {\bibinfo  {journal}
  {Physical Review Letters}\ }\textbf {\bibinfo {volume} {115}},\ \bibinfo
  {pages} {150503} (\bibinfo {year} {2015})}\BibitemShut {NoStop}%
\end{thebibliography}%


%merlin.mbs apsrev4-1.bst 2010-07-25 4.21a (PWD, AO, DPC) hacked
%Control: key (0)
%Control: author (8) initials jnrlst
%Control: editor formatted (1) identically to author
%Control: production of article title (-1) disabled
%Control: page (0) single
%Control: year (1) truncated
%Control: production of eprint (0) enabled
\begin{thebibliography}{23}%
\makeatletter
\providecommand \@ifxundefined [1]{%
 \@ifx{#1\undefined}
}%
\providecommand \@ifnum [1]{%
 \ifnum #1\expandafter \@firstoftwo
 \else \expandafter \@secondoftwo
 \fi
}%
\providecommand \@ifx [1]{%
 \ifx #1\expandafter \@firstoftwo
 \else \expandafter \@secondoftwo
 \fi
}%
\providecommand \natexlab [1]{#1}%
\providecommand \enquote  [1]{``#1''}%
\providecommand \bibnamefont  [1]{#1}%
\providecommand \bibfnamefont [1]{#1}%
\providecommand \citenamefont [1]{#1}%
\providecommand \href@noop [0]{\@secondoftwo}%
\providecommand \href [0]{\begingroup \@sanitize@url \@href}%
\providecommand \@href[1]{\@@startlink{#1}\@@href}%
\providecommand \@@href[1]{\endgroup#1\@@endlink}%
\providecommand \@sanitize@url [0]{\catcode `\\12\catcode `\$12\catcode
  `\&12\catcode `\#12\catcode `\^12\catcode `\_12\catcode `\%12\relax}%
\providecommand \@@startlink[1]{}%
\providecommand \@@endlink[0]{}%
\providecommand \url  [0]{\begingroup\@sanitize@url \@url }%
\providecommand \@url [1]{\endgroup\@href {#1}{\urlprefix }}%
\providecommand \urlprefix  [0]{URL }%
\providecommand \Eprint [0]{\href }%
\providecommand \doibase [0]{http://dx.doi.org/}%
\providecommand \selectlanguage [0]{\@gobble}%
\providecommand \bibinfo  [0]{\@secondoftwo}%
\providecommand \bibfield  [0]{\@secondoftwo}%
\providecommand \translation [1]{[#1]}%
\providecommand \BibitemOpen [0]{}%
\providecommand \bibitemStop [0]{}%
\providecommand \bibitemNoStop [0]{.\EOS\space}%
\providecommand \EOS [0]{\spacefactor3000\relax}%
\providecommand \BibitemShut  [1]{\csname bibitem#1\endcsname}%
\let\auto@bib@innerbib\@empty
%</preamble>
\bibitem [{\citenamefont {Reagor}\ \emph {et~al.}(2015)\citenamefont {Reagor},
  \citenamefont {Pfaff}, \citenamefont {Axline}, \citenamefont {Heeres},
  \citenamefont {Ofek}, \citenamefont {Sliwa}, \citenamefont {Holland},
  \citenamefont {Wang}, \citenamefont {Blumoff}, \citenamefont {Chou},
  \citenamefont {Hatridge}, \citenamefont {Frunzio}, \citenamefont {Devoret},
  \citenamefont {Jiang},\ and\ \citenamefont
  {Schoelkopf}}]{reagor_quantum_2015}%
  \BibitemOpen
  \bibfield  {author} {\bibinfo {author} {\bibfnamefont {M.}~\bibnamefont
  {Reagor}}, \bibinfo {author} {\bibfnamefont {W.}~\bibnamefont {Pfaff}},
  \bibinfo {author} {\bibfnamefont {C.}~\bibnamefont {Axline}}, \bibinfo
  {author} {\bibfnamefont {R.~W.}\ \bibnamefont {Heeres}}, \bibinfo {author}
  {\bibfnamefont {N.}~\bibnamefont {Ofek}}, \bibinfo {author} {\bibfnamefont
  {K.}~\bibnamefont {Sliwa}}, \bibinfo {author} {\bibfnamefont
  {E.}~\bibnamefont {Holland}}, \bibinfo {author} {\bibfnamefont
  {C.}~\bibnamefont {Wang}}, \bibinfo {author} {\bibfnamefont {J.}~\bibnamefont
  {Blumoff}}, \bibinfo {author} {\bibfnamefont {K.}~\bibnamefont {Chou}},
  \bibinfo {author} {\bibfnamefont {M.~J.}\ \bibnamefont {Hatridge}}, \bibinfo
  {author} {\bibfnamefont {L.}~\bibnamefont {Frunzio}}, \bibinfo {author}
  {\bibfnamefont {M.~H.}\ \bibnamefont {Devoret}}, \bibinfo {author}
  {\bibfnamefont {L.}~\bibnamefont {Jiang}}, \ and\ \bibinfo {author}
  {\bibfnamefont {R.~J.}\ \bibnamefont {Schoelkopf}},\ }\href
  {http://arxiv.org/abs/1508.05882} {\bibfield  {journal} {\bibinfo  {journal}
  {arXiv:1508.05882 [cond-mat, physics:quant-ph]}\ } (\bibinfo {year}
  {2015})}\BibitemShut {NoStop}%
\bibitem [{\citenamefont {Reagor}\ \emph {et~al.}(2013)\citenamefont {Reagor},
  \citenamefont {Paik}, \citenamefont {Catelani}, \citenamefont {Sun},
  \citenamefont {Axline}, \citenamefont {Holland}, \citenamefont {Pop},
  \citenamefont {Masluk}, \citenamefont {Brecht}, \citenamefont {Frunzio},
  \citenamefont {Devoret}, \citenamefont {Glazman},\ and\ \citenamefont
  {Schoelkopf}}]{reagor_reaching_2013}%
  \BibitemOpen
  \bibfield  {author} {\bibinfo {author} {\bibfnamefont {M.}~\bibnamefont
  {Reagor}}, \bibinfo {author} {\bibfnamefont {H.}~\bibnamefont {Paik}},
  \bibinfo {author} {\bibfnamefont {G.}~\bibnamefont {Catelani}}, \bibinfo
  {author} {\bibfnamefont {L.}~\bibnamefont {Sun}}, \bibinfo {author}
  {\bibfnamefont {C.}~\bibnamefont {Axline}}, \bibinfo {author} {\bibfnamefont
  {E.}~\bibnamefont {Holland}}, \bibinfo {author} {\bibfnamefont {I.~M.}\
  \bibnamefont {Pop}}, \bibinfo {author} {\bibfnamefont {N.~A.}\ \bibnamefont
  {Masluk}}, \bibinfo {author} {\bibfnamefont {T.}~\bibnamefont {Brecht}},
  \bibinfo {author} {\bibfnamefont {L.}~\bibnamefont {Frunzio}}, \bibinfo
  {author} {\bibfnamefont {M.~H.}\ \bibnamefont {Devoret}}, \bibinfo {author}
  {\bibfnamefont {L.}~\bibnamefont {Glazman}}, \ and\ \bibinfo {author}
  {\bibfnamefont {R.~J.}\ \bibnamefont {Schoelkopf}},\ }\href {\doibase
  10.1063/1.4807015} {\bibfield  {journal} {\bibinfo  {journal} {Applied
  Physics Letters}\ }\textbf {\bibinfo {volume} {102}},\ \bibinfo {pages}
  {192604} (\bibinfo {year} {2013})}\BibitemShut {NoStop}%
\bibitem [{\citenamefont {Axline}\ \emph {et~al.}(2016)\citenamefont {Axline},
  \citenamefont {Reagor}, \citenamefont {Reinhold}, \citenamefont {Heeres},
  \citenamefont {Wang}, \citenamefont {Pfaff}, \citenamefont {Chu},
  \citenamefont {Frunzio},\ and\ \citenamefont
  {Schoelkopf}}]{axline_coaxline_2015}%
  \BibitemOpen
  \bibfield  {author} {\bibinfo {author} {\bibfnamefont {C.}~\bibnamefont
  {Axline}}, \bibinfo {author} {\bibfnamefont {M.}~\bibnamefont {Reagor}},
  \bibinfo {author} {\bibfnamefont {P.}~\bibnamefont {Reinhold}}, \bibinfo
  {author} {\bibfnamefont {R.}~\bibnamefont {Heeres}}, \bibinfo {author}
  {\bibfnamefont {C.}~\bibnamefont {Wang}}, \bibinfo {author} {\bibfnamefont
  {W.}~\bibnamefont {Pfaff}}, \bibinfo {author} {\bibfnamefont
  {Y.}~\bibnamefont {Chu}}, \bibinfo {author} {\bibfnamefont {L.}~\bibnamefont
  {Frunzio}}, \ and\ \bibinfo {author} {\bibfnamefont {R.}~\bibnamefont
  {Schoelkopf}},\ }\href@noop {} {\bibfield  {journal} {\bibinfo  {journal} {In
  preparation}\ } (\bibinfo {year} {2016})}\BibitemShut {NoStop}%
\bibitem [{\citenamefont {Nigg}\ \emph {et~al.}(2012)\citenamefont {Nigg},
  \citenamefont {Paik}, \citenamefont {Vlastakis}, \citenamefont {Kirchmair},
  \citenamefont {Shankar}, \citenamefont {Frunzio}, \citenamefont {Devoret},
  \citenamefont {Schoelkopf},\ and\ \citenamefont
  {Girvin}}]{nigg_black-box_2012}%
  \BibitemOpen
  \bibfield  {author} {\bibinfo {author} {\bibfnamefont {S.~E.}\ \bibnamefont
  {Nigg}}, \bibinfo {author} {\bibfnamefont {H.}~\bibnamefont {Paik}}, \bibinfo
  {author} {\bibfnamefont {B.}~\bibnamefont {Vlastakis}}, \bibinfo {author}
  {\bibfnamefont {G.}~\bibnamefont {Kirchmair}}, \bibinfo {author}
  {\bibfnamefont {S.}~\bibnamefont {Shankar}}, \bibinfo {author} {\bibfnamefont
  {L.}~\bibnamefont {Frunzio}}, \bibinfo {author} {\bibfnamefont {M.~H.}\
  \bibnamefont {Devoret}}, \bibinfo {author} {\bibfnamefont {R.~J.}\
  \bibnamefont {Schoelkopf}}, \ and\ \bibinfo {author} {\bibfnamefont {S.~M.}\
  \bibnamefont {Girvin}},\ }\href {\doibase 10.1103/PhysRevLett.108.240502}
  {\bibfield  {journal} {\bibinfo  {journal} {Physical Review Letters}\
  }\textbf {\bibinfo {volume} {108}},\ \bibinfo {pages} {240502} (\bibinfo
  {year} {2012})}\BibitemShut {NoStop}%
\bibitem [{\citenamefont {Kirchmair}\ \emph {et~al.}(2013)\citenamefont
  {Kirchmair}, \citenamefont {Vlastakis}, \citenamefont {Leghtas},
  \citenamefont {Nigg}, \citenamefont {Paik}, \citenamefont {Ginossar},
  \citenamefont {Mirrahimi}, \citenamefont {Frunzio}, \citenamefont {Girvin},\
  and\ \citenamefont {Schoelkopf}}]{kirchmair_observation_2013}%
  \BibitemOpen
  \bibfield  {author} {\bibinfo {author} {\bibfnamefont {G.}~\bibnamefont
  {Kirchmair}}, \bibinfo {author} {\bibfnamefont {B.}~\bibnamefont
  {Vlastakis}}, \bibinfo {author} {\bibfnamefont {Z.}~\bibnamefont {Leghtas}},
  \bibinfo {author} {\bibfnamefont {S.~E.}\ \bibnamefont {Nigg}}, \bibinfo
  {author} {\bibfnamefont {H.}~\bibnamefont {Paik}}, \bibinfo {author}
  {\bibfnamefont {E.}~\bibnamefont {Ginossar}}, \bibinfo {author}
  {\bibfnamefont {M.}~\bibnamefont {Mirrahimi}}, \bibinfo {author}
  {\bibfnamefont {L.}~\bibnamefont {Frunzio}}, \bibinfo {author} {\bibfnamefont
  {S.~M.}\ \bibnamefont {Girvin}}, \ and\ \bibinfo {author} {\bibfnamefont
  {R.~J.}\ \bibnamefont {Schoelkopf}},\ }\href {\doibase 10.1038/nature11902}
  {\bibfield  {journal} {\bibinfo  {journal} {Nature}\ }\textbf {\bibinfo
  {volume} {495}},\ \bibinfo {pages} {205} (\bibinfo {year}
  {2013})}\BibitemShut {NoStop}%
\bibitem [{\citenamefont {Wallraff}\ \emph {et~al.}(2004)\citenamefont
  {Wallraff}, \citenamefont {Schuster}, \citenamefont {Blais}, \citenamefont
  {Frunzio}, \citenamefont {Huang}, \citenamefont {Majer}, \citenamefont
  {Kumar}, \citenamefont {Girvin},\ and\ \citenamefont
  {Schoelkopf}}]{wallraff_strong_2004}%
  \BibitemOpen
  \bibfield  {author} {\bibinfo {author} {\bibfnamefont {A.}~\bibnamefont
  {Wallraff}}, \bibinfo {author} {\bibfnamefont {D.~I.}\ \bibnamefont
  {Schuster}}, \bibinfo {author} {\bibfnamefont {A.}~\bibnamefont {Blais}},
  \bibinfo {author} {\bibfnamefont {L.}~\bibnamefont {Frunzio}}, \bibinfo
  {author} {\bibfnamefont {R.-S.}\ \bibnamefont {Huang}}, \bibinfo {author}
  {\bibfnamefont {J.}~\bibnamefont {Majer}}, \bibinfo {author} {\bibfnamefont
  {S.}~\bibnamefont {Kumar}}, \bibinfo {author} {\bibfnamefont {S.~M.}\
  \bibnamefont {Girvin}}, \ and\ \bibinfo {author} {\bibfnamefont {R.~J.}\
  \bibnamefont {Schoelkopf}},\ }\href {\doibase 10.1038/nature02851} {\bibfield
   {journal} {\bibinfo  {journal} {Nature}\ }\textbf {\bibinfo {volume}
  {431}},\ \bibinfo {pages} {162} (\bibinfo {year} {2004})}\BibitemShut
  {NoStop}%
\bibitem [{\citenamefont {Leghtas}\ \emph {et~al.}(2016)\citenamefont
  {Leghtas}, \citenamefont {Touzard}, \citenamefont {Petrenko}, \citenamefont
  {Pop}, \citenamefont {Kou}, \citenamefont {Vlastakis}, \citenamefont {Narla},
  \citenamefont {Sliwa}, \citenamefont {Shankar}, \citenamefont {Hatridge},
  \citenamefont {Albert}, \citenamefont {Jiang}, \citenamefont {Frunzio},
  \citenamefont {Schoelkopf}, \citenamefont {Mirrahimi},\ and\ \citenamefont
  {Devoret}}]{Zaki_pumps_2016}%
  \BibitemOpen
  \bibfield  {author} {\bibinfo {author} {\bibfnamefont {Z.}~\bibnamefont
  {Leghtas}}, \bibinfo {author} {\bibfnamefont {T.}~\bibnamefont {Touzard}},
  \bibinfo {author} {\bibfnamefont {A.}~\bibnamefont {Petrenko}}, \bibinfo
  {author} {\bibfnamefont {I.}~\bibnamefont {Pop}}, \bibinfo {author}
  {\bibfnamefont {A.}~\bibnamefont {Kou}}, \bibinfo {author} {\bibfnamefont
  {B.}~\bibnamefont {Vlastakis}}, \bibinfo {author} {\bibfnamefont
  {A.}~\bibnamefont {Narla}}, \bibinfo {author} {\bibfnamefont
  {K.}~\bibnamefont {Sliwa}}, \bibinfo {author} {\bibfnamefont
  {S.}~\bibnamefont {Shankar}}, \bibinfo {author} {\bibfnamefont
  {M.}~\bibnamefont {Hatridge}}, \bibinfo {author} {\bibfnamefont
  {V.}~\bibnamefont {Albert}}, \bibinfo {author} {\bibfnamefont
  {L.}~\bibnamefont {Jiang}}, \bibinfo {author} {\bibfnamefont
  {L.}~\bibnamefont {Frunzio}}, \bibinfo {author} {\bibfnamefont
  {R.}~\bibnamefont {Schoelkopf}}, \bibinfo {author} {\bibfnamefont
  {M.}~\bibnamefont {Mirrahimi}}, \ and\ \bibinfo {author} {\bibfnamefont
  {M.}~\bibnamefont {Devoret}},\ }\href@noop {} {\bibfield  {journal} {\bibinfo
   {journal} {In preparation}\ } (\bibinfo {year} {2016})}\BibitemShut
  {NoStop}%
\bibitem [{\citenamefont {Peterer}\ \emph {et~al.}(2015)\citenamefont
  {Peterer}, \citenamefont {Bader}, \citenamefont {Jin}, \citenamefont {Yan},
  \citenamefont {Kamal}, \citenamefont {Gudmundsen}, \citenamefont {Leek},
  \citenamefont {Orlando}, \citenamefont {Oliver},\ and\ \citenamefont
  {Gustavsson}}]{peterer_coherence_2015}%
  \BibitemOpen
  \bibfield  {author} {\bibinfo {author} {\bibfnamefont {M.~J.}\ \bibnamefont
  {Peterer}}, \bibinfo {author} {\bibfnamefont {S.~J.}\ \bibnamefont {Bader}},
  \bibinfo {author} {\bibfnamefont {X.}~\bibnamefont {Jin}}, \bibinfo {author}
  {\bibfnamefont {F.}~\bibnamefont {Yan}}, \bibinfo {author} {\bibfnamefont
  {A.}~\bibnamefont {Kamal}}, \bibinfo {author} {\bibfnamefont {T.~J.}\
  \bibnamefont {Gudmundsen}}, \bibinfo {author} {\bibfnamefont {P.~J.}\
  \bibnamefont {Leek}}, \bibinfo {author} {\bibfnamefont {T.~P.}\ \bibnamefont
  {Orlando}}, \bibinfo {author} {\bibfnamefont {W.~D.}\ \bibnamefont {Oliver}},
  \ and\ \bibinfo {author} {\bibfnamefont {S.}~\bibnamefont {Gustavsson}},\
  }\href {\doibase 10.1103/PhysRevLett.114.010501} {\bibfield  {journal}
  {\bibinfo  {journal} {Physical Review Letters}\ }\textbf {\bibinfo {volume}
  {114}},\ \bibinfo {pages} {010501} (\bibinfo {year} {2015})}\BibitemShut
  {NoStop}%
\bibitem [{\citenamefont {Heeres}\ \emph {et~al.}(2015)\citenamefont {Heeres},
  \citenamefont {Vlastakis}, \citenamefont {Holland}, \citenamefont
  {Krastanov}, \citenamefont {Albert}, \citenamefont {Frunzio}, \citenamefont
  {Jiang},\ and\ \citenamefont {Schoelkopf}}]{heeres_cavity_2015}%
  \BibitemOpen
  \bibfield  {author} {\bibinfo {author} {\bibfnamefont {R.~W.}\ \bibnamefont
  {Heeres}}, \bibinfo {author} {\bibfnamefont {B.}~\bibnamefont {Vlastakis}},
  \bibinfo {author} {\bibfnamefont {E.}~\bibnamefont {Holland}}, \bibinfo
  {author} {\bibfnamefont {S.}~\bibnamefont {Krastanov}}, \bibinfo {author}
  {\bibfnamefont {V.~V.}\ \bibnamefont {Albert}}, \bibinfo {author}
  {\bibfnamefont {L.}~\bibnamefont {Frunzio}}, \bibinfo {author} {\bibfnamefont
  {L.}~\bibnamefont {Jiang}}, \ and\ \bibinfo {author} {\bibfnamefont {R.~J.}\
  \bibnamefont {Schoelkopf}},\ }\href {\doibase 10.1103/PhysRevLett.115.137002}
  {\bibfield  {journal} {\bibinfo  {journal} {Physical Review Letters}\
  }\textbf {\bibinfo {volume} {115}},\ \bibinfo {pages} {137002} (\bibinfo
  {year} {2015})}\BibitemShut {NoStop}%
\bibitem [{\citenamefont {Vlastakis}\ \emph {et~al.}(2013)\citenamefont
  {Vlastakis}, \citenamefont {Kirchmair}, \citenamefont {Leghtas},
  \citenamefont {Nigg}, \citenamefont {Frunzio}, \citenamefont {Girvin},
  \citenamefont {Mirrahimi}, \citenamefont {Devoret},\ and\ \citenamefont
  {Schoelkopf}}]{vlastakis_deterministically_2013}%
  \BibitemOpen
  \bibfield  {author} {\bibinfo {author} {\bibfnamefont {B.}~\bibnamefont
  {Vlastakis}}, \bibinfo {author} {\bibfnamefont {G.}~\bibnamefont
  {Kirchmair}}, \bibinfo {author} {\bibfnamefont {Z.}~\bibnamefont {Leghtas}},
  \bibinfo {author} {\bibfnamefont {S.~E.}\ \bibnamefont {Nigg}}, \bibinfo
  {author} {\bibfnamefont {L.}~\bibnamefont {Frunzio}}, \bibinfo {author}
  {\bibfnamefont {S.~M.}\ \bibnamefont {Girvin}}, \bibinfo {author}
  {\bibfnamefont {M.}~\bibnamefont {Mirrahimi}}, \bibinfo {author}
  {\bibfnamefont {M.~H.}\ \bibnamefont {Devoret}}, \ and\ \bibinfo {author}
  {\bibfnamefont {R.~J.}\ \bibnamefont {Schoelkopf}},\ }\href {\doibase
  10.1126/science.1243289} {\bibfield  {journal} {\bibinfo  {journal}
  {Science}\ }\textbf {\bibinfo {volume} {342}},\ \bibinfo {pages} {607}
  (\bibinfo {year} {2013})}\BibitemShut {NoStop}%
\bibitem [{\citenamefont {Leghtas}\ \emph {et~al.}(2013)\citenamefont
  {Leghtas}, \citenamefont {Kirchmair}, \citenamefont {Vlastakis},
  \citenamefont {Devoret}, \citenamefont {Schoelkopf},\ and\ \citenamefont
  {Mirrahimi}}]{leghtas_deterministic_2013}%
  \BibitemOpen
  \bibfield  {author} {\bibinfo {author} {\bibfnamefont {Z.}~\bibnamefont
  {Leghtas}}, \bibinfo {author} {\bibfnamefont {G.}~\bibnamefont {Kirchmair}},
  \bibinfo {author} {\bibfnamefont {B.}~\bibnamefont {Vlastakis}}, \bibinfo
  {author} {\bibfnamefont {M.~H.}\ \bibnamefont {Devoret}}, \bibinfo {author}
  {\bibfnamefont {R.~J.}\ \bibnamefont {Schoelkopf}}, \ and\ \bibinfo {author}
  {\bibfnamefont {M.}~\bibnamefont {Mirrahimi}},\ }\href {\doibase
  10.1103/PhysRevA.87.042315} {\bibfield  {journal} {\bibinfo  {journal}
  {Physical Review A}\ }\textbf {\bibinfo {volume} {87}},\ \bibinfo {pages}
  {042315} (\bibinfo {year} {2013})}\BibitemShut {NoStop}%
\bibitem [{\citenamefont {Chuang}\ \emph {et~al.}(1997)\citenamefont {Chuang},
  \citenamefont {Leung},\ and\ \citenamefont {Yamamoto}}]{Chuang_1997}%
  \BibitemOpen
  \bibfield  {author} {\bibinfo {author} {\bibfnamefont {I.}~\bibnamefont
  {Chuang}}, \bibinfo {author} {\bibfnamefont {D.~W.}\ \bibnamefont {Leung}}, \
  and\ \bibinfo {author} {\bibfnamefont {Y.}~\bibnamefont {Yamamoto}},\
  }\href@noop {} {\bibfield  {journal} {\bibinfo  {journal} {Physical Review
  A}\ }\textbf {\bibinfo {volume} {56}} (\bibinfo {year} {1997})}\BibitemShut
  {NoStop}%
\bibitem [{\citenamefont {Michael}\ \emph {et~al.}(2016)\citenamefont
  {Michael}, \citenamefont {Silveri}, \citenamefont {Brierley}, \citenamefont
  {Albert}, \citenamefont {Salmilehto}, \citenamefont {Jiang},\ and\
  \citenamefont {Girvin}}]{Silveri_kittenCode_2016}%
  \BibitemOpen
  \bibfield  {author} {\bibinfo {author} {\bibfnamefont {M.}~\bibnamefont
  {Michael}}, \bibinfo {author} {\bibfnamefont {M.}~\bibnamefont {Silveri}},
  \bibinfo {author} {\bibfnamefont {R.}~\bibnamefont {Brierley}}, \bibinfo
  {author} {\bibfnamefont {V.~V.}\ \bibnamefont {Albert}}, \bibinfo {author}
  {\bibfnamefont {J.}~\bibnamefont {Salmilehto}}, \bibinfo {author}
  {\bibfnamefont {L.}~\bibnamefont {Jiang}}, \ and\ \bibinfo {author}
  {\bibfnamefont {S.~M.}\ \bibnamefont {Girvin}},\ }\href@noop {} {\bibfield
  {journal} {\bibinfo  {journal} {In preparation}\ } (\bibinfo {year}
  {2016})}\BibitemShut {NoStop}%
\bibitem [{\citenamefont {Sun}\ \emph {et~al.}(2014)\citenamefont {Sun},
  \citenamefont {Petrenko}, \citenamefont {Leghtas}, \citenamefont {Vlastakis},
  \citenamefont {Kirchmair}, \citenamefont {Sliwa}, \citenamefont {Narla},
  \citenamefont {Hatridge}, \citenamefont {Shankar}, \citenamefont {Blumoff},
  \citenamefont {Frunzio}, \citenamefont {Mirrahimi}, \citenamefont {Devoret},\
  and\ \citenamefont {Schoelkopf}}]{sun_tracking_2014}%
  \BibitemOpen
  \bibfield  {author} {\bibinfo {author} {\bibfnamefont {L.}~\bibnamefont
  {Sun}}, \bibinfo {author} {\bibfnamefont {A.}~\bibnamefont {Petrenko}},
  \bibinfo {author} {\bibfnamefont {Z.}~\bibnamefont {Leghtas}}, \bibinfo
  {author} {\bibfnamefont {B.}~\bibnamefont {Vlastakis}}, \bibinfo {author}
  {\bibfnamefont {G.}~\bibnamefont {Kirchmair}}, \bibinfo {author}
  {\bibfnamefont {K.~M.}\ \bibnamefont {Sliwa}}, \bibinfo {author}
  {\bibfnamefont {A.}~\bibnamefont {Narla}}, \bibinfo {author} {\bibfnamefont
  {M.}~\bibnamefont {Hatridge}}, \bibinfo {author} {\bibfnamefont
  {S.}~\bibnamefont {Shankar}}, \bibinfo {author} {\bibfnamefont
  {J.}~\bibnamefont {Blumoff}}, \bibinfo {author} {\bibfnamefont
  {L.}~\bibnamefont {Frunzio}}, \bibinfo {author} {\bibfnamefont
  {M.}~\bibnamefont {Mirrahimi}}, \bibinfo {author} {\bibfnamefont {M.~H.}\
  \bibnamefont {Devoret}}, \ and\ \bibinfo {author} {\bibfnamefont {R.~J.}\
  \bibnamefont {Schoelkopf}},\ }\href {\doibase 10.1038/nature13436} {\bibfield
   {journal} {\bibinfo  {journal} {Nature}\ }\textbf {\bibinfo {volume}
  {511}},\ \bibinfo {pages} {444} (\bibinfo {year} {2014})}\BibitemShut
  {NoStop}%
\bibitem [{\citenamefont {Khaneja}\ \emph {et~al.}(2005)\citenamefont
  {Khaneja}, \citenamefont {Reiss}, \citenamefont {Kehlet}, \citenamefont
  {Schulte-Herbrüggen},\ and\ \citenamefont {Glaser}}]{khaneja_optimal_2005}%
  \BibitemOpen
  \bibfield  {author} {\bibinfo {author} {\bibfnamefont {N.}~\bibnamefont
  {Khaneja}}, \bibinfo {author} {\bibfnamefont {T.}~\bibnamefont {Reiss}},
  \bibinfo {author} {\bibfnamefont {C.}~\bibnamefont {Kehlet}}, \bibinfo
  {author} {\bibfnamefont {T.}~\bibnamefont {Schulte-Herbrüggen}}, \ and\
  \bibinfo {author} {\bibfnamefont {S.~J.}\ \bibnamefont {Glaser}},\ }\href
  {\doibase 10.1016/j.jmr.2004.11.004} {\bibfield  {journal} {\bibinfo
  {journal} {Journal of Magnetic Resonance}\ }\textbf {\bibinfo {volume}
  {172}},\ \bibinfo {pages} {296} (\bibinfo {year} {2005})}\BibitemShut
  {NoStop}%
\bibitem [{\citenamefont {Leonhardt}(1997)}]{leonhardt_1997}%
  \BibitemOpen
  \bibfield  {author} {\bibinfo {author} {\bibfnamefont {U.}~\bibnamefont
  {Leonhardt}},\ }\href@noop {} {\emph {\bibinfo {title} {Measuring the Quantum
  State of Light}}}\ (\bibinfo  {publisher} {Cambridge University Press},\
  \bibinfo {year} {1997})\BibitemShut {NoStop}%
\bibitem [{\citenamefont {Rigetti}\ \emph {et~al.}(2012)\citenamefont
  {Rigetti}, \citenamefont {Gambetta}, \citenamefont {Poletto}, \citenamefont
  {Plourde}, \citenamefont {Chow}, \citenamefont {Córcoles}, \citenamefont
  {Smolin}, \citenamefont {Merkel}, \citenamefont {Rozen}, \citenamefont
  {Keefe}, \citenamefont {Rothwell}, \citenamefont {Ketchen},\ and\
  \citenamefont {Steffen}}]{rigetti_superconducting_2012}%
  \BibitemOpen
  \bibfield  {author} {\bibinfo {author} {\bibfnamefont {C.}~\bibnamefont
  {Rigetti}}, \bibinfo {author} {\bibfnamefont {J.~M.}\ \bibnamefont
  {Gambetta}}, \bibinfo {author} {\bibfnamefont {S.}~\bibnamefont {Poletto}},
  \bibinfo {author} {\bibfnamefont {B.~L.~T.}\ \bibnamefont {Plourde}},
  \bibinfo {author} {\bibfnamefont {J.~M.}\ \bibnamefont {Chow}}, \bibinfo
  {author} {\bibfnamefont {A.~D.}\ \bibnamefont {Córcoles}}, \bibinfo {author}
  {\bibfnamefont {J.~A.}\ \bibnamefont {Smolin}}, \bibinfo {author}
  {\bibfnamefont {S.~T.}\ \bibnamefont {Merkel}}, \bibinfo {author}
  {\bibfnamefont {J.~R.}\ \bibnamefont {Rozen}}, \bibinfo {author}
  {\bibfnamefont {G.~A.}\ \bibnamefont {Keefe}}, \bibinfo {author}
  {\bibfnamefont {M.~B.}\ \bibnamefont {Rothwell}}, \bibinfo {author}
  {\bibfnamefont {M.~B.}\ \bibnamefont {Ketchen}}, \ and\ \bibinfo {author}
  {\bibfnamefont {M.}~\bibnamefont {Steffen}},\ }\href {\doibase
  10.1103/PhysRevB.86.100506} {\bibfield  {journal} {\bibinfo  {journal}
  {Physical Review B}\ }\textbf {\bibinfo {volume} {86}},\ \bibinfo {pages}
  {100506} (\bibinfo {year} {2012})}\BibitemShut {NoStop}%
\bibitem [{\citenamefont {Mirrahimi}\ \emph {et~al.}(2014)\citenamefont
  {Mirrahimi}, \citenamefont {Leghtas}, \citenamefont {Albert}, \citenamefont
  {Touzard}, \citenamefont {Schoelkopf}, \citenamefont {Jiang},\ and\
  \citenamefont {Devoret}}]{mirrahimi_dynamically_2014}%
  \BibitemOpen
  \bibfield  {author} {\bibinfo {author} {\bibfnamefont {M.}~\bibnamefont
  {Mirrahimi}}, \bibinfo {author} {\bibfnamefont {Z.}~\bibnamefont {Leghtas}},
  \bibinfo {author} {\bibfnamefont {V.~V.}\ \bibnamefont {Albert}}, \bibinfo
  {author} {\bibfnamefont {S.}~\bibnamefont {Touzard}}, \bibinfo {author}
  {\bibfnamefont {R.~J.}\ \bibnamefont {Schoelkopf}}, \bibinfo {author}
  {\bibfnamefont {L.}~\bibnamefont {Jiang}}, \ and\ \bibinfo {author}
  {\bibfnamefont {M.~H.}\ \bibnamefont {Devoret}},\ }\href {\doibase
  10.1088/1367-2630/16/4/045014} {\bibfield  {journal} {\bibinfo  {journal}
  {New Journal of Physics}\ }\textbf {\bibinfo {volume} {16}},\ \bibinfo
  {pages} {045014} (\bibinfo {year} {2014})}\BibitemShut {NoStop}%
\bibitem [{\citenamefont {Motzoi}\ \emph {et~al.}(2009)\citenamefont {Motzoi},
  \citenamefont {Gambetta}, \citenamefont {Rebentrost},\ and\ \citenamefont
  {Wilhelm}}]{motzoi_simple_2009}%
  \BibitemOpen
  \bibfield  {author} {\bibinfo {author} {\bibfnamefont {F.}~\bibnamefont
  {Motzoi}}, \bibinfo {author} {\bibfnamefont {J.~M.}\ \bibnamefont
  {Gambetta}}, \bibinfo {author} {\bibfnamefont {P.}~\bibnamefont
  {Rebentrost}}, \ and\ \bibinfo {author} {\bibfnamefont {F.~K.}\ \bibnamefont
  {Wilhelm}},\ }\href {\doibase 10.1103/PhysRevLett.103.110501} {\bibfield
  {journal} {\bibinfo  {journal} {Physical Review Letters}\ }\textbf {\bibinfo
  {volume} {103}},\ \bibinfo {pages} {110501} (\bibinfo {year}
  {2009})}\BibitemShut {NoStop}%
\bibitem [{\citenamefont {Chen}\ \emph {et~al.}(2016)\citenamefont {Chen},
  \citenamefont {Kelly}, \citenamefont {Quintana}, \citenamefont {Barends},
  \citenamefont {Campbell}, \citenamefont {Chen}, \citenamefont {Chiaro},
  \citenamefont {Dunsworth}, \citenamefont {Fowler}, \citenamefont {Lucero},
  \citenamefont {Jeffrey}, \citenamefont {Megrant}, \citenamefont {Mutus},
  \citenamefont {Neeley}, \citenamefont {Neill}, \citenamefont {O’Malley},
  \citenamefont {Roushan}, \citenamefont {Sank}, \citenamefont {Vainsencher},
  \citenamefont {Wenner}, \citenamefont {White}, \citenamefont {Korotkov},\
  and\ \citenamefont {Martinis}}]{chen_measuring_2016}%
  \BibitemOpen
  \bibfield  {author} {\bibinfo {author} {\bibfnamefont {Z.}~\bibnamefont
  {Chen}}, \bibinfo {author} {\bibfnamefont {J.}~\bibnamefont {Kelly}},
  \bibinfo {author} {\bibfnamefont {C.}~\bibnamefont {Quintana}}, \bibinfo
  {author} {\bibfnamefont {R.}~\bibnamefont {Barends}}, \bibinfo {author}
  {\bibfnamefont {B.}~\bibnamefont {Campbell}}, \bibinfo {author}
  {\bibfnamefont {Y.}~\bibnamefont {Chen}}, \bibinfo {author} {\bibfnamefont
  {B.}~\bibnamefont {Chiaro}}, \bibinfo {author} {\bibfnamefont
  {A.}~\bibnamefont {Dunsworth}}, \bibinfo {author} {\bibfnamefont
  {A.}~\bibnamefont {Fowler}}, \bibinfo {author} {\bibfnamefont
  {E.}~\bibnamefont {Lucero}}, \bibinfo {author} {\bibfnamefont
  {E.}~\bibnamefont {Jeffrey}}, \bibinfo {author} {\bibfnamefont
  {A.}~\bibnamefont {Megrant}}, \bibinfo {author} {\bibfnamefont
  {J.}~\bibnamefont {Mutus}}, \bibinfo {author} {\bibfnamefont
  {M.}~\bibnamefont {Neeley}}, \bibinfo {author} {\bibfnamefont
  {C.}~\bibnamefont {Neill}}, \bibinfo {author} {\bibfnamefont
  {P.}~\bibnamefont {O’Malley}}, \bibinfo {author} {\bibfnamefont
  {P.}~\bibnamefont {Roushan}}, \bibinfo {author} {\bibfnamefont
  {D.}~\bibnamefont {Sank}}, \bibinfo {author} {\bibfnamefont {A.}~\bibnamefont
  {Vainsencher}}, \bibinfo {author} {\bibfnamefont {J.}~\bibnamefont {Wenner}},
  \bibinfo {author} {\bibfnamefont {T.}~\bibnamefont {White}}, \bibinfo
  {author} {\bibfnamefont {A.}~\bibnamefont {Korotkov}}, \ and\ \bibinfo
  {author} {\bibfnamefont {J.~M.}\ \bibnamefont {Martinis}},\ }\href {\doibase
  10.1103/PhysRevLett.116.020501} {\bibfield  {journal} {\bibinfo  {journal}
  {Physical Review Letters}\ }\textbf {\bibinfo {volume} {116}},\ \bibinfo
  {pages} {020501} (\bibinfo {year} {2016})}\BibitemShut {NoStop}%
\bibitem [{\citenamefont {Banaszek}\ and\ \citenamefont
  {Wódkiewicz}(1999)}]{banaszek_testing_1999}%
  \BibitemOpen
  \bibfield  {author} {\bibinfo {author} {\bibfnamefont {K.}~\bibnamefont
  {Banaszek}}\ and\ \bibinfo {author} {\bibfnamefont {K.}~\bibnamefont
  {Wódkiewicz}},\ }\href {\doibase 10.1103/PhysRevLett.82.2009} {\bibfield
  {journal} {\bibinfo  {journal} {Physical Review Letters}\ }\textbf {\bibinfo
  {volume} {82}},\ \bibinfo {pages} {2009} (\bibinfo {year}
  {1999})}\BibitemShut {NoStop}%
\bibitem [{\citenamefont {Milman}\ \emph {et~al.}(2005)\citenamefont {Milman},
  \citenamefont {Auffeves}, \citenamefont {Yamaguchi}, \citenamefont {Brune},
  \citenamefont {Raimond},\ and\ \citenamefont
  {Haroche}}]{milman_proposal_2005}%
  \BibitemOpen
  \bibfield  {author} {\bibinfo {author} {\bibfnamefont {P.}~\bibnamefont
  {Milman}}, \bibinfo {author} {\bibfnamefont {A.}~\bibnamefont {Auffeves}},
  \bibinfo {author} {\bibfnamefont {F.}~\bibnamefont {Yamaguchi}}, \bibinfo
  {author} {\bibfnamefont {M.}~\bibnamefont {Brune}}, \bibinfo {author}
  {\bibfnamefont {J.~M.}\ \bibnamefont {Raimond}}, \ and\ \bibinfo {author}
  {\bibfnamefont {S.}~\bibnamefont {Haroche}},\ }\href {\doibase
  10.1140/epjd/e2004-00171-6} {\bibfield  {journal} {\bibinfo  {journal} {The
  European Physical Journal D - Atomic, Molecular, Optical and Plasma Physics}\
  }\textbf {\bibinfo {volume} {32}},\ \bibinfo {pages} {233} (\bibinfo {year}
  {2005})}\BibitemShut {NoStop}%
\bibitem [{\citenamefont {Vlastakis}\ \emph {et~al.}(2015)\citenamefont
  {Vlastakis}, \citenamefont {Petrenko}, \citenamefont {Ofek}, \citenamefont
  {Sun}, \citenamefont {Leghtas}, \citenamefont {Sliwa}, \citenamefont {Liu},
  \citenamefont {Hatridge}, \citenamefont {Blumoff}, \citenamefont {Frunzio},
  \citenamefont {Mirrahimi}, \citenamefont {Jiang}, \citenamefont {Devoret},\
  and\ \citenamefont {Schoelkopf}}]{vlastakis_characterizing_2015}%
  \BibitemOpen
  \bibfield  {author} {\bibinfo {author} {\bibfnamefont {B.}~\bibnamefont
  {Vlastakis}}, \bibinfo {author} {\bibfnamefont {A.}~\bibnamefont {Petrenko}},
  \bibinfo {author} {\bibfnamefont {N.}~\bibnamefont {Ofek}}, \bibinfo {author}
  {\bibfnamefont {L.}~\bibnamefont {Sun}}, \bibinfo {author} {\bibfnamefont
  {Z.}~\bibnamefont {Leghtas}}, \bibinfo {author} {\bibfnamefont
  {K.}~\bibnamefont {Sliwa}}, \bibinfo {author} {\bibfnamefont
  {Y.}~\bibnamefont {Liu}}, \bibinfo {author} {\bibfnamefont {M.}~\bibnamefont
  {Hatridge}}, \bibinfo {author} {\bibfnamefont {J.}~\bibnamefont {Blumoff}},
  \bibinfo {author} {\bibfnamefont {L.}~\bibnamefont {Frunzio}}, \bibinfo
  {author} {\bibfnamefont {M.}~\bibnamefont {Mirrahimi}}, \bibinfo {author}
  {\bibfnamefont {L.}~\bibnamefont {Jiang}}, \bibinfo {author} {\bibfnamefont
  {M.~H.}\ \bibnamefont {Devoret}}, \ and\ \bibinfo {author} {\bibfnamefont
  {R.~J.}\ \bibnamefont {Schoelkopf}},\ }\href {\doibase 10.1038/ncomms9970}
  {\bibfield  {journal} {\bibinfo  {journal} {Nature Communications}\ }\textbf
  {\bibinfo {volume} {6}},\ \bibinfo {pages} {8970} (\bibinfo {year}
  {2015})}\BibitemShut {NoStop}%
\end{thebibliography}%

\end{document}

\title{Supplementary Material:\\
A Schr\"odinger Cat living in two boxes}

\author{Chen Wang}
\author{Yvonne Y.~Gao}
\author{Philip Reinhold}
\author{R.~W.~Heeres}
\author{Nissim Ofek}
\author{Kevin Chou}
\author{Christopher Axline}
\author{Matthew Reagor}
\author{Jacob Blumoff}
\author{K.~M.~Sliwa}
\author{L.~Frunzio}
\author{S.~M.~Girvin}
\author{Liang Jiang}
\affiliation{Department of Applied Physics and Physics, Yale University, New Haven, Connecticut 06511, USA}
\author{M.~Mirrahimi}
\affiliation{Department of Applied Physics and Physics, Yale University, New Haven, Connecticut 06511, USA}
\affiliation{INRIA Paris-Rocquencourt, Domaine de Voluceau, B.~P.~105, 78153 Le Chesnay cedex, France}
\author{M.~H.~Devoret}
\author{R.~J.~Schoelkopf}
\affiliation{Department of Applied Physics and Physics, Yale University, New Haven, Connecticut 06511, USA}

\date{\today}
\maketitle

\section*{Materials and Methods}

\subsection{Device Architecture}

Our cQED system includes two 3D cavities, a quasi-planar linear resonator, and a Y-shaped transmon.  A single block of high-purity (5N5) aluminum is machined to form a 3D structure that contains both superconducting cavity resonators and also functions as a package for the sapphire chip with deposited Josephson junction.  Each of the two cavities can be considered a 3D version of a $\lambda/4$ transmission line resonator between a center stub 3.2 mm in diameter and a cylindrical wall (outer conductor) 9.5 mm in diameter~\cite{reagor_quantum_2015}.  The heights of the stubs control the resonance frequency, and are about 12.2 mm and 16.3 mm respectively for Alice and Bob. 
A tunnel (with a maximum width of 5.8 mm and a maximum height of 3.9 mm) is opened from the outside towards the middle wall between the two cavities, creating a three way joint between the tunnel and the two cavities (see photo image, Fig.~S1).  The whole package is chemically etched by about 80 $\mu$m after machining to improve the surface quality of the cavity resonators~\cite{reagor_reaching_2013}.

The superconducting transmon is on a 5.5 mm $\times$ 27.5 mm chip, which is diced from a 430 $\mu$m-thick c-plane sapphire wafer after fabrication.  The fabrication process uses electron-beam lithography and the standard shadow-mask evaporation of Al/AlOx/Al Josephson junction.  The sapphire chip is inserted into the tunnel, with the antenna pads of the transmon slightly intruding into the coaxial cavities to provide mode coupling. The chip is mechanically held at one end with an aluminum clamping structure and indium seal~\cite{axline_coaxline_2015}.

\begin{figure}[tbp]
    \centering
    \includegraphics[scale=0.38]{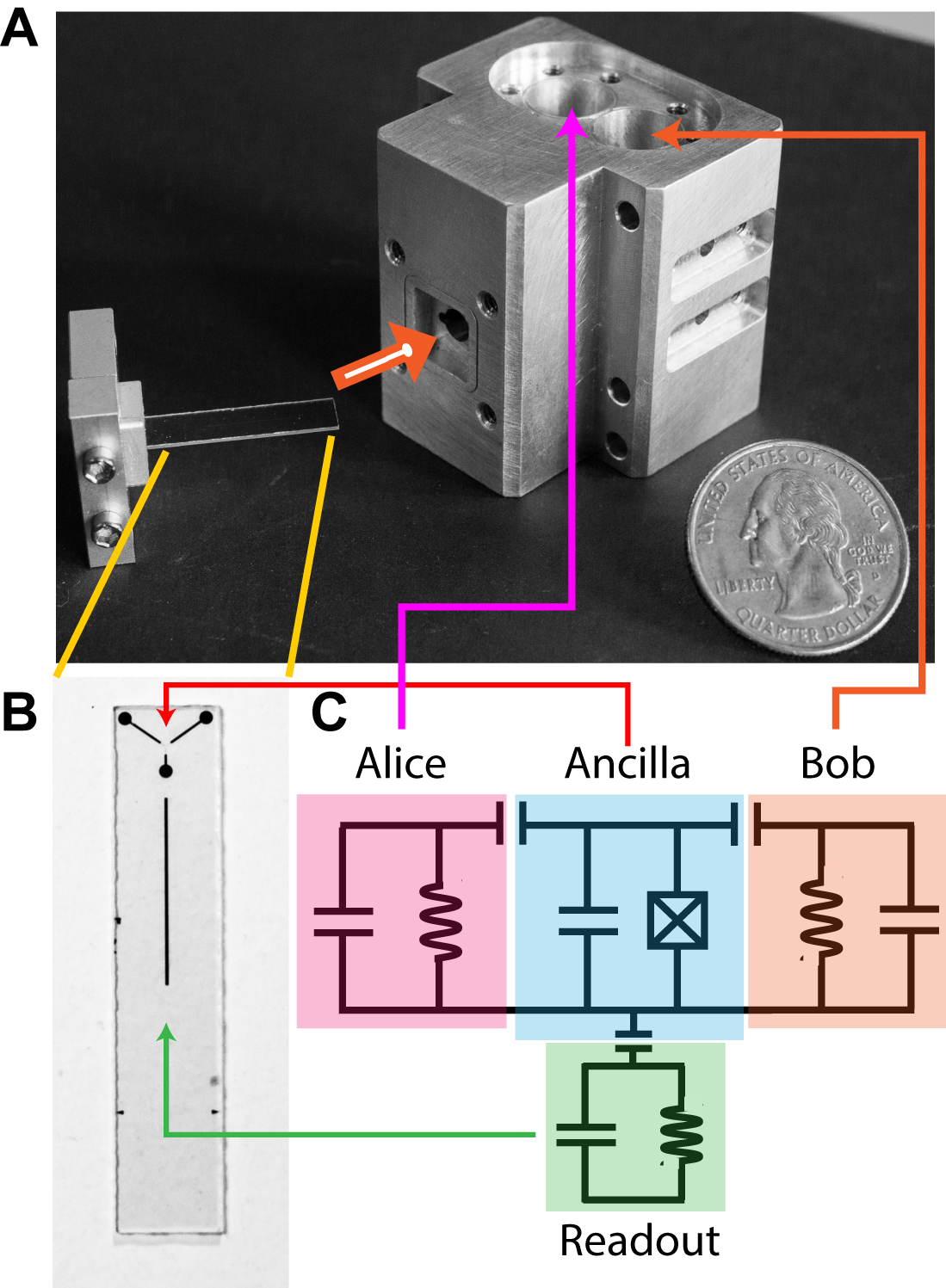}%[width=8.0cm]
    \caption{\textbf{Overview of the 3D cQED device. (A)} A photograph of the full assembly of the device used for this study.  The machined aluminum package contains two coaxial stub cavity resonators.  A sapphire chip hosting the transmon ancilla, clamped by an aluminum chip-holder, is inserted through a tunnel to be coupled to the two cavities from the side.  The sapphire chip contains an extra strip of aluminum, forming a stripline resonator with the tunnel wall.  \textbf{(B)} A micrograph image of the Y-shaped transmon.  \textbf{(C)} A schematic effective circuit of the cQED system containing three LC oscillators and one anharmonic oscillator (artificial atom) coupled together.}
\end{figure}

During the transmon fabrication process, a 100 $\mu$m$\times 9.8$ mm strip of aluminum film is also deposited on the sapphire chip.  This metal strip and the wall of the tunnel form a planar-3D hybrid $\lambda/2$ stripline resonator.  This resonator design has the advantages of both lithographic dimensional control and low surface/radiation loss, and is systematically studied in Ref.~\citen{axline_coaxline_2015}.  Here it is capacitively coupled to the transmon, and strongly coupled to a 50 $\Omega$ transmission line for readout.

\subsection{Sytem Hamiltonian Model}

The cQED system has four bosonic modes involved in this experiment: the two 3D cavities, the readout resonator and the transmon ancilla.  The transmon can be understood as an LC oscillator with much larger anharmonicity compared with the other modes, and is treated explicitly as a three-level artificial atom.  Following Black-box quantization of superconducting circuits~\cite{nigg_black-box_2012}, the other cavity/resonator modes are modeled as near-harmonic oscillators with weak nonlinearity inherited from coupling to the Josephson junction. 

The full system Hamiltonian can then be written in the following form up to the fourth order in the coupling of the resonators to the transmon:
\begin{align}
H/\hbar & = \omega_{A}(a^{\dagger}a+\frac{1}{2})+\omega_{B}(b^{\dagger}b++\frac{1}{2}) + \omega_{R}(r^{\dagger}r+\frac{1}{2})\nonumber\\
		& +\omega_{ge}|e\rangle\langle e|+(\omega_{ge}+\omega_{ef})|f\rangle\langle f|\nonumber\\
		& -\chi_A^{ge} a^{\dagger}a|e\rangle\langle e|-(\chi_A^{ge}+\chi_A^{ef}) a^{\dagger}a|f\rangle\langle f|\nonumber\\
        & -\chi_B^{ge} b^{\dagger}b|e\rangle\langle e|-(\chi_B^{ge}+\chi_B^{ef}) b^{\dagger}b|f\rangle\langle f|\nonumber\\
        & -\chi_R^{ge} r^{\dagger}r|e\rangle\langle e|-(\chi_R^{ge}+\chi_R^{ef}) r^{\dagger}r|f\rangle\langle f|\nonumber\\
        & -\frac{K_A}{2}a^{\dagger}a^{\dagger}aa -\frac{K_B}{2}b^{\dagger}b^{\dagger}bb -\frac{K_R}{2}r^{\dagger}r^{\dagger}rr\nonumber\\
        & -K_{AB}a^{\dagger}ab^{\dagger}b-K_{AR}a^{\dagger}ar^{\dagger}r-K_{BR}b^{\dagger}br^{\dagger}r
    \label{Hamiltonian}
\end{align}
Eq.~(\ref{Hamiltonian}) is an expanded version of Eq.~(3) of the main text, which now includes the readout resonator (with subscript $R$ and operators $r$ and $r^{\dagger}$) as well as Kerr nonlinearities of the cavities.  The first two rows represent the excitation energy of the all modes, explicitly including the transmon anharmonicity of $\omega_{ge}-\omega_{ef}=115.17$ MHz.  The next three rows are second order terms ($\sim1$ MHz) representing the dispersive interactions ($\chi$'s) between the transmon and each of the three resonators.  The last two rows are the fourth order terms ($\sim10$ kHz), including the self-Kerr energies ($K_A, K_B, K_R$) of the resonators and the cross-Kerr interactions between any pairs of resonators ($K_{AB}, K_{AR}, K_{BR}$).  All Hamiltonian parameters of our device are listed in Table S1. 

\begin{table}[b]
\caption{Hamiltonian parameters of all cQED components, including the transmon ancilla, the two cavity resonators (Alice and Bob) and the readout resonator.  The measured parameters include all transition frequencies ($\omega/2\pi$), dispersive shifts between each resonator and each transmon transition ($\chi/2\pi$), the self-Kerr of Alice ($K_A/2\pi$) and Bob ($K_B/2\pi$), and the cross-Kerr interaction between Alice and Bob ($K_{AB}/2\pi$). The Kerr parameters and $\chi^{ef}$ associated with the readout resonator are theoretical estimates based on the other measured parameters.}
\centering  
\begin{tabular}{c c c c c} % centered columns (4 columns)
\hline\hline\\[-2ex]
		& Frequency	&	 Nonlinear &interactions&versus:\,\,\,\,\,\,\,\,	 \\
		& $\omega/2\pi$	& Alice		& Bob	& Readout \\
\hline\\[-2ex]
$|e\rangle\rightarrow|g\rangle$	& 4.87805 GHz	& 0.71 MHz	& 1.41 MHz & 1.74 MHz\\
$|f\rangle\rightarrow|e\rangle$	& 4.76288 GHz	& 1.54 MHz	& 0.93 MHz & 1.63 MHz\\
Alice	& \,4.2196612 GHz\,	& 0.83 kHz	&  -9 kHz	&	5 kHz\\
Bob		& 5.4467677 GHz		& -9 kHz	& 5.6 kHz	&	12 kHz\\
Readout	& 7.6970 GHz		& 5 kHz		& 12 kHz	&	7 kHz\\[0.5ex]
% [1ex] adds vertical space
\hline
\end{tabular}
\end{table}

The key Hamiltonian terms that enable the cat state generation and joint parity measurement are the dispersive shifts, $\chi_A^{ge},\chi_A^{ef},\chi_B^{ge},\chi_B^{ef}$, highlighted in the main text.  Since the readout resonator is always kept in the vacuum state until a final measurement is needed, its Hamiltonian terms do not affect the quantum control and quantum state evolution in this experiment.  The fourth-order Hamiltonian terms of Alice and Bob give a minor contribution to the infidelity of the experiment.  The self-Kerr terms ($K_A$ and $K_B$) induce distortion of the Gaussian probability distribution of the coherent state components of the cat state, and will cause state collapse and revival at long time scales~\cite{kirchmair_observation_2013}.  The cross-Kerr interaction $K_{AB}$ induces spontaneous entTrueanglement between Alice and Bob over long time scales in ways not included in our simple analysis.

\subsection{Measurement Setup and Protocol}

Fig.~S2 shows the diagram of our measurement setup.  The device package is installed inside a Cryoperm magnetic shield and thermalized to the mixing chamber of a dilution refrigerator with a base temperature of 20 mK.  Low-pass filters and infrared (eccosorb) filters are used to reduce stray radiation and photon shot noise.  A Josephson parametric converter (JPC) is also mounted to the 20 mK stage, connected to the output port of the device package via circulators, providing near-quantum-limited amplification, with a power gain of 20dB and bandwidth of 5MHz.

\begin{figure*}
\floatbox[{\capbeside\thisfloatsetup{capbesideposition={right,center},capbesidewidth=6.5cm}}]{figure}[\FBwidth]
{\caption{\textbf{Circuit diagram of the measurement setup.} We use a field programmable gate array (FPGA) to control the experiment.  The FPGA has a total of 4 analogue channels, which are used to provide I-Q control of the classical cavity drives via sideband modulation of the outputs of microwave generators labeled ``Alice" and ``Bob".  These drives realize arbitrary cavity displacement operations $D_{\alpha}$ on the two high-Q cavities (Alice and Bob).  A digital channel from the FPGA triggers an arbitrary waveform generator (AWG), whose two analogue channels provide I-Q control of the ancilla drives for both $|g\rangle-|e\rangle$ and $|e\rangle-|f\rangle$ rotation, modulating the microwave tone labeled ``ancilla".  The readout pulse is generated by a ``RO" generator gated by a FPGA digital pulse, which is transmitted through the readout resonator for measuring the ancilla state.  The transmitted signal is amplified by a Josephson parametric converter (JPC) at 15 mK, a high electron mobility transistor (HEMT) at 4K, a standard (Mini-circuit) RF amplifier at room temperature, and then mixed with a local oscillator (``LO") to produce a 50 MHz signal to be digitized and recorded by the FPGA.  Another two analogue channels of the AWG controls an off-resonant pump on the readout resonator (``RO pump").  This drive, together with off-resonant pumps on Alice and Bob, allows fast reset of the high-Q cavities through 4-wave mixing processes.}}
{\includegraphics[scale=0.35]{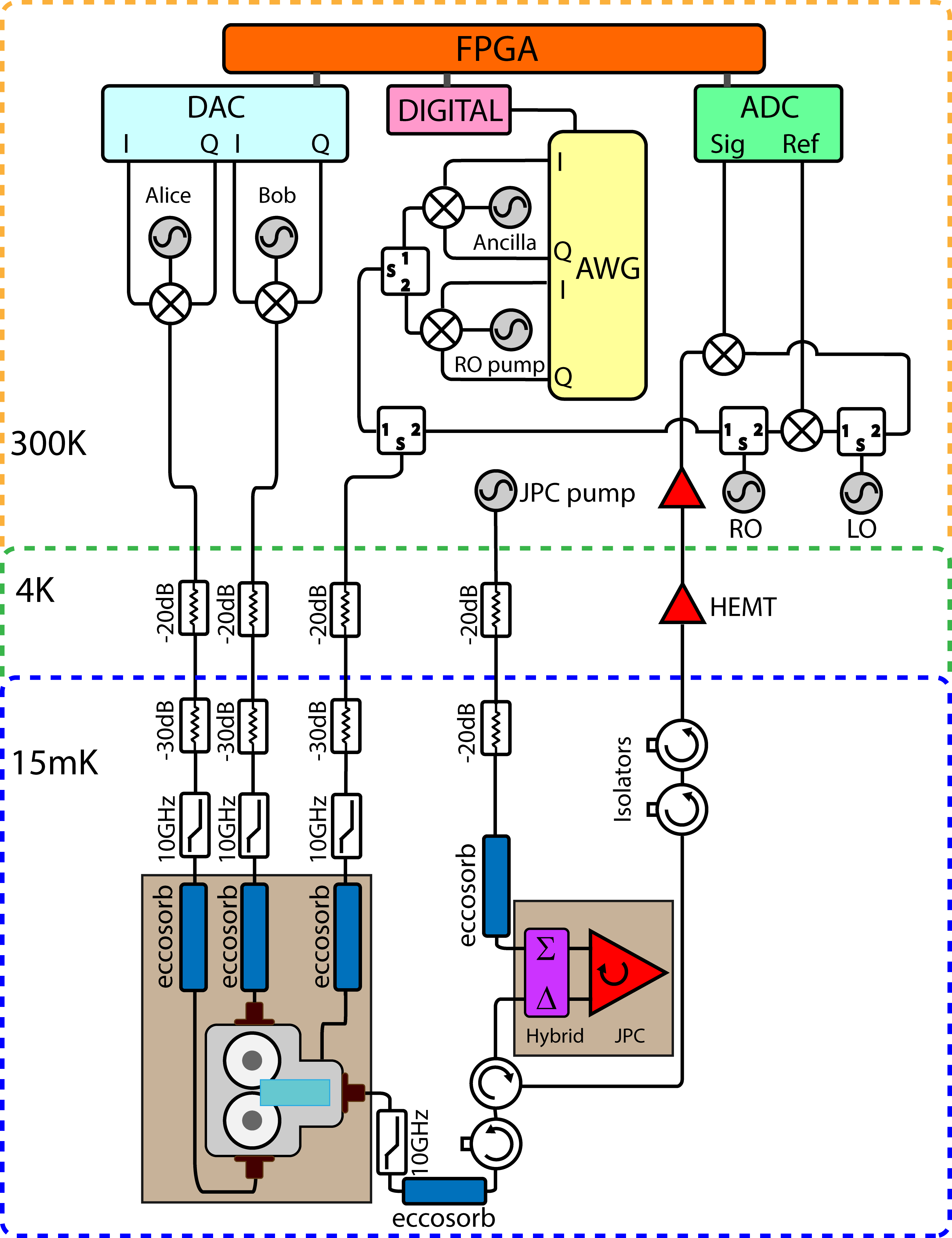}}%[width=10.8cm]
\end{figure*}

\begin{figure*}[btp]
    \centering
    \includegraphics[scale=0.32]{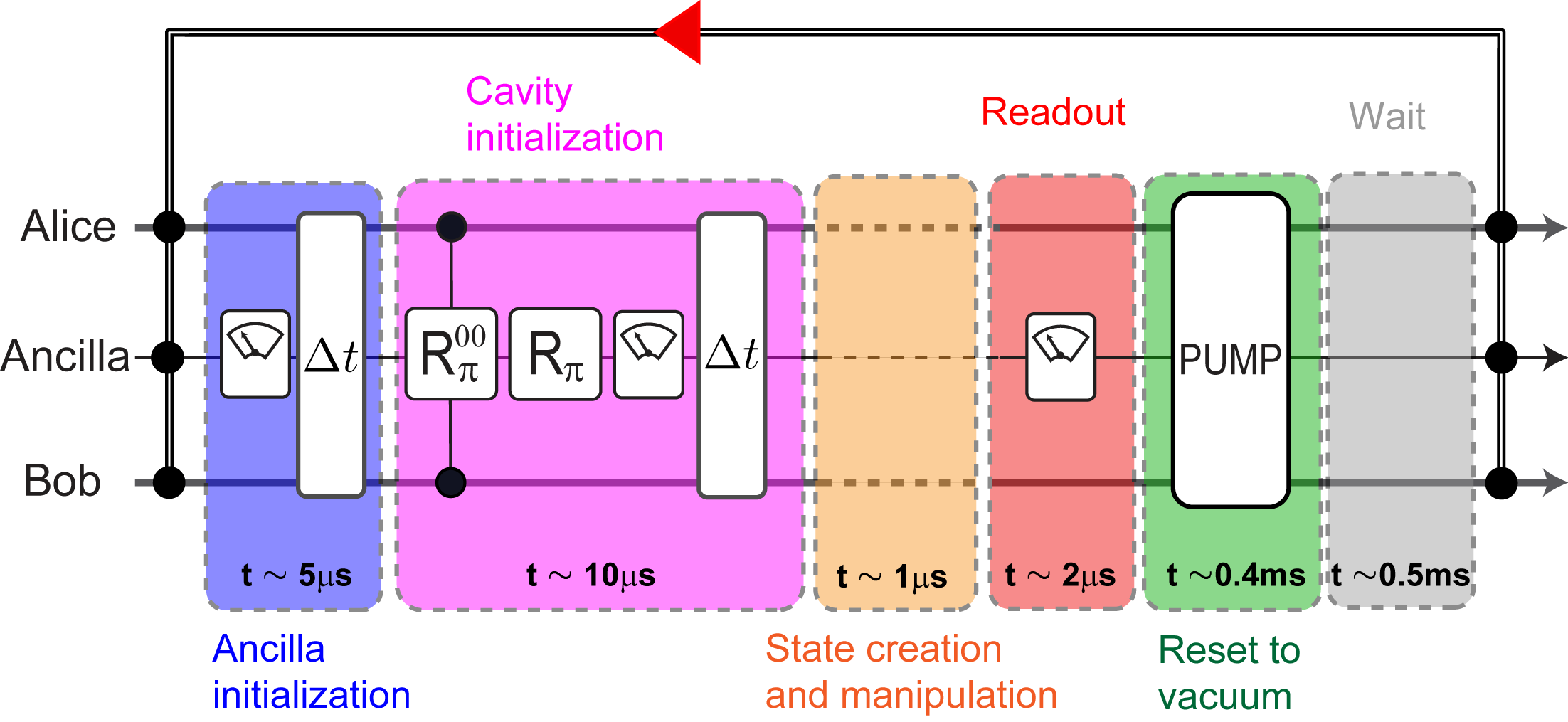}%[width=10.5cm]
    \caption{\textbf{Data acquisition flow chart with state initialization and reset.} Purification of the initial ground state against equilibrium excitations is implemented by performing two additional readouts of the ancilla state preceding each run of the actual experiment.  The first readout is performed without any ancilla/cavity manipulation, intended to filter out the cases where the ancilla starts in $|e\rangle$.  Then two $\pi$-pulses of ancilla $|g\rangle$-$|e\rangle$ rotations are applied, with the first conditional on $|0\rangle_A|0\rangle_B$, and the second unconditional.  The second readout of the ancilla state discriminates the presence or absence of photons in the two cavities, and allows filtering out the cases where either Alice or Bob starts with 1 or more photons.  The pulse sequences for the cat state generation and tomography follow this initialization sequence, and we select data from runs where both readouts return the outcome of $|g\rangle$.  We use a four-wave mixing method to reset the cavity state close to vacuum by applying three microwave pump tones at frequencies detuned from Alice, Bob and ancilla respectively for a duration of 400 $\mu s$.}
\end{figure*}

We use a field programmable gate array (FPGA) to operate both the quantum-control pulse sequences and the data acquisition process.  Our experiment does not rely on the real-time feedback capability of the FPGA, but our in-house programmed FPGA offers an important practical advantage in its ability to compute rather than store sideband-modulation waveforms.  This leads to minimal usage of waveform memory for a large number of different cavity displacements, allowing us to measure the joint Wigner function over a large number of points in the phase space in a single run.

Both cavity drives and transmon drives are generated by sideband-modulation of continuous-wave (CW) carrier tones produced by respective microwave generators.  The 4 FPGA analogue channels are used as 2 IQ-pairs that control the cavity drives to implement arbitrary cavity displacements.  Rotations of the transmon ancilla are controlled by another pair of IQ channels provided by an arbitrary waveform generator (AWG) synchronized to the FPGA via a digital marker.  This IQ pair controls both $|g\rangle$-$|e\rangle$ and $|e\rangle$-$|f\rangle$ transitions by using different intermediate frequencies (IF). 

Ancilla readout is performed by heterodyne measurement of the microwave transmission of a readout pulse through the two ports of the quasi-planar readout resonator near its resonance frequency.  Using the well-established cQED dispersive readout~\cite{wallraff_strong_2004}, the amplitude and phase of the transmitted signal depends on the quantum state of the ancilla.  This readout pulse is produced by a microwave generator (RO) gated by a FPGA digital channel.  The transmitted signal, after being amplified by the JPC, is further amplified by a high electron mobility transistor (HEMT) at 4K and a regular RF amplifier at room temperature.  The amplified signal is then mixed down to 50 MHz with the output of a ``local oscillator" (LO) microwave generator, and analyzed by the FPGA.  A split copy of the readout pulse is directly mixed with the LO without entering the refrigerator to provide a phase reference for the measured transmission.  

The long lifetimes of the cavities allow preparation of highly coherent cavity quantum states, but severely limits the rate one can repeat the measurement process.  (With $T_1\approx3$ ms for Alice, it takes 15-20 ms for the cavity photon number to naturally decay to the order of 0.01.)  Since tomographic measurement of the two-cavity quantum state requires large amounts of measurements, we implement four-wave mixing processes to realize fast reset for both cavities\cite{Zaki_pumps_2016}.  These processes effectively convert photons in Alice or Bob into photons in the short-lived readout resonator mode using three parametric pumping tones. %The relevant Hamiltonian terms are:
%\begin{align}
%\frac{H}{\hbar} = & \omega_a a^{\dagger}a + \omega_b b^{\dagger}b + \omega_c c^{\dagger}c\nonumber\\
%&+ g_a a^{\dagger}c + g_a^* a c^{\dagger} + g_b b^{\dagger}c + g_b^* b c^{\dagger} 
%\end{align}
%where $c^{\dagger}$ and $c$ are the creation and anihilation operators in the readout resonator.  $g_a$ and $g_b$ are the coupling strength induced by four-wave mixing, which is realized by three parametric pumping tones detuned by about 85 MHz below $\omega_a$, $\omega_b$ and $\omega_c$ respectively.  The reset method reduces the photon decay time of Alice and Bob to 10's of $\mu s$.  
We apply this reset operation for 400 $\mu$s, and acquire our experimental data with a repetition cycle of about 900 $\mu$s.

Ideally, in thermal equilibrium at our base temperature (20 mK) all modes (Alice, Bob, or ancilla) of our quantum system should be in their ground state.  However, in our experiment there is a non-negligible probability that any of the three modes is found in an excited state possibly due to insufficient thermalization.  These erroneous excited state populations, about 8$\%$ for the ancilla and 2-3$\%$ for Alice and Bob, can reduce the fidelity of the subsequently-prepared quantum state and the parity measurement.  To eliminate these effects, we perform two measurements as shown in Fig.~S3 to ``purify" the initial state of the system before we start each run of the experiment.  This is implemented by post-selecting the cases where our experiment starts from the ground state, $|g\rangle|0\rangle_A|0\rangle_B$, before any non-trivial quantum operation is performed, which amounts to about 80\% of the all the data acquired.  It should be noted that the use of post-selection here is purely for experimental convenience and does not compromise the deterministic nature of the generation of cat states.  One can in principle use real-time feedback to prepare the initial state and achieve a slightly higher data rate.  No post-selection beyond the ground state initialization is applied in our analyses of the two-mode cat state.

\subsection{Device Characterization}

\begin{figure*}
\floatbox[{\capbeside\thisfloatsetup{capbesideposition={right,center},capbesidewidth=5.0cm}}]{figure}[\FBwidth]
{\caption{\textbf{Measurement of cavity coherence.} \textbf{(A, B)} Measurement of the relaxation of a coherent state in Alice and Bob over time, which determine cavity relaxation times ($T_1$).  The vertical axes represent the overlap of the (coherent) state with the vacuum state.  \textbf{(C, D)} Ramsey interference experiment of a $\mathcal{N}(|0\rangle+|1\rangle)$ Fock state in Alice and Bob, which determine cavity coherence times ($T^{*}_2$). \textbf{(E)} Cavity resonance frequency of Bob extracted from Ramsey interference experiments over the course of eight months, showing long-term stability on the order of 100 Hz.}}
{\includegraphics[scale=0.34]{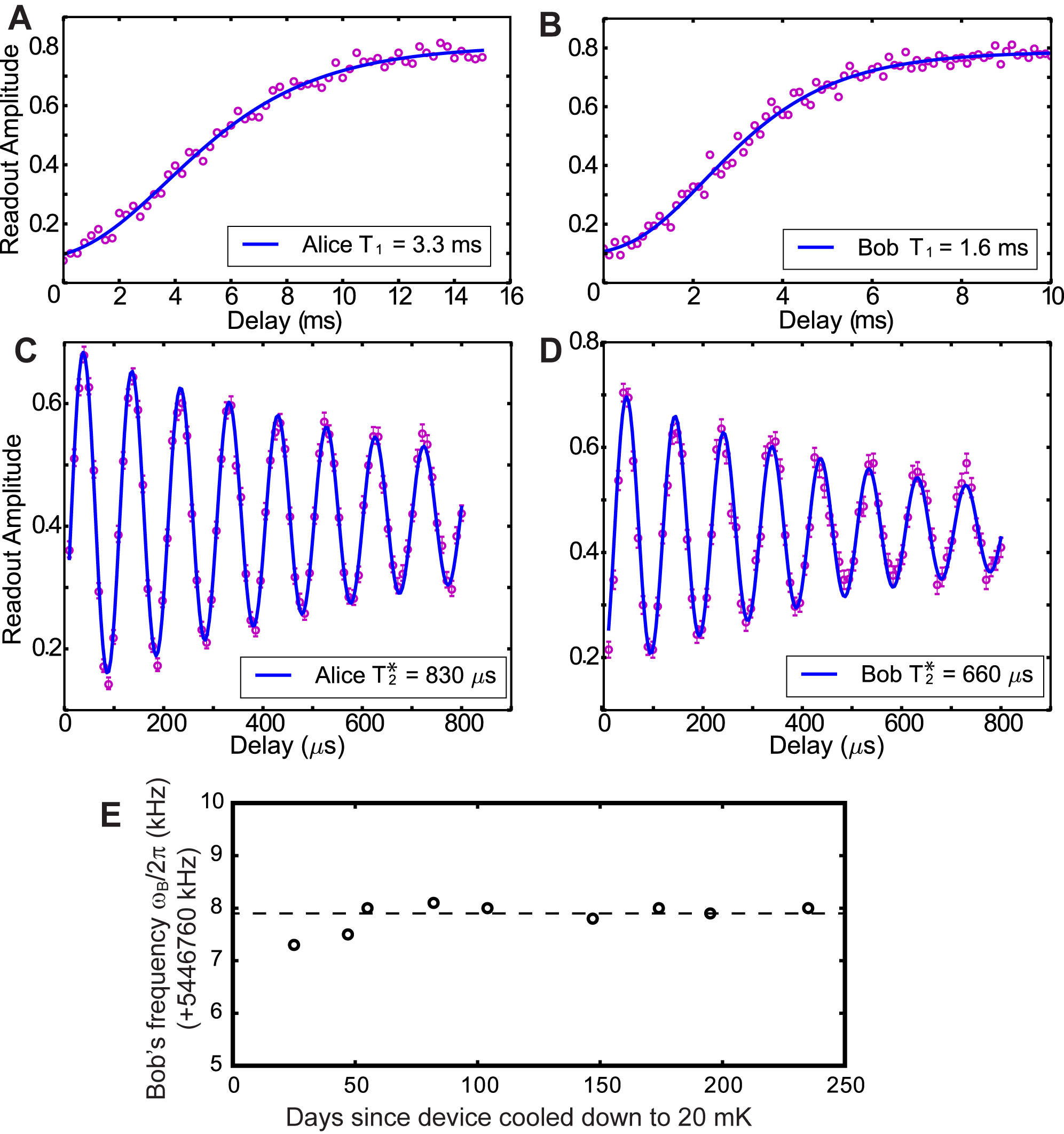}}%[width=10.8cm]
\end{figure*}

The Hamiltonian parameters and coherence properties of the device are mostly characterized by adapting established techniques, in particular, various forms of Ramsey interferometry.  The measured coherence times of the Alice, Bob and the ancilla are listed in Table I of the main text, and a more complete list is provided in Table S2.  We briefly comment on a few characterization methods that are noteworthy.  

Measurement of the coherence of higher excited states of a superconducting transmon is a relatively new topic recently reported in Ref.~\citen{peterer_coherence_2015}.  In addition to using a similar sequence to determine $T_{2,ef}^{*}$ between $|f\rangle$ and $|e\rangle$, we note that $T_{2,gf}^{*}$ (between $|f\rangle$ and $|g\rangle$) is more relevant to our experiment.  This $T_{2,gf}^{*}$ time cannot be simply derived from $T_{2,ge}^{*}$ and $T_{2,ef}^{*}$, but can be determined by a Ramsey experiment starting from a $\frac{1}{\sqrt2}(|g\rangle+|f\rangle)$ superposition.  

\begin{table}[b]
\centering  
\begin{tabular}{c c c c c} % centered columns (4 columns)
\hline\hline\\[-2ex]
		& $T_1$	& $T_2^{*}$	& $T_{2E}$	& $P_e$	 \\
\hline\\[-2ex]
Ancilla $|e\rangle$	& 65-75 $\mu$s	& 30-45 $\mu$s	& 55-65 $\mu$s & 7.5\%\\
Ancilla $|f\rangle$	& 26-32 $\mu$s	& 12-24 $\mu$s\footnote{$T_2^{*}$=11-19 $\mu$s for $|f\rangle$ vs.~$|g\rangle$, $T_{2E}$=18-26 $\mu$s for $|f\rangle$ vs.~$|g\rangle$.}	& 20-30 $\mu$s$^{a}$	 & 0.5\%\\
Alice	& 2.2-3.3 ms	& 0.8-1.1 ms	& N/A	& 2-3\%	\\
Bob		& 1.2-1.7 ms	& 0.6-0.8 ms	& N/A	& 2-3\%\\
Readout	& 260-290 ns	& N/A			& N/A	& $<$0.2\%\\[0.5ex]
% [1ex] adds vertical space
\hline
\end{tabular}
\caption{Energy relaxation time ($T_1$), Ramsey coherence time ($T_{2}^{*}$) , coherence time with Hahn echo ($T_{2E}$), and thermal population of the excited state ($P_e$) of all components when applicable.}
\end{table}

The coherence times of Alice and Bob are measured using Ramsey interference of $|0\rangle$ and $|1\rangle$ Fock states following a Selective Number Arbitrary Phase (SNAP) gate~\cite{heeres_cavity_2015} that prepares the Fock state superposition.  This method, which is described in Ref.~\citen{reagor_quantum_2015}, extracts the $T_2^{*}$ without being affected by high-order nonlinearities (Fig.~S4).  Together with the measurement of the cavity $T_1$, we find the cavity pure dephasing time $T_{\phi}=1/(\frac{1}{T_2^{*}}-\frac{1}{2T_1})=1.1\pm0.2$ ms for Alice and $0.9\pm0.2$ ms for Bob.  Such pure dephasing times can be fully explained by their dispersive frequency shifts due to the thermal excitation of the transmon ancilla~\cite{reagor_quantum_2015}.  The rate of the $|g\rangle\rightarrow|e\rangle$ transition of the transmon, $\Gamma_\uparrow$, can be determined from its $T_1$ and thermal population of the $|e\rangle$ state, $P_e$: $\Gamma_\uparrow = P_e/T_1 = 7.5\%/(70$ $\mu\rm{s})=1/(0.9$ ms).  In addition, the cavity Ramsey experiment also allows extraction of the precise resonance frequency of the cavity.  Over the course of 8 months while the device was continuously operated at 20 mK, we observed no slow drift of cavity frequency exceeding its linewidth ($\approx 200$ Hz).

The dispersive frequency shifts ($\chi$'s) are measured both in the frequency domain by spectroscopy techniques and in the time domain by a Ramsey-type qubit state revival experiment~\cite{vlastakis_deterministically_2013}) (Fig.~S5).  These techniques established previously for 2-level qubits, can both be extended for the 3-level artificial atom considered in this study.  The latter measurement is also a valuable procedure for tuning up the joint parity measurement.  

\begin{figure*}
\floatbox[{\capbeside\thisfloatsetup{capbesideposition={right,center},capbesidewidth=5.0cm}}]{figure}[\FBwidth]
{\caption{\textbf{Characterization of transmon-cavity dispersive coupling.} \textbf{(A)} Photon-number-splitting of the transmon $|g\rangle$-$|e\rangle$ transition frequency for a coherent state in Alice (blue) or Bob (red).  The vertical axis represents the probability of exciting the $|g\rangle$-$|e\rangle$ transition with a microwave tone at a frequency marked by the horizontal axis.  \textbf{(B)} Photon-number-splitting of the transmon $|e\rangle$-$|f\rangle$ transition frequency for a coherent state in Alice (blue) or Bob (red).  \textbf{(C, D)} Revival of a transmon state $\mathcal{N}(|g\rangle+|e\rangle)$ in the presence of a coherent state in (C) Alice or (D) Bob.  \textbf{(E, F)} Revival of a transmon state $\mathcal{N}(|e\rangle+|f\rangle)$ in the presence of a coherent state in (E) Alice or (F) Bob.}}
{\includegraphics[scale=0.34]{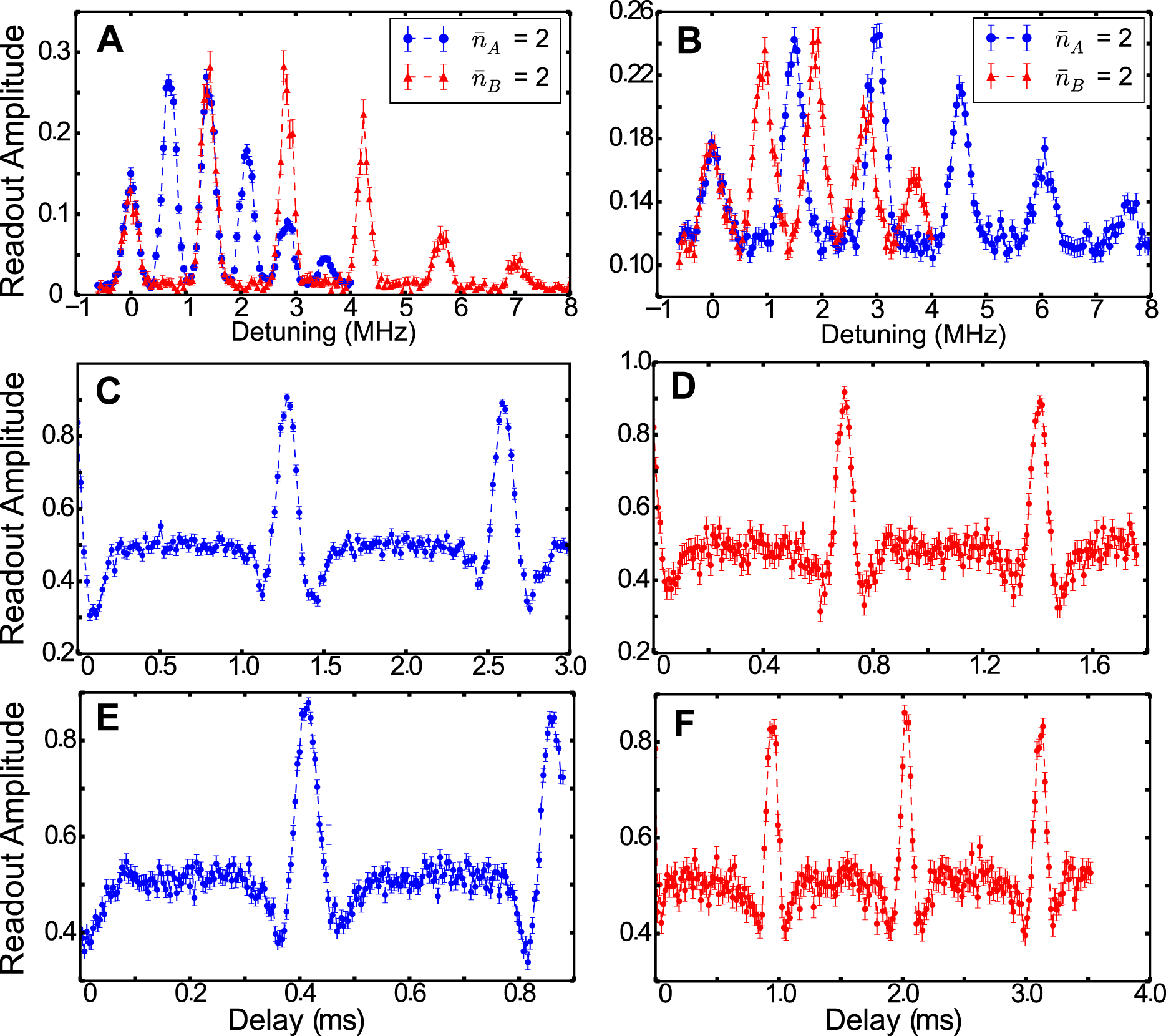}}%[width=10.8cm]
\end{figure*}

The self-Kerr effect of Alice or Bob ($K_A$ or $K_B$ terms in the Hamiltonian) can be characterized by the collapse and revival of a single-cavity coherent state~\cite{kirchmair_observation_2013}.  In addition, with two long-lived cavities, we are able to directly measure the inter-cavity cross-Kerr effect ($K_{AB}$) by observing the frequency shift of Bob in response to the presence of photons in Alice.

\subsection{Cat State Generation}
\label{sec:qcmap}

\begin{figure*}
\floatbox[{\capbeside\thisfloatsetup{capbesideposition={right,center},capbesidewidth=5.0cm}}]{figure}[\FBwidth]
{\caption{\textbf{Experimental protocol for deterministic generation of the two-mode cat state.} (A) Microwave control pulse sequences for state generation, written in a general form that allows different amplitude in two cavities.  (B) Cartoon representation of the step-by-step state evolution of Alice and Bob in their respective IQ planes.  Blue and red indicate the cavity photon probability distributions associated with the ancilla in $|g\rangle$ and in $|e\rangle$ respectively.  The parameters used for generating the two-mode cat state presented in the main text are $\alpha_1=\alpha_2=2.25$, $\Delta t=444$ ns, $\alpha'_1=2.25\times e^{-1.03i}$, $\alpha'_2=2.25\times e^{1.03i}$, $\alpha''_1=1.93\times e^{-0.48i}$, $\alpha''_2=1.93\times e^{0.65i}$.  After arriving in the final state $|\psi_{\pm}\rangle$, we rotate the IQ reference frame so that $\alpha\approx1.92$ is a real number.}}
{\includegraphics[scale=0.35]{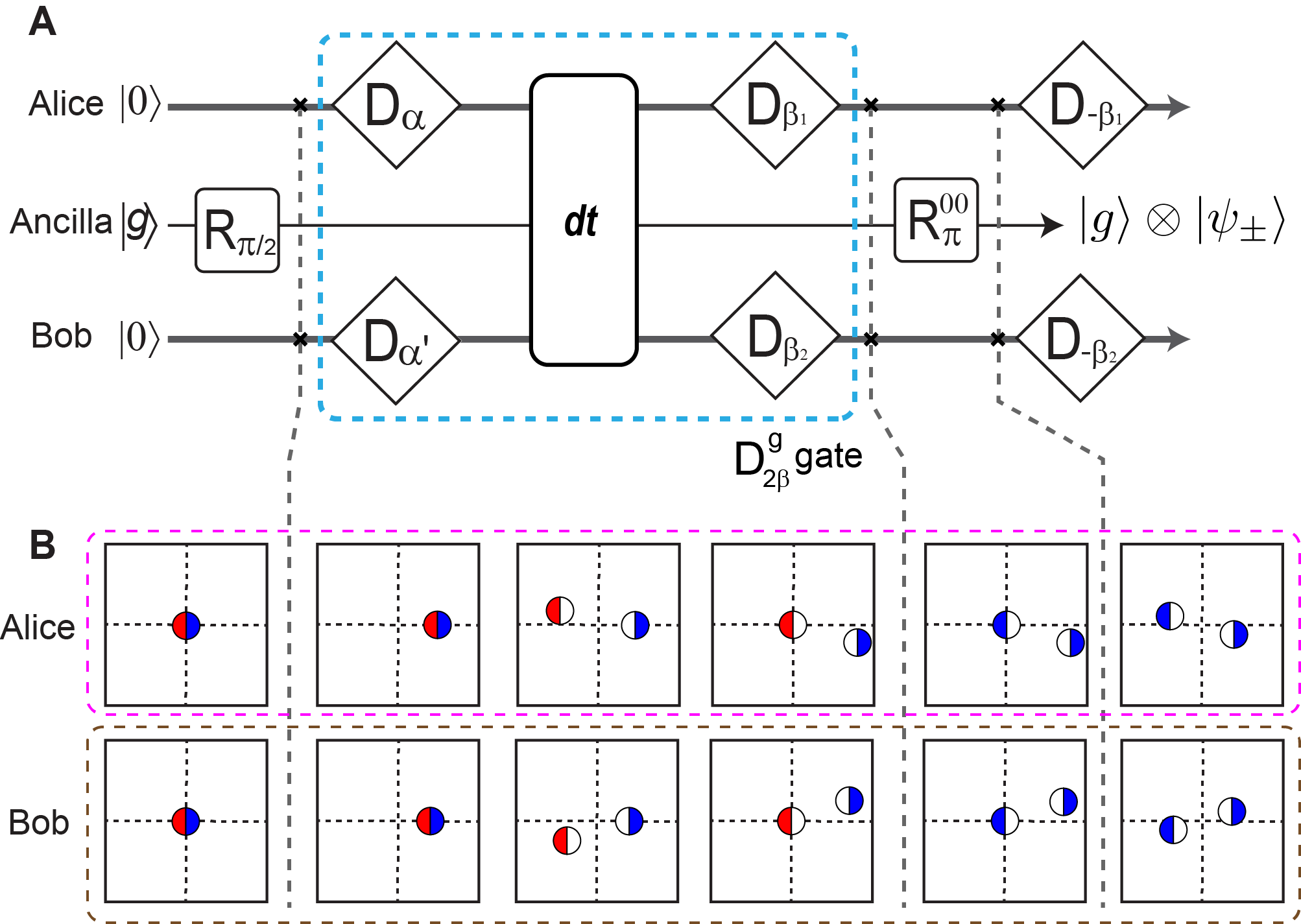}}%[width=10.8cm]
\end{figure*}

The two-mode cat state is generated deterministically using conditional operations between the ancilla and the two cavities.  As presented in the main text (Fig.~1C), the generation sequence is composed of the following steps~\cite{leghtas_deterministic_2013}: 1) preparing ancilla superposition ($R^{ge}_{\pi/2}$), 2) displacements of Alice and Bob conditional on the ancilla state ($D^{g}_{2\alpha}$), realizing a three-way entangling gate, 3) conditional flip (reset) of ancilla ($R^{00}_{\pi}$), disentangling it from the cavity state, 4) unconditional displacements of Alice and Bob ($D_{-\alpha}$) to center the cat state in the phase space (which is a trivial step purely for convenience of presentation). 

The conditional displacement ($D^{g}_{2\alpha}$) is the key step in this state generation process.  Although this operation can be directly implemented using cavity drives with a bandwidth smaller than the dispersive interaction strength ($\chi^{ge}_i$, $i$=A or B), such a method requires a rather long pulse duration (and therefore higher infidelity due to decoherence and Kerr effects).  Alternatively, for each cavity we use two unconditional displacements separated by a wait time $\Delta t$ in between to effectively realize $D^{g}_{2\alpha}$.  

During the wait time $\Delta t$, due to the dispersive interaction, cavity coherent states in both cavities accumulate conditional phases of $\phi_i=\chi^{ge}_i\Delta t$ if the ancilla is in $|e\rangle$:
\begin{equation}
U(\Delta t) =  \mathbb{I}_A\otimes\mathbb{I}_B\otimes|g\rangle\langle g| + e^{i\phi_A a^{\dagger}a}\otimes e^{i\phi_B b^{\dagger}b}\otimes|e\rangle\langle e|
\end{equation}
Using the IQ plane to describe the photon probability distribution in each cavity in the rotating frame, a coherent state $|\alpha'\rangle_i$ can be represented by a (Gaussian) circle that stays stationary when the ancilla is in $|g\rangle$, and rotates with the angular velocity $\chi^{ge}_{i}$ when the ancilla is in $|e\rangle$: $|\alpha'\rangle_i\rightarrow|\alpha'e^{i\chi_i^{ge}\Delta t}\rangle_i$.  Therefore, this conditional phase gate can split the cavity coherent state in phase space when the ancilla is prepared in $\frac{1}{\sqrt2}(|g\rangle+|e\rangle)$, effectively realizing a conditional displacement. A similar strategy has been previously implemented for a single cavity~\cite{vlastakis_deterministically_2013}.  The actual pulse sequences for creating the two-mode cat state and the resultant state evolution for the two cavity system are illustrated in Fig.~S6.  

%Therefore, after we prepare the system in $\frac{1}{2}(|g\rangle+|e\rangle)|\alpha'\rangle_A|\alpha'\rangle_B$, the cavity coherent state will split into two components, resulting in a global quantum state $\frac{1}{2}(|g\rangle|\alpha'\rangle_A|\alpha'\rangle_B+|e\rangle|\alpha'e^{i}\rangle_A|\alpha'\rangle_B)$

\begin{figure}[tbp]
    \centering
    \includegraphics[scale=0.35]{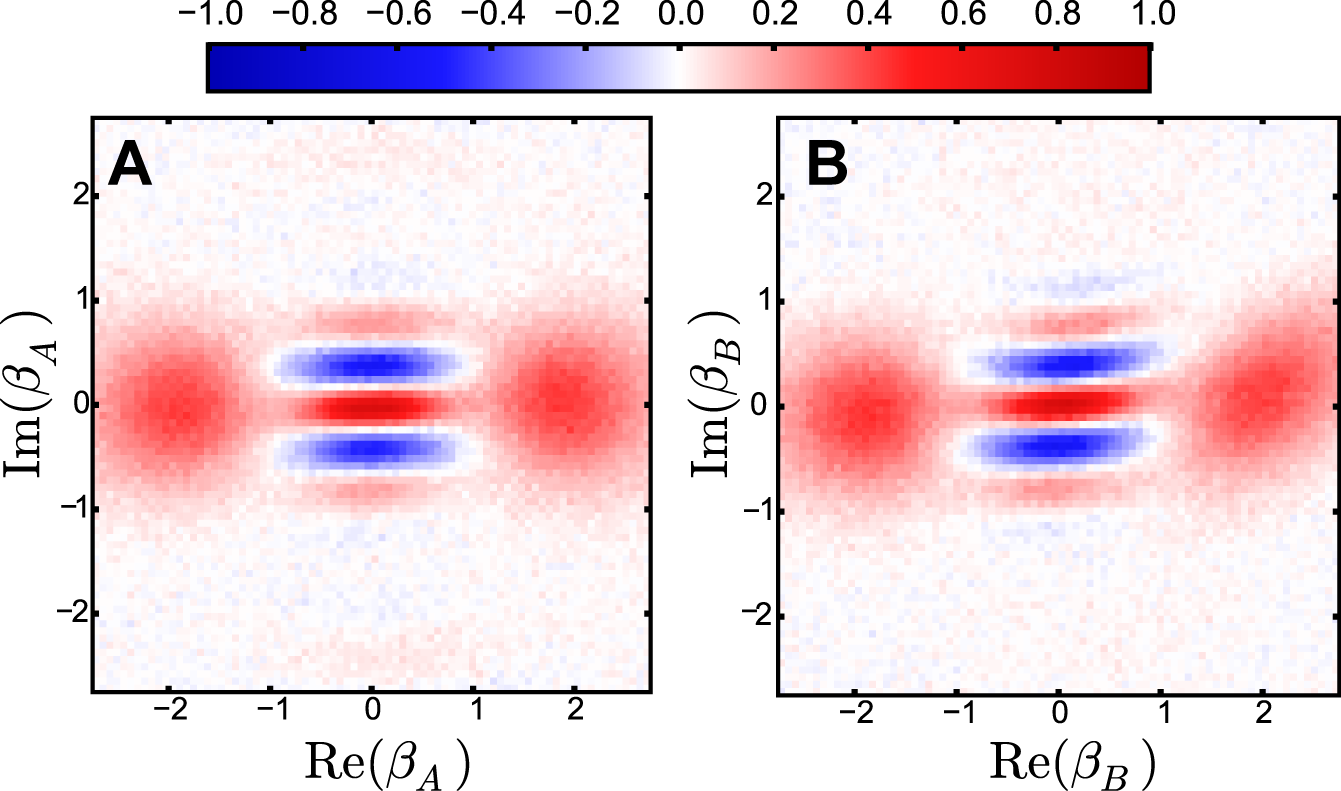}%[width=8.8cm]
    \caption{\textbf{Wigner tomography of single-mode cat states.} \textbf{(A)} Scaled single-cavity Wigner function of Alice, $\frac{\pi}{2}W_A(\beta_A)$, of the state $\mathcal{N}(|\alpha\rangle_A+|-\alpha\rangle_A)\otimes|0\rangle_B$. \textbf{(B)} Scaled single-cavity Wigner function of Bob of the state $|0\rangle_A\otimes\mathcal{N}(|\alpha\rangle_B+|-\alpha\rangle_B)$.}
\end{figure}

In this work we have focused on studying the two-mode cat state $|\psi_{\pm}\rangle$.  This state is equivalent to a cat state of the form $\mathcal{N}(|\sqrt{2}\alpha\rangle_C \pm |-\sqrt{2}\alpha\rangle_C)$ in a hybridized mode basis of $c^{\dagger}=\frac{1}{\sqrt2}(a^{\dagger}+b^{\dagger})$. Cat-state control over the $c^{\dagger}$ mode alone would not be fundamentally more useful than controlling a single local mode (as in previous works) if we cannot access modes other than $c^{\dagger}$.  However, it is easy to see that we can create cat states in Alice (or Bob) alone by omitting all pulses on Bob (or Alice) in Fig.~S6.  The resultant conventional single-mode cat state can be characterized by Wigner tomography of individual cavities (Fig.~S7).  Furthermore, we can reverse the cavity displacements in either Alice or Bob in Fig.~S6 to realize a cat state in the $\frac{1}{\sqrt2}(a^{\dagger}-b^{\dagger})$ basis.  The important message from the present experiment is that we can create cat states in an arbitrary basis spanned by two modes. In fact, it can be shown that the dispersive Hamiltonian permits universal quantum control of this system. \cite{Chuang_1997} \cite{Silveri_kittenCode_2016}

\subsection{Joint Parity Measurement}
\label{sec:parity}

Measurement of the joint photon number parity, $P_J$, is critical to our study of two-cavity quantum states.  Here we explain in detail the quantum operations and required system parameters to realize such measurements.

Photon parity measurement of a single-cavity quantum state using an ancilla qubit (using only $|g\rangle$ and $|e\rangle$ levels) has been previous demonstrated in Ref.~\citen{sun_tracking_2014} (see Fig.~1 therein).  This single-cavity protocol is applicable to either one of our cavities when the other cavity is in the vacuum state.  It uses the dispersive interaction $\chi^{ge}_i$ to map even-photon-number and odd-photon-number states in the cavity of interest (i=A or B) to different qubit levels.  This is realized by two $\pi/2$ rotations of the qubit, $R^{ge}_{\pi/2}$ (around the same X-axis), separated by a wait time of $\pi/\chi^{ge}_i$.  For example, if Bob is in the vacuum state ($b^{\dagger}b=0$), the conditional phase shift described in Eq.~(S2) over the time $\Delta t=\pi/\chi^{ge}_A$ is:
\begin{equation}
U(\pi/\chi^{ge}_A)= C^A_{\pi} = \mathbb{I}\otimes|g\rangle\langle g| + e^{i\pi a^{\dagger}a}\otimes|e\rangle\langle e|
\end{equation}
This is equivalent to a qubit Z-rotation of $\pi$ conditioned on the photon number in Alice being odd because $e^{i\pi a^{\dagger}a} = P_A$.  Therefore the whole sequence $R^{ge}_{\pi/2}C^A_{\pi}R^{ge}_{\pi/2}$ flips the qubit if and only if the photon number parity in Alice is even, and subsequent readout of the qubit state measures the parity.

The control and measurement sequence described above can in principle be directly implemented in our experiment to measure the joint photon number parity if $\chi_A^{ge}$ is exactly equal to $\chi_B^{ge}$.  This is because for a wait time of $\Delta t= \pi/\chi_A^{ge}(=\pi/\chi_B^{ge})$, from Eq.~(S2) we have:
\begin{align}
U(\pi/\chi_i^{ge})
&=C^{A}_{\pi}C^B_{\pi}\nonumber\\
&=\mathbb{I}\otimes|g\rangle\langle g| + P_AP_B\otimes|e\rangle\langle e|
\label{eq:PJunitary}
\end{align}
Noting $P_J=P_AP_B$, an identical control sequence of $R^{ge}_{\pi/2}U(\Delta t)R^{ge}_{\pi/2}$ followed by a qubit readout would achieve the joint parity measurement.  However, without strictly identical $\chi_A^{ge}$ and $\chi_B^{ge}$, the phase accumulation in one cavity is faster than the other, and it is in general not possible to realize parity operators in both cavities simultaneously using this simple protocol.  Moreover, for a general two-cavity quantum state, this sequence can not measure a  single-cavity parity operator ($P_A$ or $P_B$) due to inevitable entanglement between the ancilla and the photons in the other cavity during the process. 

As noted in the main text, we introduce a technique for measuring $P_J$ with less stringent requirements on Hamiltonian parameters by exploiting the $|f\rangle$-level of the ancilla.  This method is most helpful when the $|e\rangle\rightarrow|g\rangle$ transition of the ancilla shows stronger interaction with Bob ($\chi_B^{ge}>\chi_A^{ge}$), while the $|f\rangle\rightarrow|e\rangle$ transition shows stronger interaction with Alice ($\chi_A^{ef}>\chi_B^{ef}$).  This is physically realized by engineering the ancilla frequency to lie between the two cavities, \textit{i.e.}$\omega_A<\omega_{ef}<\omega_{ge}<\omega_B$.

%By manipulating the ancilla in different superposition states among the three levels, we can realize conditional phase gate associated with $\chi_i^{ge}$, $\chi_i^{ef}$, $\chi_i^{gf}$ or a combination of the three with arbitrary weight.  This additional degree of freedom not only allows for exact joint parity measurement $P_j$, but also enables parity measurement of each cavity, ($P_a$ and $P_b$) individually without affecting the other.

Considering the quantum state with two cavities and three ancilla levels in general, the unitary evolution for any wait time $\Delta t$ is:
\begin{align}
U(\Delta t) = & \mathbb{I}_A\otimes\mathbb{I}_B\otimes|g\rangle\langle g| + e^{i\phi_A a^{\dagger}a}\otimes e^{i\phi_B b^{\dagger}b}\otimes|e\rangle\langle e|\nonumber\\
& + e^{i\phi'_A a^{\dagger}a}\otimes e^{i\phi'_B b^{\dagger}b}\otimes|f\rangle\langle f|
\end{align}
where
\begin{align}
&\phi_A = \chi_A^{ge}\Delta t, &\phi_B = \chi_B^{ge}\Delta t \nonumber\\
&\phi'_A = \chi_A^{gf}\Delta t, &\phi'_B = \chi_B^{gf}\Delta t
\end{align}
Here we define $\chi_A^{gf}\equiv\chi_A^{ge}+\chi_A^{ef}$ and $\chi_B^{gf}\equiv\chi_B^{ge}+\chi_B^{ef}$.
 Therefore, the two cavities simultaneously acquire conditional phases in their coherent state components at relative rates that differ for $|e\rangle$ and $|f\rangle$.

\begin{figure*}
\floatbox[{\capbeside\thisfloatsetup{capbesideposition={right,center},capbesidewidth=6.0cm}}]{figure}[\FBwidth]
{\caption{\textbf{Two experimental protocols for joint parity measurement.} (A) The same control pulse sequences shown in Fig.~1(C) of the main text. The conditional phase gates are realized by the wait time $\Delta t_1$ and $\Delta t_2$ while the ancilla is in a superposition state.  Here $\phi_A = \chi_A^{ge}\Delta t, \phi_B = \chi_B^{ge}\Delta t, \phi'_A = \chi_A^{gf}\Delta t, \phi'_B = \chi_B^{gf}\Delta t$.  We use $\Delta t_1=0$ ns, and $\Delta t_2=184$ ns for $P_J$ measurements presented in the main text, which involves a small systematic phase error in the parity mapping operation.  One can choose larger $\Delta t_1$ and $\Delta t_2$ to avoid this error but at the cost of more decoherence and Kerr effects.  (B) An alternative parity measurement sequence involving the $|e\rangle$-$|f\rangle$ superposition of the ancilla, where $\phi_A = \chi_A^{ef}\Delta t, \phi_B = \chi_B^{ef}\Delta t$ instead.  We use $\Delta t_1=28$ ns, and $\Delta t_2=168$ ns for $P_J$ measurements presented in Fig.~S10.  This method reduces the parity mapping phase error at the cost of more ancilla pulse errors.}}
{\includegraphics[scale=0.36]{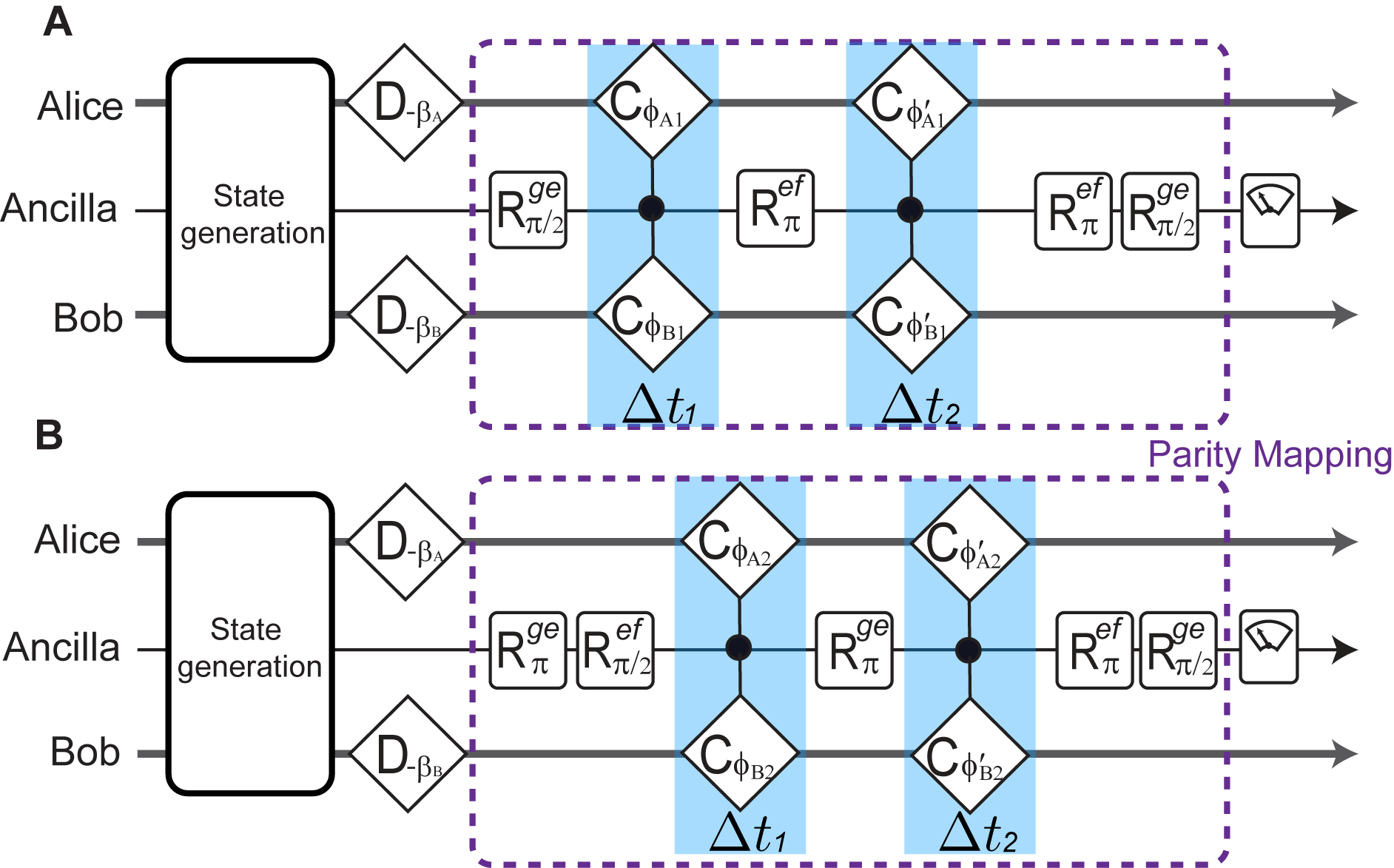}%[width=10.8cm]
\label{paritymapping}}
\end{figure*}

One possible pulse sequence for such $P_J$ measurement using three ancilla levels is shown in Fig.~1C of the main text and reproduced as Fig.~\ref{paritymapping}A.  For a given two-cavity quantum state $\Psi_{AB}$, We first use a $R_{\pi/2}^{ge}$ rotation to prepare the ancilla in $\frac{1}{\sqrt2}(|g\rangle+|e\rangle)$.  Then a wait time $\Delta t_1$ imparts phases $\phi_{A1}=\chi_A^{ge}\Delta t_1$ and $\phi_{B1}=\chi_B^{ge}\Delta t_1$ to the two cavities for the $|e\rangle$ component of the state:
\begin{align}
&\Psi_{AB}\otimes\frac{1}{\sqrt2}(|g\rangle+|e\rangle)\nonumber\\
&\Rightarrow\frac{1}{\sqrt2}\big[\Psi_{AB}\otimes|g\rangle+e^{i\phi_{A1} a^{\dagger}a} e^{i\phi_{B1} b^{\dagger}b}\Psi_{AB}\otimes|e\rangle\big]
\end{align}
Next, the $|e\rangle$ component in this intermediate state is converted to $|f\rangle$ by a $\pi$ rotation in the $|e\rangle$-$|f\rangle$ space, $R_{\pi}^{ef}$.  Subsequently a second wait time $\Delta t_2$ leads to a second simultaneous conditional phase gate, imparting  phases $\phi_{A2}=\chi_A^{gf}\Delta t_2$ and $\phi_{B2}=\chi_B^{gf}\Delta t_2$ to the two cavities for the now $|f\rangle$ component of the state:
\begin{align}
&\frac{1}{\sqrt2}\big[\Psi_{AB}\otimes|g\rangle+e^{i\phi_{A1} a^{\dagger}a} e^{i\phi_{B1} b^{\dagger}b}\Psi_{AB}\otimes|f\rangle\big]\Rightarrow\nonumber\\
&\frac{1}{\sqrt2}\big[\Psi_{AB}\otimes|g\rangle+e^{i(\phi_{A1}+\phi_{A2}) a^{\dagger}a} e^{i(\phi_{B1}+\phi_{B2}) b^{\dagger}b}\Psi_{AB}\otimes|f\rangle\big]
\end{align}
The $|f\rangle$ component is then converted back to $|e\rangle$ by another $R_{\pi}^{ef}$ pulse.  If we can find $\Delta t_1$ and $\Delta t_2$ so that:
\begin{align}
\phi_{A1}+\phi_{A2}=\chi_A^{ge} \Delta t_1 + \chi_A^{gf}\Delta t_2 = \pi \nonumber\\
\phi_{B1}+\phi_{B2}=\chi_B^{ge} \Delta t_1 + \chi_B^{gf}\Delta t_2 = \pi
\label{eq:phasesimul}
\end{align}
the obtained quantum state is:
\begin{align}
\frac{1}{\sqrt2}\big[\Psi_{AB}\otimes|g\rangle+P_J\Psi_{AB}\otimes|e\rangle\big]
\end{align}
effectively realizing the simultaneous controlled $\pi$-phase gate ($C^A_\pi C^B_\pi$) in Eq.~(\ref{eq:PJunitary}). 
Finally a $R_{\pi/2}^{ge}$ pulse completes the projection of joint parity to the ancilla $|g\rangle$, $|e\rangle$ levels, ready for readout through the readout resonator. 

The condition for finding non-negative solutions for $\Delta t_1$ and $\Delta t_2$ in Eq.~(\ref{eq:phasesimul}) is that $\chi_A^{ge}-\chi_B^{ge}$ and $\chi_A^{gf}-\chi_B^{gf}$ have opposite signs.  In essence, the cavity that acquires phase slower than the other at $|e\rangle$ due to smaller $\chi_i^{ge}$ is allowed to catch up at $|f\rangle$ using its larger $\chi_i^{gf}$. 

\begin{figure}[tbp]
    \centering
    \includegraphics[scale=0.35]{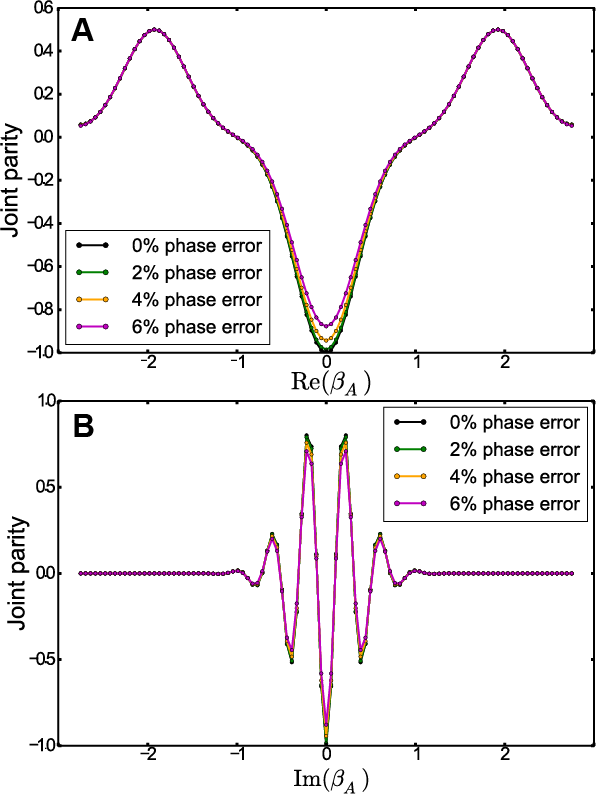}%[width=8.0cm]
    \caption{\textbf{Sensitivity of Wigner tomography to the phase error of joint parity mapping.} By numerical simulation, we consider non-ideal joint parity mapping where an operator $\mathbb{I}\otimes|g\rangle\langle g|+e^{-i\epsilon\pi a^{\dagger}a}\otimes e^{i\epsilon\pi b^{\dagger}b}P_J\otimes|e\rangle\langle e|$ is applied.  ($\epsilon=0$ corresponds to perfect joint parity mapping.)  \textbf{(A)} The Re$(\beta_A)$=Re$(\beta_B)$ and \textbf{(B)} the Im$(\beta_A)$=Re$(\beta_B)$ line-cuts of the simulated scaled ``joint Wigner function" $\langle e^{i\epsilon\pi(b^{\dagger}b-a^{\dagger}a)} P_J\rangle$ are plotted for the two-mode cat state $|\psi_{-}\rangle$ ($\alpha=1.92$) for various $\epsilon$.  No other non-ideality is included in this simulation.  Our measurement condition using the protocol in Fig.~\ref{paritymapping}A corresponds to $\epsilon\approx0.03$.}
\end{figure}

It should be noted that such relative relation of the $\chi$'s is just a practically preferred condition rather than an absolute mathematical requirement.  This is because parity mapping can be achieved whenever both cavities acquire a conditional phase of $\pi$ modulo $2\pi$.  It is always possible to allow extra multiples of $2\pi$ phases applied to the cavity with stronger dispersive coupling to the ancilla, although it increases the total gate time and incurs more decoherence.  The essential ingredient in engineering the $P_J$ operator is the extra tuning parameter $\Delta t_2$ (in addition to $\Delta t_1$) that allows two equations such as Eq.~(\ref{eq:phasesimul}) to be simultaneously satisfied.  

This extra degree of freedom also enables measurement of the photon number parity of a single cavity, $P_A$ or $P_B$, for an arbitrary two-cavity quantum state.  This can be realized with the same control sequences (Fig.~\ref{paritymapping}A), choosing wait times such that one cavity acquires a conditional $\pi$ phase (modulo 2$\pi$) while the other acquires 0 phase (modulo 2$\pi$).  For example, to measure $P_A$ we use $\Delta t_1$ and $\Delta t_2$ satisfying:
\begin{align}
\phi_{A1}+\phi_{A2}=\chi_A^{ge} \Delta t_1 + \chi_A^{gf}\Delta t_2 = \pi \,(\rm{mod}\, 2\pi) \nonumber\\
\phi_{B1}+\phi_{B2}=\chi_B^{ge} \Delta t_1 + \chi_B^{gf}\Delta t_2 = 0 \,(\rm{mod}\, 2\pi)
\end{align}

Fig.~\ref{paritymapping}B shows an alternative version of joint parity mapping protocol, which uses more ancilla operations, but is better adapted to to a larger parameter space of $\chi$'s.  In this protocol, the ancilla spends time at  the $|e\rangle$-$|f\rangle$ superposition so that conditional phases proportional to $\chi_i^{ef}$ are applied to the cavities.  To achieve joint parity mapping, the two time intervals $\Delta  t_1$ and $\Delta t_2$ should satisfy:
\begin{align}
\phi_{A1}+\phi_{A2}=\chi_A^{ef} \Delta t_1 + \chi_A^{gf}\Delta t_2 = \pi \,(\rm{mod}\, 2\pi)\nonumber\\
\phi_{B1}+\phi_{B2}=\chi_B^{ef} \Delta t_1 + \chi_B^{gf}\Delta t_2 = \pi \,(\rm{mod}\, 2\pi)
\end{align}
which can avoid the use of extra $2\pi$ phases to $\chi_A^{ef}-\chi_B^{ef}$ has opposite sign versus $\chi_A^{gf}-\chi_B^{gf}$.

\begin{figure}
\includegraphics[scale=0.34]{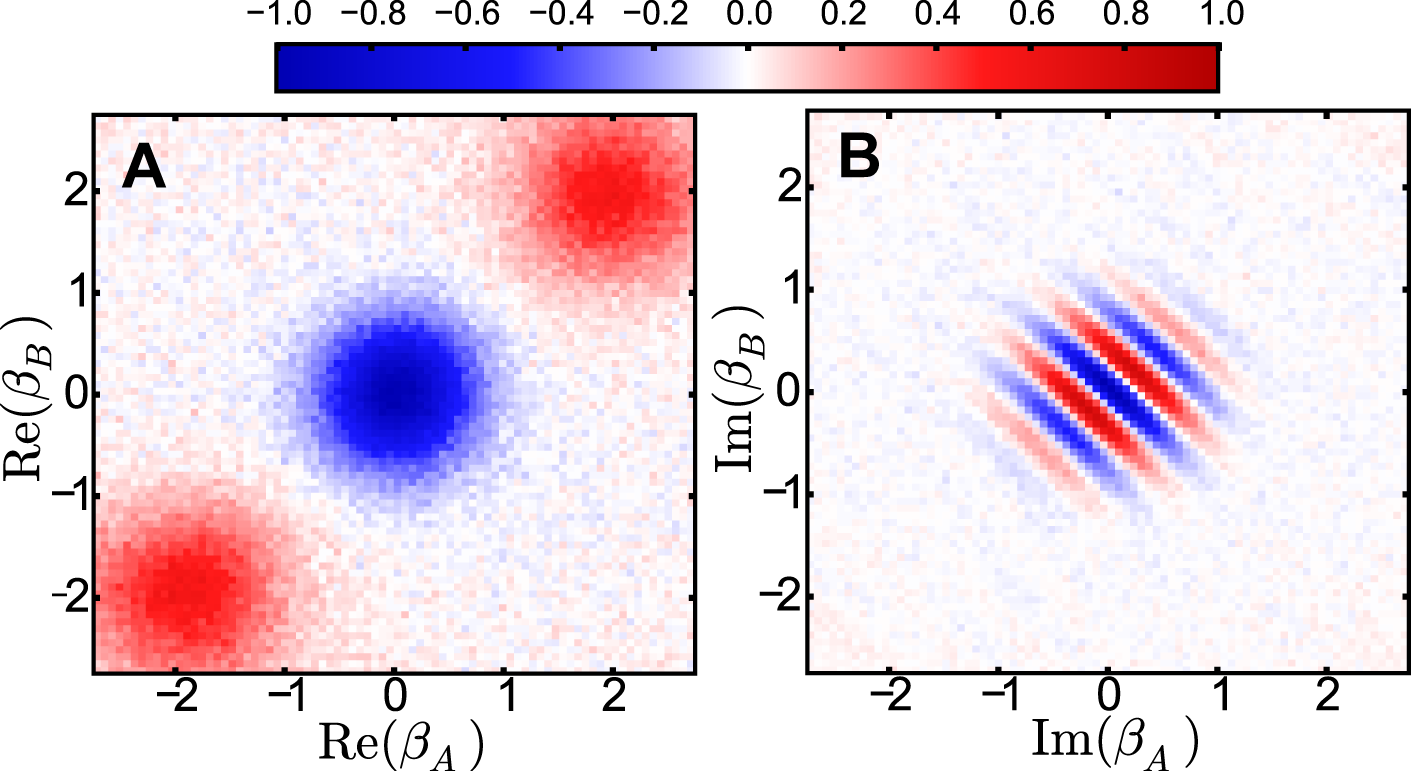}%[width=11.0cm]
%\floatbox[{\capbeside\thisfloatsetup{capbesideposition={right,center},capbesidewidth=6cm}}]{figure}[\FBwidth]
{\caption{\textbf{Joint Wigner function remeasured using an alternative parity mapping protocol.} To perform the same joint Wigner tomography of the same state ($|\psi_{-}\rangle$ with $\alpha=1.92$) as shown in Fig.~3 of the main text, we can also use an alternative joint parity mapping protocol (shown in Fig.~\ref{paritymapping}B).  The resultant 2D plane cuts of the scaled Wigner function, $\langle P_J(\beta_A,\beta_B)\rangle$ are almost identical to Fig.~3, where the measured minimum joint parity is about -0.80.}}
%{\includegraphics[scale=0.35]{fig_gef_cat.png}}%[width=11.0cm]
\label{gef_cat}
\end{figure}

Experimentally, choices of the parity mapping sequence and gate times involve trade-offs in various aspects such as pulse speed/bandwidth and coherence time.  We have measured joint parity (and subsequently Wigner functions) using both protocols.  For the sequence of Fig.~\ref{paritymapping}A, $\Delta t_1=0$, $\Delta t_2=184$ ns was experimentally implemented.  For the sequence of Fig.~\ref{paritymapping}B, $\Delta t_1=28$ ns, $\Delta t_2=168$ ns was used.  The actual effective wait time was longer due to the non-zero duration (16 ns) of each ancilla rotation.  The first protocol, with this choice of wait times, does not yield the exact $\pi$ phases required for exact parity mapping (We estimate $\phi_{A1}+\phi_{A2}=0.97\pi$ and $\phi_{B1}+\phi_{B2}=1.03\pi$.  %(By minimizing $\Delta t_1$, we take advantage of the closeness of the two $\chi^{gf}$'s, $\chi^{gf}_A=2.25$ MHz and $\chi^{gf}_B=2.34$ MHz.) 
These phase errors lead to an estimated infidelity of the joint parity measurement of about 3\% for the two-cavity states in this study (Fig.~S9).  Exact phases can be achieved with longer wait times so that $\phi_{A1}+\phi_{A2}=3\pi$ and $\phi_{B1}+\phi_{B2}=5\pi$, but the infidelity due to decoherence and high-order Hamiltonian terms outweighs the benefits.  In principle, the second protocol that achieves exact $\pi$ phases at relatively short total gate time should be more advantageous.  However, using the second protocol, we observe visibly identical results of joint Wigner tomography of the two-mode cat states with fidelity nearly equal to the first protocol (Fig.~S10).  This is attributed to the extra infidelity from the more complicated ancilla rotations involved in the second protocol (due to pulse bandwidth limitations and unwanted ancilla population mixing, see Section~\ref{sec:fidelity} of this Supplementary).  All joint Wigner tomography shown in the main text are measured using the first protocol (matching Fig.~1C).

Single cavity Wigner tomography is also performed using the protocol of Fig.~\ref{paritymapping}A, with $\Delta t_1=688$ ns, $\Delta t_2=0$ for $P_A$, and $\Delta t_1=660$ ns, $\Delta t_2=204$ ns for $P_B$.  Moreover, for $P_A$ measurement the $R^{ef}_\pi$ pulses were skipped, taking advantage of the fact that $\chi_B^{ge}\approx 2\chi_A^{ge}$.

\section*{Supplementary Text}

\subsection{Extended Data for Two-mode Cat State}

The joint tomography of a two-cavity quantum state is described by a four dimensional joint Wigner function $W_J(\beta_A,\beta_B)$.  The most informative plane cuts of $W_J$, along Re$(\beta_A)$-Re$(\beta_B)$ and Im$(\beta_A)$-Im$(\beta_B)$, have been presented in the main text.  This section provides additional data of our measurement on the joint quantum state of Alice and Bob. 

Fig.~S11 shows the joint Wigner function of $|\psi_{+}\rangle$ ($\alpha=1.92$) measured in the natural single-cavity IQ planes, or plane cuts of $W_J$ along Re$(\beta_A)$-Im$(\beta_A)$ and Re$(\beta_B)$-Im$(\beta_B)$ respectively (and through the origin).  Both plane cuts contain interference fringes along the imaginary axis as expected.  No features of Gaussian coherent states are observed in either figure because the underlying coherent states $|\alpha\rangle_A$$|\alpha\rangle_B$ and $|-\alpha\rangle_A$$|-\alpha\rangle_B$ are located outside these plane cuts.

To assist visualization of the two-mode cat state, the joint Wigner function of $|\psi_{-}\rangle$ ($\alpha=1.92$) is further presented as a movie, which is a direct extension of Fig.~3 of the main text.  It shows 21 frames of Re$(\beta_A)$-Re$(\beta_B)$ plane cuts as a function of the Im$(\beta_A)$,Im$(\beta_B)$ coordinates.

\begin{figure}[tbp]
    \centering
    \includegraphics[scale=0.35]{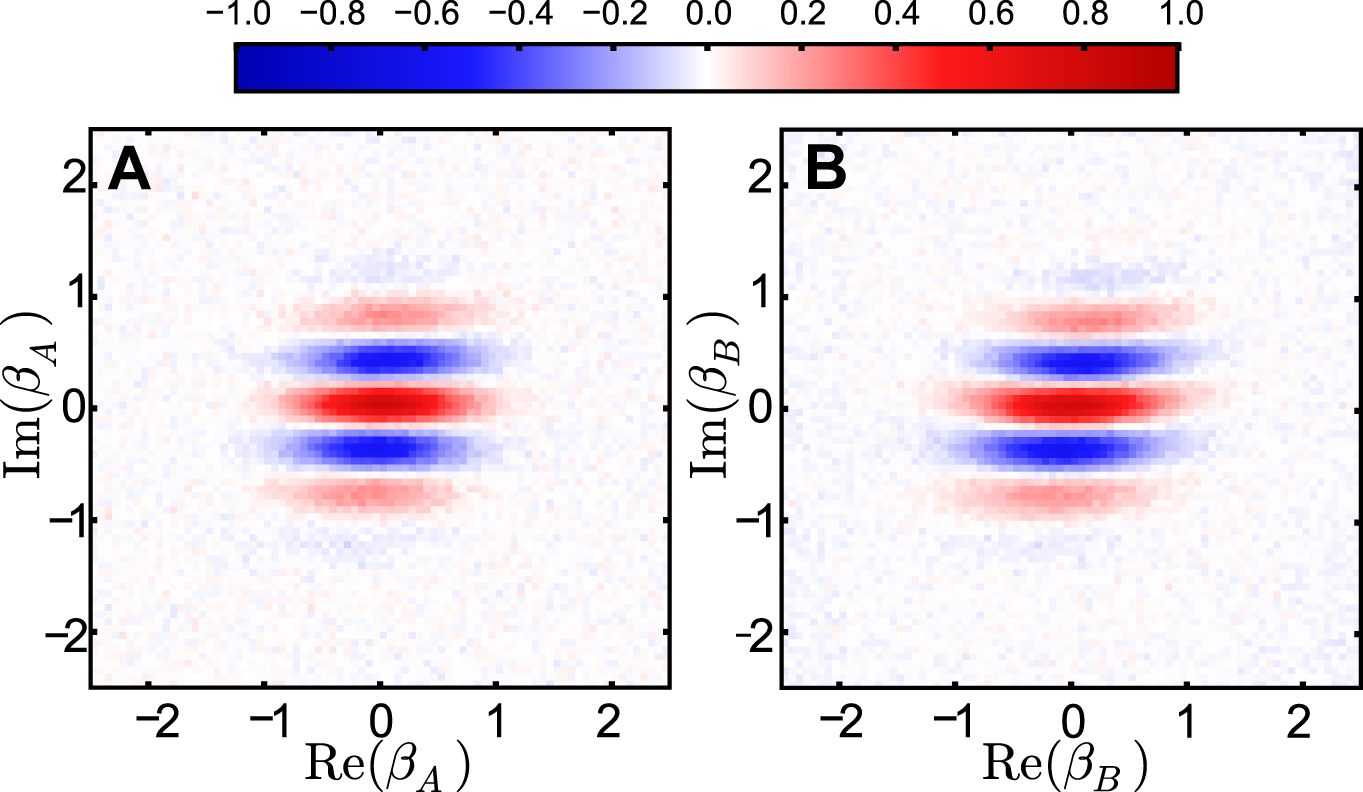}%[width=8.8cm]
    \caption{\textbf{Joint Wigner function of the two-mode cat state along the single-cavity I-Q plane.} 2D plane cuts of the scaled joint Wigner function, $\langle P_J(\beta_A,\beta_B)\rangle$, of the state $|\psi_{+}\rangle$ ($\alpha=1.92$) along \textbf{(A)} the axes Re($\beta_A$)-Im($\beta_A$) and \textbf{(B)} the axes Re($\beta_B$)-Im($\beta_B$). }
\end{figure}

\begin{figure}[tbp]
    \centering
    \includegraphics[scale=0.35]{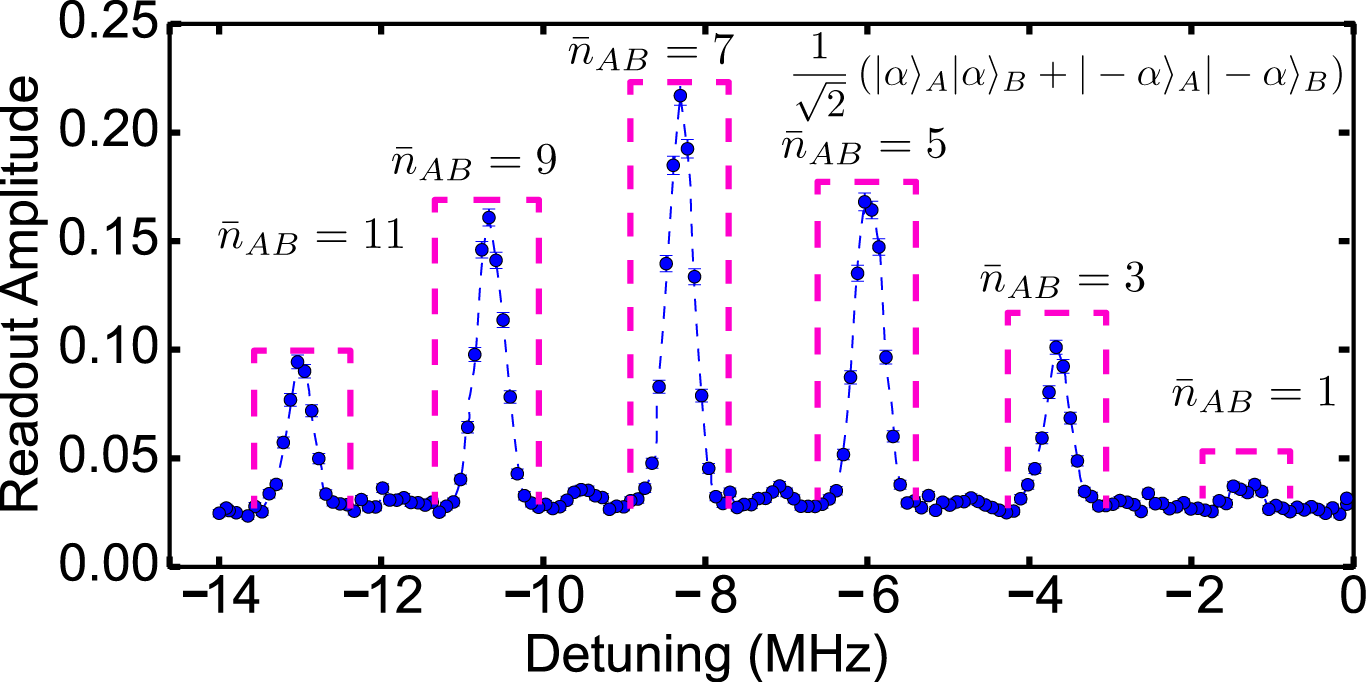}%[width=8.8cm]
    \caption{\textbf{Spectroscopic measurement of the two-mode cat state.} The data points show the probability of exciting the $|g\rangle\rightarrow|f\rangle$ two-photon transition of the transmon ancilla as a function of the drive frequency $\omega_d$ after an odd-parity two-mode cat state $|\psi_{-}\rangle$ ($\alpha=1.92$) is prepared in Alice and Bob.  Such two-photon transitions can be excited at $\omega_{gf/2}=(\omega_{ge}+\omega_{ef})/2=4.82047$ GHz if both cavities are in the vacuum state, but shifts towards lower frequency by $\chi^{gf}_A/2$ for each photon in Alice and or $\chi^{gf}_B/2$ for each photon in Bob.  %The x-axis is the detuning of the drive, $\omega_d-\omega_{gf/2}$.  
Because $\chi^{gf}_A\approx\chi^{gf}_B$, each peak in the spectrum can be identified with a total photon number in two cavities.  This measurement therefore shows the probability distribution of total photon numbers.  Only peaks associated with odd total number of photons are pronounced in the measurement, confirming the odd joint parity of the state $|\psi_{-}\rangle$.}
    \label{catspec}
\end{figure}

\begin{figure}
\includegraphics[scale=0.35]{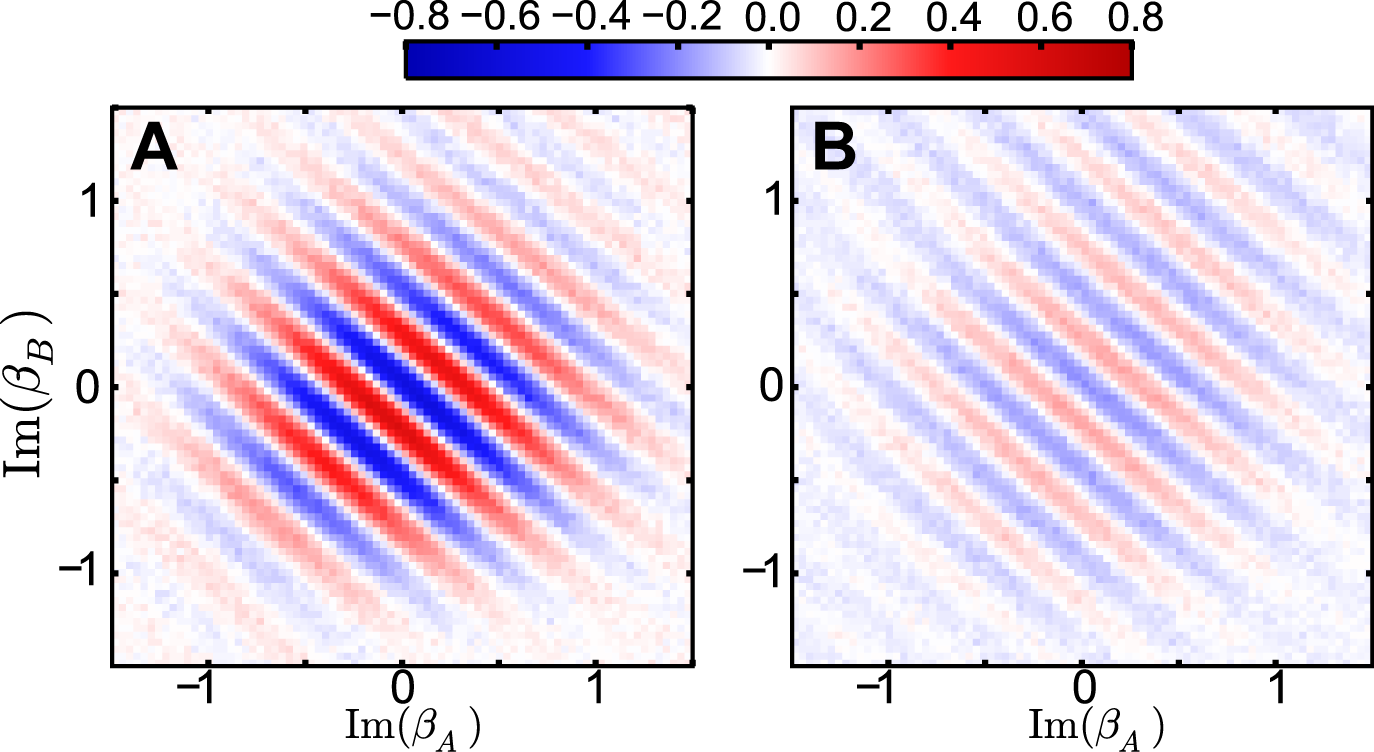}%[width=11.5cm]
%\floatbox[{\capbeside\thisfloatsetup{capbesideposition={right,center},capbesidewidth=6cm}}]{figure}[\FBwidth]
{\caption{\textbf{Interference fringes of two-mode cat states of larger size.} 2D plane-cuts of the scaled Wigner function $\langle P_J(\beta_A,\beta_B)\rangle$ along $\mathrm{Im}(\beta_A)-\mathrm{Im}(\beta_B)$ axes for two-mode cat states $|\psi_{-}\rangle$ with \textbf{(A)} $\alpha=2.7$ in Alice and 3.1 in Bob, and \textbf{(B)} $\alpha=3.0$ in Alice and 3.3 in Bob.}}
%{\includegraphics[scale=0.35]{fig_largeCat.png}}%[width=11.5cm]
\label{largecat}
\end{figure}

Via measurement of the joint photon number parity, we have inferred that the state $|\psi_{+}\rangle$ (or $|\psi_{-}\rangle$) has an even (or odd) number of photons in the two cavities combined.  Taking advantage of the fact that $\chi^{gf}_A\approx\chi^{gf}_B$ in our device, this joint parity property can be directly illustrated by measuring the statistical distribution of the total photon number in the two-mode cat state.  The measurement is realized by spectroscopic probe of the $|g\rangle\rightarrow|f\rangle$ two-photon transition (which is a second-order process occurring at much higher driving power than $|g\rangle\rightarrow|e\rangle$ or $|e\rangle\rightarrow|f\rangle$ transitions). As shown in Fig.~\ref{catspec} for the odd parity state $|\psi_{-}\rangle$ ($\alpha$=1.92), the $(|g\rangle\rightarrow|f\rangle)/2$ spectral line of the ancilla is split into multiple peaks corresponding to different cavity photon numbers, but only peaks associated with odd total number of photons are present.

\subsection{Two-mode Cat State of Larger Size}
The methods to generate and measure cat states shared in two cavities can in principle be applicable to arbitrary photon numbers.  We briefly measured the core features in the joint Wigner functions of 2-mode cat states with larger and generally different numbers of photons in Alice and Bob, \textit{i.~e.~}$\mathcal{N}\big(|\alpha_1\rangle_A|\alpha_2\rangle_B-|-\alpha_1\rangle_A|-\alpha_2\rangle_B\big)$. Fig.~S13 shows the Im$(\beta_A)$-Im$(\beta_B)$ plane-cuts of $\langle P(\beta_A,\beta_B)\rangle$ for two such states.  The number or the density of interference fringes increases with the total photon number, proportional to $\sqrt{\alpha_1^2+\alpha_2^2}$.  The largest state we have measured has a cat size of $S=(2\alpha_1)^2+(2\alpha_2)^2=80$ photons, limited mostly by room temperature electronics.  The contrast of the measured Wigner function decreases with increasing cat size, indicating lower fidelity in the prepared two-mode cat state and the joint parity measurement.  The decreased fidelity is due to a combination of stronger decoherence and ancilla rotation infidelity due to bandwidth constraints. 

\begin{figure}[tbp]
    \centering
    \includegraphics[scale=0.35]{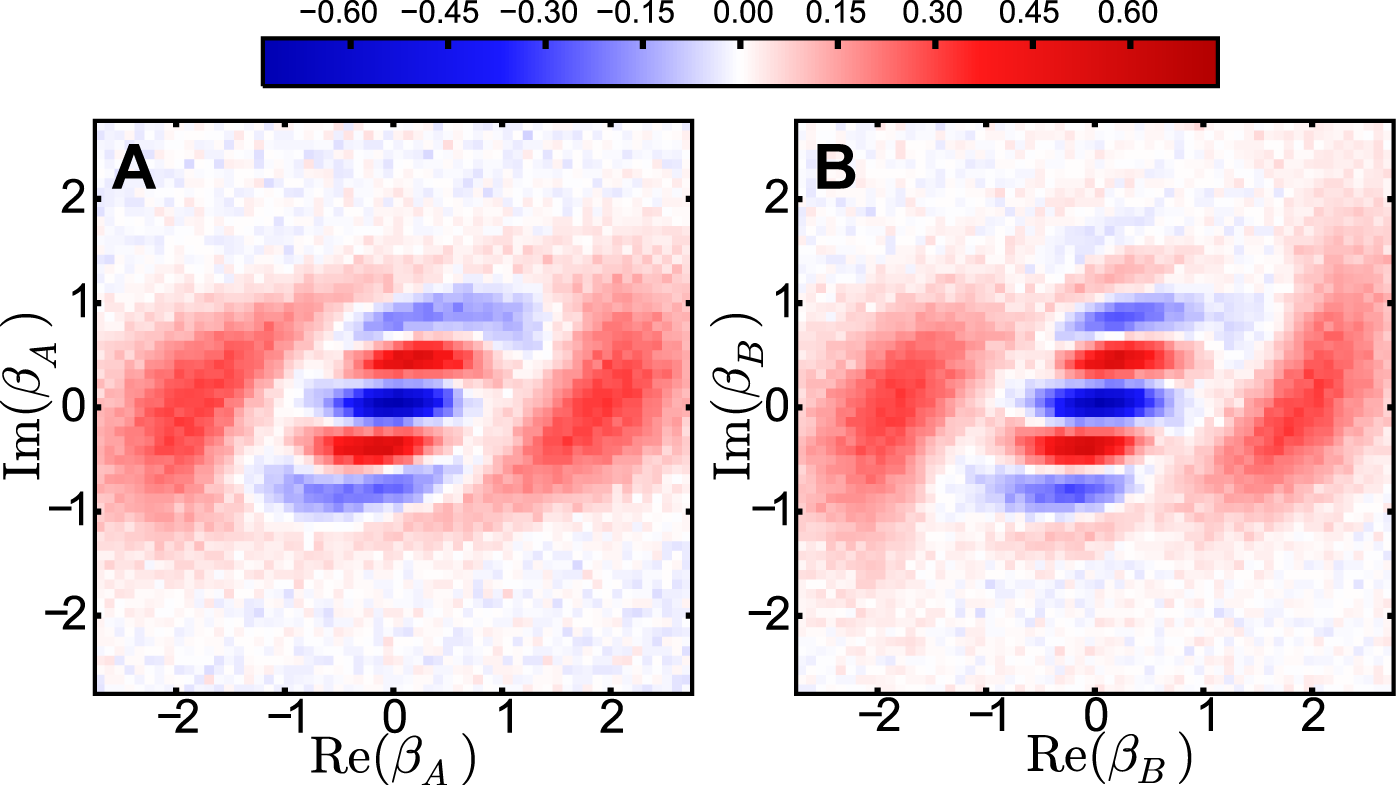}%[width=8.8cm]
    \caption{\textbf{Single-cavity Wigner tomography of the product cat state.} The scaled single-cavity Wigner function, $W_A(\beta_A)$ and $W_B(\beta_B)$, of the same quantum state presented in Fig.~4B of the main text.  This state is intended to be a product state of a cat state in Alice and a cat state in Bob, or $\mathcal{N'}(|\alpha\rangle_A-|-\alpha\rangle_A)\otimes(|\alpha\rangle_B-|-\alpha\rangle_B)$.}
    \label{productcat}
\end{figure}

\subsection{Extended Data for Product Cat State}

In an illustrative comparison with the two-mode cat state, in Fig.~4B of the main text we have shown joint tomography of an approximate product state of two independent cat states, $\mathcal{N'}(|\alpha\rangle_A-|-\alpha\rangle_A)\otimes(|\alpha\rangle_B-|-\alpha\rangle_B)$. 
We generate this ``product cat state" using post-selection after QND parity measurement:  A single-mode cat state in Bob is first created while Alice is in vacuum state (using the protocol described near the end of Section~ref{sec:qcmap}).  Then a coherent state is prepared in Alice, followed by a single-cavity parity measurement of Alice, $P_A$.  This measurement projects Alice to even or odd cat state ($\mathcal{N}(|\alpha\rangle_A\pm|-\alpha\rangle_A)$)~\cite{sun_tracking_2014} while the cat state in Bob stays intact.  We post-select odd parity from the outcome of $P_A$ measurement, obtaining the product cat state for subsequent tomography.  One can confirm the presence of independent single-mode cat states by performing Wigner tomography of individual cavities, $W_A(\beta_A)$ and $W_B(\beta_B)$ (Fig.~\ref{productcat}).  Indeed, the Wigner function of each cavity is similar to a cat state (\textit{i.e.}~Fig.~S7) containing two coherent state components with interference fringes in between.  This is in striking contrast to $W_A(\beta_A)$ and $W_B(\beta_B)$ of a two-mode (entangled) cat state (Fig.~2 of the main text), where each cavity, when analyzed on its own, only contains a statistical mixture of two coherent states.

Fig.~\ref{productcat} also shows that the coherent state components in the product cat state are significantly distorted.  This is due to the Kerr effects from the higher-order Hamiltonian terms that accumulates between the measurement-based state generation and the next measurement for tomography.  Our device parameters are not optimal for such repetitive measurements due to the relatively slow readout speed (unable to repeat faster than $\sim3$ $\mu$s, or about 10$\times$ the lifetime of the readout resonator).  It should be noted that it is possible to create product cat state deterministically (without reliance on readout) while compensating for Kerr effects using numerically optimized control pulses~\cite{khaneja_optimal_2005}.  In addition, future experiments can add separate ancillae coupled to Alice and Bob to further facilitate independent quantum operations of individual cavities.

\subsection{State Reconstruction}

Although features of the measured joint Wigner function can be compared intuitively with the ideal two-mode cat state, a full density matrix reconstruction is required to rigorously evaluate the fidelity of the quantum state.  We perform this state reconstruction using maximum likelihood estimation for an over-complete data set of $W_J(\beta_k^{(A)},\beta_k^{(B)})$ ($k\in\{1,\ldots,N_\text{disp}\}$) at $N_\text{disp}$ different sampling points of the 4D phase space.% with $\beta_k^{(A)}, \beta_k^{(A)} \leq 3$.

%Given a joint two-cavity density matrix $\rho$, and displacements $\beta_A$, $\beta_B$ for each of the cavities, the value of the joint Wigner function is the expected joint parity after displacing each cavity.

% In order to estimate this function, we first specify a set of displacements $\vec{\beta}^{(A)},\vec{\beta}^{(B)}\in\mathbb{C}^{N_\text{disp}}$ at which we sample the function value. Then, for each $k\in\{1,\ldots,N_\text{disp}\}$, we prepare $N_\text{rep}$ copies of our state, perform the displacements $D_k \equiv D_A(\beta^{(A)}_k) D_B(\beta^{(B)}_k)$, and measure $P_AP_B$.  Let $n_k$ then be the number of times we observed $P_A P_B=1$. It will be convenient to represent the measurement we perform as a POVM element $\Pi = (P_A P_B + 1)/2$ which has eigenvalue 1 when $P_A P_B = 1$ and eigenvalue 0 when $P_A P_B = -1$, and consequently whose expectation value can be interpreted as a probability.

For each point of the joint Wigner function, we prepare $N_\text{rep}$ copies of our state, perform the displacements
$D_k \equiv D_A(\beta^{(A)}_k) D_B(\beta^{(B)}_k)$, and measure the joint parity $P_J=P_A P_B$, so that:
\begin{align}
	W_J(\rho, \beta^{(A)}_k,\beta^{(B)}_k) &= \frac{4}{\pi^2}\Tr[
	\rho D_k P_J D_k^\dagger]
\end{align}
It is convenient to represent the measurement we perform as a POVM element $\Pi = (P_J + 1)/2$ which has eigenvalue 1 when $P_J = 1$ and 
eigenvalue 0 when $P_J = -1$, and consequently whose expectation value 
can be interpreted as the probability to observe even joint parity.

Note that these measurement outcomes can be written in two equivalent ways:
\begin{align}
	\Tr[\rho (D_k \Pi D_k^\dagger)] \equiv \Tr[\rho \Pi_k],
\end{align}
and 
\begin{align}
	\Tr[(D_k^\dagger \rho D_k) \Pi] \equiv \Tr[\rho_k \Pi].
\end{align}
This is to say that our experiment can be considered as a set of measurements
${\Pi_k}$ used to characterize a state $\rho$, or as a set of states $\rho_k$
used to characterize a measurement $\Pi$. Ideally, one would like to characterize
both $\Pi$ and $\rho$ simultaneously, since the expected infidelity in these
two operations are comparable. However, doing so would require a set of trusted
operations beyond just displacements, as well as a squaring in number of measurements required.  In light of this, we have performed state tomography
assuming that our measurement operator is as designed, and acknowledging that
the infidelity reported is a combination of the state preparation infidelity
and measurement operator infidelity.

Let $n_k$ be the number of times we observe $P_J=1$ at the $k$-th sampling point (out of a total of $N_\text{rep}$ repetitions). Our state reconstruction looks for the density matrix $\hat{\rho}_{ML}$ that maximizes the likelihood function:
\begin{align}
    \hat{\rho}_\text{ML} &= \underset{\rho}{\text{argmax}}\; \Lc(\rho)
\end{align}
Here the likelihood $\Lc(\rho)$ is the probability of seeing the data $n_k$ assuming
$\rho$. For a fixed $k$, $n_k$ should follow a binomial distribution:
\begin{align}
    \Lc(\rho)
    &= \prod_k P(n_k|\rho)\nonumber\\
    &= \prod_k {{N_\text{rep}}\choose{n_k}}(p_k(\rho))^{n_k}(1-p_k(\rho))^{N_\text{rep} - n_k}
\end{align}
$p_k$ is computed from the joint Wigner of $\rho$:
\begin{align}
    p_k(\rho) &= \Tr[\rho_k \Pi]\nonumber\\
    &= \frac{1}{2}(\Tr[\rho_k P_AP_B] + 1)\nonumber\\
    &= \frac{1}{2}\left(\frac{\pi^2}{4}W_J(\rho,\beta_{k_A},\beta_{k_B}) + 1\right)
\end{align}

What remains is to find an efficient method of calculating
$W_J(\rho, \beta_A, \beta_B)$. To do so, write down $\rho$
in the tensor Fock state basis, truncated to some maximum photon number $N_\text{cutoff}$:

\begin{align}
    \rho = \sum_{i,j,k,l=1}^{N_\text{cutoff}} \rho_{ijkl} \ket{ij}\bra{kl}
\end{align}

Next use the linearity of $W_J$ in $\rho$ to identify the contribution
from each component $\rho_{ijkl}$:

\begin{align}
&W_J(\rho,\beta^{(A)}_k,\beta^{(B)}_k) = \frac{4}{\pi^2}\Tr[\rho D_k P_A P_B D_k^\dagger]\nonumber\\
&= \frac{4}{\pi^2}\sum_{ijmn} \rho_{ijmn} \bra{mn}D_k P_A P_B D_k^\dagger\ket{ij}\nonumber\\
&= \frac{4}{\pi^2}\sum_{ijmn} \rho_{ijmn} K_{mi}(\beta^{(A)}_k) K_{nj}(\beta^{(B)}_k),
\end{align}

We can compute the matrix elements $K_{mn}(\beta) \equiv \bra{m} D(\beta) P D(\beta)^\dagger\ket{n}$ in
the same way one would in standard Wigner state tomography~\cite{leonhardt_1997}, i.e.
\begin{align}
        K_{mn}(\beta) &\equiv\bra{m} D(\beta) P D(\beta)^\dagger\ket{n} \nonumber\\
        &= e^{-|\beta|^2} (-1)^m (2\beta) ^ {(n - m)}
        \sqrt{\frac{m!}{n!}} L_m^{(n-m)}(|\beta|)
\end{align}
where $L_m^{(n-m)}$ is a generalized Laguerre polynomial.

In order to be a physical solution, $\rho$ must be positive semidefinite with
$\Tr[\rho] = 1$. To account for this, we adjust the optimization problem

\begin{align}
    \hat{\rho}_\text{ML} &= \hat{A}\hat{A}^\dagger\nonumber\\
    \hat{A} &= \underset{A}{\text{argmax}}\;f(A)\nonumber\\
    f(A) &= \ln\Lc(A A^\dagger) - \lambda(\Tr[A A^\dagger] - 1)^2
\end{align}
where now $A$ can be any complex matrix, and $\lambda$ is a Lagrange multiplier
whose value must be greater than some threshold in order for the trace constraint
to be satisfied. In practice, the value of $\lambda$ can simply be increased
until the deviation of the trace from unity is sufficiently small.

To solve the optimization problem for the two-mode cat state, $|\psi_{+}\rangle$, we first specify the photon number truncation
$N_\text{cutoff}$.  Based on the expected state with $\alpha \approx 1.92$,
$\bar{n} = |\alpha|^2 \approx 3.69$, according to Poissonian statistics we must
use $N_\text{cutoff}>10$ to make the probability of having more than
$N_\text{cutoff}$ photons less than $0.1\%$. In practice, we take
$N_\text{cutoff}=12$, resulting in a system dimension $d=N_\text{cutoff}^2 =
144$, and $d^2 = 20736$ real parameters in the density matrix.  (Note we have measured at $N_\text{disp}=155,600$ different sampling points to form an over-complete data set for the reconstruction.)  Because the number
of parameters is so large, it is necessary to compute the gradient
$\frac{\partial f}{\partial A_{ij}}$ of the cost function with respect to the parameters, and to use a gradient-aware optimization routine, such as the BFGS algorithm.\cite{}

After performing the reconstruction, we can extract several metrics about the state.
The largest pure state overlap is given by the largest eigenvalue
$\lambda_\text{max}(\rho) \approx 0.824$.  The purity is $\Tr[\rho^2] \approx
0.68$.  The state of form $\ket{\psi} = \ket{\alpha,\alpha} +
\ket{-\alpha,-\alpha}$ with highest fidelity is $\alpha=1.903$, with
$\bra{\psi}\rho\ket{\psi}\approx 0.803$. The state of form $\ket{\psi} = \ket{\alpha_A,\alpha_B}
+ e^{i\phi}\ket{-\alpha_A,-\alpha_B}$ with highest fidelity is 
$\alpha_A = 1.881$, $\alpha_B = 1.922$, $\phi=-0.1$ with $\bra{\psi}\rho\ket{\psi}\approx0.805$.
Given the measured joint parity of 0.81 for the two-mode cat state, the highest possible fidelity one could expect is approximately 0.9, which would arise from a 90\%-10\% mixture 
of the ideal target state and a state of opposite parity (such as produced by a single-photon loss). Our reconstruction finds the parity of the dominant eigenvector to be 0.97, and the parity of the next few eigenvectors to be small and positive. This indicates that single-photon loss is not the dominant error mechanism affectubg the state generation, as will be discussed in Section~\ref{sec:fidelity}.

\subsection{Parity Decay}

\begin{figure}[tbp]
    \centering
    \includegraphics[scale=0.3]{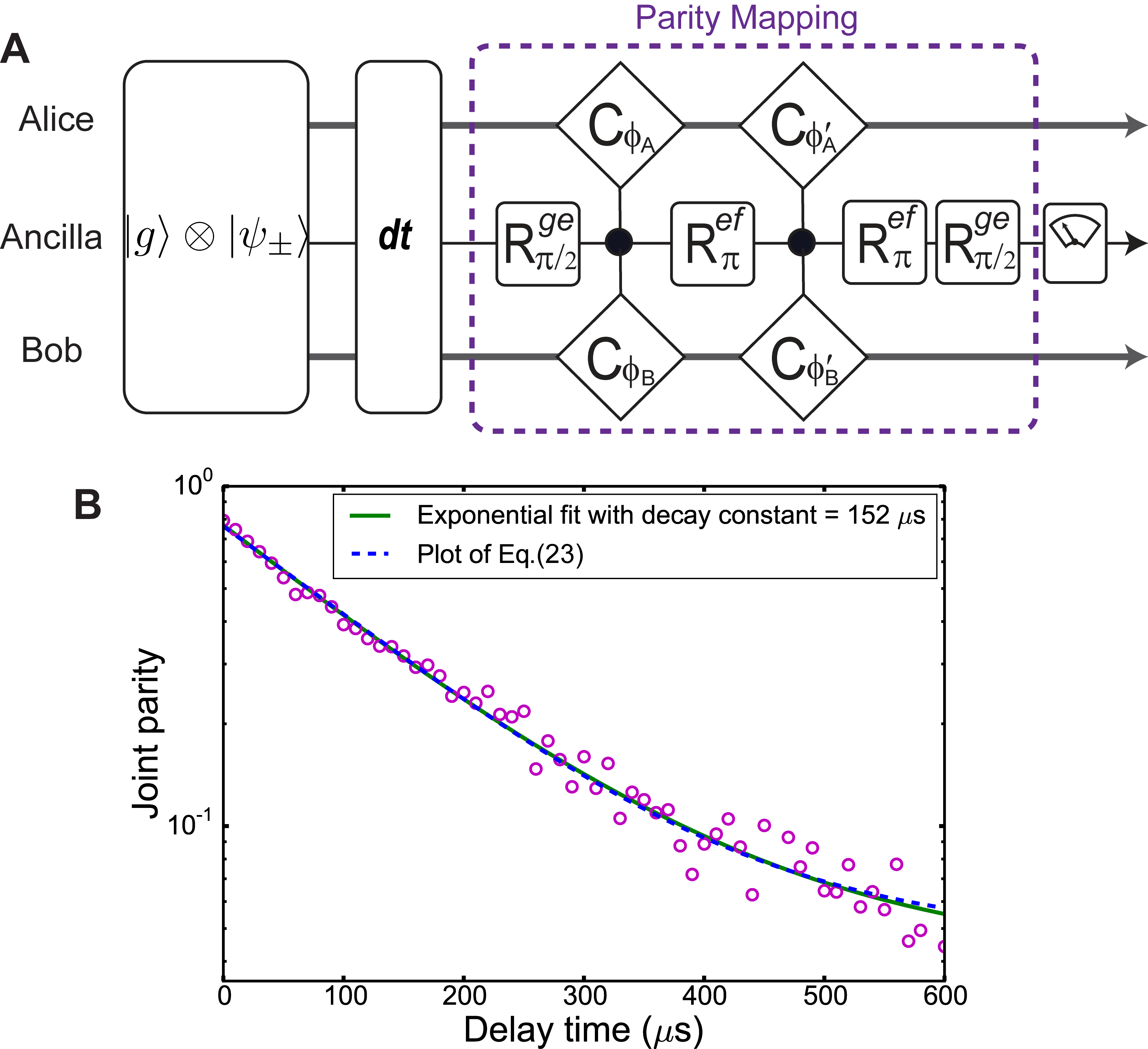}%[width=8.5cm]
    \caption{\textbf{Joint parity decay.} \textbf{(A)} Measurement sequence used to monitor the decay of joint parity. After preparing the two-mode cat state, a variable delay is implemented before reading out the joint parity at the origin. \textbf{(B)} The measured joint parity at the origin as a function of the variable delay time. Green curve is a fit to a simple exponential decay, which gives a decay constant of 152 $\mu$s. Blue dashed line shows the plot according to the functional form described in Eq.~(\ref{eq:paritydecay}) using $\alpha = 1.92$, $\tau_A$ = 2.6 ms, and $\tau_B$ = 1.5 ms.}
    \label{paritydecay}
\end{figure}

We have briefly studied the decoherence of the two-mode cat state by measuring the decay of the joint photon number parity over time (Fig.~\ref{paritydecay}).  This is a convenient method without performing full tomography to gain insight into the major decoherence mechanism that occurs after the two-mode cat state has been created: photon loss in either of the two cavities.  The observed decay of the joint parity is consistent with the combined photon loss in the two high-Q superconducting cavities:
\begin{equation}\label{eq:paritydecay}
P_J(t) = P_J(0) \exp{[-2\alpha^2(2-e^{-t/\tau_A}-e^{-t/\tau_B})]}
\end{equation}
The parity decay takes the form of an ``exponential with exponential" because the rate of parity decay is proportional to the photon numbers and therefore decreases over time.  This equation is only applicable when the total photon number is far from zero.  (The parity eventually approaches +1 as the cavities decay to the vacuum state.)

Since the parity initially decays at twice the total photon loss rate, it is still informative to consider a characteristic time of parity decay during a time span much shorter than the cavity lifetimes so that the cavity photon numbers are approximately constant.  We observe a decay time of about 150 $\mu$s for the state $|\psi_+\rangle$ with $\alpha=1.92$ based on a single exponential fit of the parity decay.  This can be well explained by taking the (average) measured cavity lifetimes $\tau_A=2.6$ ms, $\tau_B=1.5$ ms and considering the average photon number over a span of 600 $\mu$s after the initial state generation $\bar{N}_A\approx 3.3$, $\bar{N}_B\approx 3.0$, so that $2(\frac{\bar{N}_A}{\tau_{A}}+\frac{\bar{N}_B}{\tau_{B}})\approx 1/(150$ $\mu$s).  

The parity decay measurement is not sensitive to other decoherence processes such as cavity frequency shifts due to transmon thermal jumps ($|g\rangle\rightarrow|e\rangle$ and subsequent $|e\rangle\rightarrow|g\rangle$ jumps).  However, this effect is expected to induce a cavity dephasing with a relatively long time constant of about 900 $\mu$s, and should be further improved with better thermalization of the device.

It is worth noting that the coherence time of this complex two-mode cat state (at $\alpha=1.92$ with a cat size of 30 photons) is longer than the most coherent superconducting qubit reported so far~\cite{rigetti_superconducting_2012}, owing to the superior coherence property of the 3D cavities.  This illustrates an important advantage in using the cavity states as quantum memories in cQED~\cite{reagor_quantum_2015} \cite{Chuang_1997} or as logical qubits~\cite{mirrahimi_dynamically_2014} in addition to the potential simplification of error correction operations.

\subsection{Error Sources}
\label{sec:fidelity}

\begin{table*}[btp]
\caption{Estimated contribution to the loss of contrast in the measured joint parity of the two-mode cat state, $|\psi_{\pm}\rangle$ ($\alpha=1.92$), from various error sources.}
\centering
\begin{tabular}{c c c c c} % centered columns (4 columns)
\hline\hline\\[-2ex]
		& Assessment	&  Estimated infidelity	 \\
\hline\\[-2ex]
ancilla initialization  & $\sim0.5\%$ probability not in $|g\rangle$	& $\sim1\%$\\
cavity initialization	& $\sim0.5\%$ probability not in $|0\rangle_A|0\rangle_B$	& $\sim1\%$\\
readout infidelity		& 1.0-1.5\% error rate				& $2.5\%$\\
ancilla decoherence in state generation	& $|g\rangle$-$|e\rangle$ superposition for 0.65 $\mu$s & 2.2\%\\
pulse error in state generation	& imperfect spectral selectivity of $R_\pi^{00}$	& $\sim1\%$\\
ancilla decoherence in parity mapping	& $|g\rangle$-$|e\rangle$ \& $|g\rangle$-$|f\rangle$ superposition for 0.25 $\mu$s	& 2.2\%\\
timing (phase) error in parity mapping	& $\pm3\%$ phase error in $C_\pi^A$ and $C_\pi^B$	&$\sim3\%$ 	\\
pulse error in parity mapping	& population mixing in $|g\rangle$-$|e\rangle$-$|f\rangle$ rotations & $\sim5\%$	\\
photon loss	in two cavities & 3.7-7.3 photons in each cavity for 0.9 $\mu$s &
0.9\% \\
\hline\\[-2ex]
Total	&	&	$\sim19\%$\\[0.3ex]
% [1ex] adds vertical space
\hline
\end{tabular}
\end{table*}

The fidelity of the generation and measurement of cat states is limited by various factors as listed in Table S3.  Because the joint parity of the two-mode cat state is the simplest figure of merit for evaluating the overall fidelity, we further focus on analyzing the errors contributing to the loss of contrast in the measured $P_J=\pm0.81$ for $|\psi_{\pm}\rangle$ with $\alpha=1.92$ (compared with the ideal value of $P_J = \pm1$).  The measuredvalue corresponds to a single point in the scaled joint Wigner function at the origin, $(\pi^2/4)W_J(\beta_A=0,\beta_B=0)$, but is the most representative point.  We note again that there are mechanisms affecting the fidelity of the entire quantum state without directly contributing to the measured parity of the cat state, most notably the Kerr effects.  However, via density matrix reconstruction we found that the fidelity of the full state is also close to 81\%, indicating that the most significant errors can be understood by analyzing $P_J$ of the state alone.

Many contributions to the loss of parity contrast can be estimated from system parameters and tested by controlled experiments.  Contributions from ancilla and cavity decoherence are estimated to be about 5\% in total from their respective coherence times and the gate times.  Contrast loss in measured parity due to infidelity of the single-shot readout is twice the readout error rate and determined to be about 2.5\%.  We estimate about 2\% loss of parity contrast due to state initialization errosr (experiments starting not from $|g\rangle|0\rangle_A|0\rangle_B$), which is primarily due to $|g\rangle\rightarrow|e\rangle$ thermal transition of the ancilla during the relatively slow initial state purification protocol.  These estimates are consistent with measurements of ancilla Rabi oscillations between ($|g\rangle$ and $|e\rangle$) (96-97\% of full contrast) and the parity of single-cavity cat states in either Alice or Bob ($\sim$90\% of full contrast).

Additional infidelity arises in measurement of the two-mode cat state in our experiment, which can be primarily attributed to imperfections associated with mapping the joint photon number parity to the ancilla state.  One source of error as discussed in Section~\ref{sec:paritymapping} is the non-ideal waiting time $\Delta t_1$ and $\Delta t_2$ used in the controlled-phase gate, where Alice and Bob acquire phases different from $\pi$ by $\pm3\%$, causing a joint parity measurement infidelity of about 3\%.  

Another major source of error is from the bandwidth constraint on the ancilla pulses.  Ideally, the ancilla operations described in our joint parity mapping (Fig.~\ref{paritymapping}) require both infinite bandwidth (in order to be completely independent of cavity photon numbers) and no spectral overlap with unwanted ancilla transitions (in order to minimize state leakage out of the intended ancilla levels), which are conflicting requirements.  We have used Gaussian pulses with $\sigma_\omega=2\pi\cdot40$ MHz and duration of 16 ns for ancilla rotations as a compromise.  The influence of multiple levels of a transmon under fast microwave drive has been studied before, but mostly limited to $|g\rangle$-$|e\rangle$ operations, where state leakage to $|f\rangle$ is a second-order effect on the computation (proportional to leakage population, or amplitude squared).  In our control pulses, we have implemented derivative removal via adiabatic gate (DRAG)~\cite{motzoi_simple_2009} developed for two-level qubits to correct for the presence of the third level.  However, it is known that the standard DRAG technique does not fully address state leakage~\cite{chen_measuring_2016}, which has a first-order effect to our joint parity mapping utilizing three computational  levels.  Furthermore, high-fidelity rotations in the $|e\rangle$-$|f\rangle$ space require correction for the presence of both $|g\rangle$ and $|h\rangle$ (the fourth transmon level) not yet considered in the literature.  Last but not least, optimal ancilla rotations in the presence of cavity photons also remain to be developed.  We attribute the unaccounted loss of contrast in $P_J$ (about 5\%) to such control pulse errors in parity mapping.  This is consistent with a controlled test that measures Ramsey interference of the ancilla $|g\rangle$-$|f\rangle$ superposition in the presence of cavity coherent states.  This experiment uses very similar pulse sequences to joint parity mapping but does not incur errors associated with cat state generation, parity mapping phase and cavity initialization errors, and shows a contrast of 89\%.  

Based on these semi-quantitative analyses of error contributions listed in Table S3, one can further categorize the total 19\% loss of contrast in the joint parity to be about 6\% due to imperfection of the state preparation and 13\% from the infidelity (or loss of visibility) of the joint parity measurement.  Because there is no simple way to independently determine the visibility of the parity measurement (which is photon-number-dependent due to pulse errors), we do not attempt to draw quantitative conclusions as to the quantum state fidelity.  %As an estimate, there may be 4-5\% infidelity due to decoherence and initial thermal population (note the thermal population and cavity photon loss affects the parity contrast 2$\times$ as much as the state fidelity).
However, one can find evidence in spectroscopy of the two-mode cat state (Fig.~\ref{catspec}) that the magnitude of joint parity is indeed higher than 0.9.

\subsection{Bell's Inequality}

\begin{figure}[tbp]
    \centering
    \includegraphics[scale=0.38]{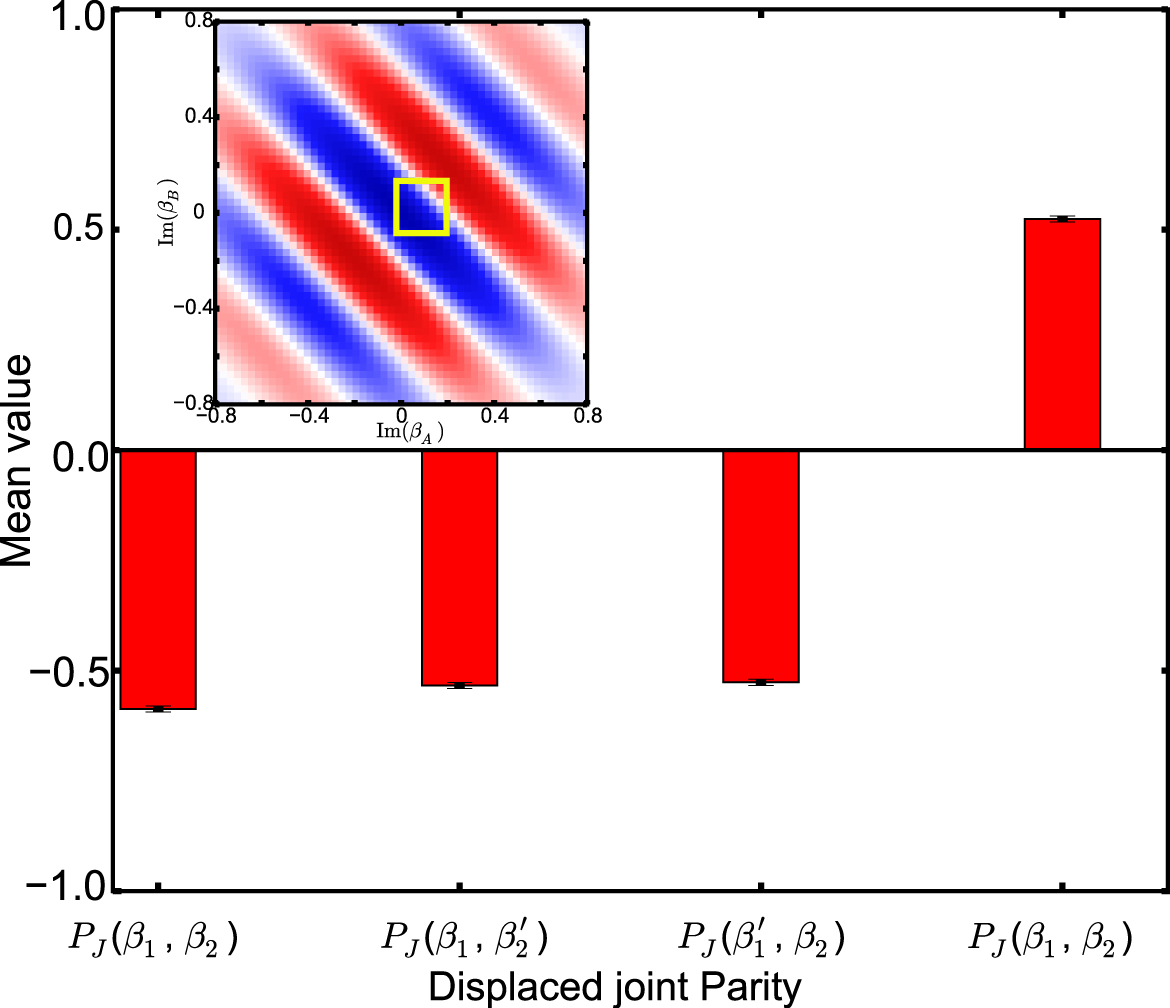}%[width=7.5cm]
    \caption{\textbf{Bell signal of an odd two-mode cat state.} Red bars show the amplitude of scaled joint Wigner function at the selected sampling points in the phase space defined by the imaginary amplitudes of $\beta_A, \beta_B$. Inset shows the location of the chosen sampling points in phase space}
    \label{fig:bell}
\end{figure}

The two-mode cat state is a quantum mechanical system consisting of two continuous-variable subsystems.  For two continuous-variable systems, the correlation between their individual parities after independent displacements has a classical upper bound, which can be described by a CHSH-type Bell's inequality using the formulation proposed in Ref.~\citen{banaszek_testing_1999} and discussed in Ref.~\citen{milman_proposal_2005}. Arbitrarily choosing two test displacements $\beta_A,\beta'_A$ in Alice and two test displacements $\beta_B,\beta'_B$ in Bob, the Bell signal $\mathcal{B}$ can be constructed from parity correlations after all four combinations of displacements in the two cavities:
\begin{align}
\mathcal{B} = &\big|\langle P_A(\beta_A)P_B(\beta_B)\rangle+ \langle P_A(\beta'_A)P_B(\beta_B)\rangle \nonumber\\ 
&+ \langle P_A(\beta_A)P_B(\beta'_B)\rangle-\langle P_A(\beta'_A)P_B(\beta'_B)\rangle \big|\leq 2 
\end{align}
where $P_i(\beta)\equiv D_{\beta_i}P_i D^{\dagger}_{\beta_i}$ ($i=$A, B) is the displaced parity operator.  Here measuring parity after different displacements is analogous to measuring $\sigma_z$ of a spin-$\frac{1}{2}$ system after different rotations.

Equivalently, this Bell signal is represented by the values of joint Wigner function (or displaced joint parity) at the four vertices of a rectangle:
\begin{align}
\mathcal{B}=\frac{\pi^2}{4}\big|W_J(\beta_A, \beta_B)+ W_J(\beta'_A, \beta_B)+ W_J(\beta_A, \beta'_B) \nonumber\\-W_J(\beta'_A, \beta'_B)\big|\leq 2
\end{align}

For a quantum state with entanglement between the two subsystems, this Bell's inequality can be violated.  For near-optimal violation, we choose a square in the Im$(\beta_A)$-Im$(\beta_B)$ plane with prominent interference fringes.  The square is positioned to have three of the vertices close to the minimum of the central negative fringe and one in the vicinity of the maximum of the adjacent positive fringe, using $\beta_A = \beta_B = -i\pi/(16\beta)=-0.102$ and $\beta'_A = \beta'_B = 3i\pi/(16\beta)=0.307$ (Fig.~\ref{fig:bell}).  Given these sampling points, the measured amplitude gives a Bell signal $\mathcal{B} = 2.17 \pm 0.01$, surpassing the classical threshold by more than 10 standard deviations.  This indicates the non-classical nature of the two-mode cat state and the presence of quantum correlations between the two modes.  This also demonstrates the robustness of our experimental technique in both the creation of the quantum mechanical two-mode cat state as well as the joint parity measurement procedure.

\subsection{Encoded Two-qubit Tomography}

Complete joint-Wigner tomography of the two-cavity quantum state requires large numbers of measurements.  However, if we are restricted to a particular coherent state basis of the two cavities to encode two logical qubits (as described in the main text), efficient Pauli tomography can be performed for this logical subspace with a total of only 16 measurements.  

Using the encoding scheme $|\alpha\rangle_i\rightarrow|0\rangle_j$ and $|-\alpha\rangle_j\rightarrow|1\rangle_i$ (i, j= A or B), the single-qubit Pauli operators are:
\begin{align}
X_j &=|-\alpha\rangle_j\langle\alpha|_j+|\alpha\rangle_j\langle-\alpha|_j\nonumber\\
Y_j &= i|-\alpha\rangle_j\langle\alpha|_j-j|\alpha\rangle_j\langle-\alpha|_j\nonumber\\
Z_j &=|\alpha\rangle_j\langle\alpha|_j-|-\alpha\rangle_j\langle-\alpha|_j\nonumber\\
I_j &=|\alpha\rangle_j\langle\alpha|_j+|-\alpha\rangle_j\langle-\alpha|_j
\end{align}
Following the derivation in Ref.~\citen{vlastakis_characterizing_2015}, these single-qubit operators can be linked to the displaced parity operators of a cavity ($P_j(\beta)\equiv D_{\beta_j}P_j D^{\dagger}_{\beta_j}$):
\begin{align}
X_j &\approx P_j(0)\nonumber\\
Y_j &\approx P_j\big(\frac{j\pi}{8\alpha}\big) \nonumber\\
Z_j &\approx P_j(\alpha)-P_j(-\alpha)\nonumber\\
I_j &\approx P_j(\alpha)+P_j(-\alpha)
\end{align}
These relations can be verified by projecting the displaced parity operators onto the encoded subspace using the projector $M_j=|\alpha\rangle_j\langle\alpha|_j+|-\alpha\rangle_j\langle-\alpha|_j$ (for more details, see supplementary notes in Ref.~\citen{vlastakis_characterizing_2015}). 

The 16 two-qubit observables are products of single-qubit Pauli operators, and can all be expressed in the form of displaced joint parities.  Since operators in different cavities commute,
\begin{equation}
P_A(\beta_A)P_B(\beta_B)=D_{\beta_A} D_{\beta_B} P_J D^{\dagger}_{\beta_A} D^{\dagger}_{\beta_B} \equiv P_J(\beta_A,\beta_B)
\end{equation}
we have,
\begin{align}
&I_A I_B =P_J(\alpha,\alpha)+P_J(\alpha,-\alpha)+P_J(-\alpha,\alpha)+P_J(-\alpha,-\alpha)\nonumber\\
&I_A X_B =P_J(\alpha,0)+P_J(-\alpha,0)\nonumber\\
&I_A Y_B =P_J\big(\alpha,\frac{i\pi}{8\alpha}\big)+P_J\big(-\alpha,\frac{i\pi}{8\alpha}\big)\nonumber\\
&I_A Z_B =P_J(\alpha,\alpha)-P_J(\alpha,-\alpha)+P_J(-\alpha,\alpha)-P_J(-\alpha,-\alpha)\nonumber
\end{align}
\begin{align}
&X_A I_B =P_J(0,\alpha,)+P_J(0,-\alpha)\nonumber\\
&Y_A I_B =P_J\big(\frac{i\pi}{8\alpha},\alpha\big)+P_J\big(\frac{i\pi}{8\alpha},-\alpha\big)\nonumber\\
&Z_A I_B =P_J(\alpha,\alpha)+P_J(\alpha,-\alpha)-P_J(-\alpha,\alpha)-P_J(-\alpha,-\alpha)\nonumber\\
&X_A X_B =P_J(0,0)\nonumber\\
&X_A Y_B =P_J\big(0,\frac{i\pi}{8\alpha}\big)\nonumber\\
&X_A Z_B =P_J(0,\alpha)-P_J(0,-\alpha)\nonumber\\
&Y_A X_B =P_J\big(\frac{i\pi}{8\alpha},0\big)\nonumber\\
&Y_A Y_B =P_J\big(\frac{i\pi}{8\alpha},\frac{i\pi}{8\alpha}\big) \nonumber\\
&Y_A Z_B =P_J\big(\frac{i\pi}{8\alpha},\alpha\big)-P_J\big(\frac{i\pi}{8\alpha},-\alpha\big)\nonumber\\
&Z_A X_B =P_J(\alpha,0)-P_J(-\alpha,0)\nonumber\\
&Z_A Y_B =P_J\big(\alpha,\frac{i\pi}{8\alpha}\big)-P_J\big(-\alpha,\frac{i\pi}{8\alpha}\big)\nonumber\\
&Z_A Z_B =P_J(\alpha,\alpha)-P_J(\alpha,-\alpha)-P_J(-\alpha,\alpha)+P_J(-\alpha,-\alpha)\nonumber\\
\label{eq:pauli}
\end{align}
To obtain a full set of two-qubit tomography measurements, we perform joint parity measurements following 16 different cavity displacement combinations ($\beta=0$, $\frac{i\pi}{8\alpha}$, $\alpha$ and $-\alpha$ in each cavity), and the 16 two-qubit Pauli operators can be computed from Eq.~(\ref{eq:pauli}). 

\bibliography{Zotero}